\documentclass[12pt,a4paper]{article}
\pdfoutput=1

\usepackage[utf8]{inputenc}
\usepackage[english]{babel}
\usepackage[T1]{fontenc}
\usepackage{newtxtext}

\usepackage{amsmath}
\usepackage{amsfonts}
\usepackage{amssymb}
\usepackage{mathtools}
\usepackage{newtxmath}
\usepackage{bbm}

\usepackage{graphicx}
\usepackage[margin=1in]{geometry}
\usepackage{textcase}
\usepackage[usenames,dvipsnames]{xcolor}

\usepackage[]{longtable}
\usepackage{tabu}
\usepackage{hhline}
\usepackage{makecell}

\usepackage[tableposition=top,figureposition=bottom,font=footnotesize, labelsep=endash, skip=3pt]{caption}
\usepackage[font=footnotesize, labelfont=bf, labelsep=endash]{caption}
\usepackage[export]{adjustbox}  % enables valign option in \includegraphics
\usepackage{standalone}		% load after adjustbox
\usepackage{tikz}
\usepackage{tikz-3dplot}
\usetikzlibrary{backgrounds,plotmarks}
\usetikzlibrary{arrows.meta}
\usetikzlibrary{patterns}
\usetikzlibrary{decorations.pathmorphing}
\usetikzlibrary{calc}
\tdplotsetmaincoords{60}{110}

\usepackage{varioref}
\usepackage{hyperref}
\usepackage{cleveref}
\crefname{SI-sec}{SI section}{SI sections}
\crefname{SI-subsec}{SI subsection}{SI subsections}
\crefname{app}{appendix}{appendices}
\crefname{subapp}{sub-appendix}{sub-appendices}

\usepackage[backend=biber, doi=false, url=false, giveninits=true, isbn=false, maxbibnames=99, sorting=none, sortcites=true, style=nature]{biblatex}
\usepackage{etoc}
\usepackage[title,titletoc]{appendix}

\usepackage[version=3]{mhchem}  % chemistry
\usepackage{siunitx}  		% for units
\usepackage{pdflscape}
\usepackage{afterpage}
\usepackage{caption}
\usepackage{titling}
\usepackage{needspace}

\newcommand{\proba}[1]{\mathrm{Pr} \left(#1\right)}
\newcommand{\ud}[1]{\mathrm{d} #1 \,}

\renewcommand{\vec}[1]{\mathbf{#1}}

\providecommand{\myref}[1]{\textbf{(\uppercase{#1})}}
\crefname{equation}{Eq.}{Eqs.}
\crefname{figure}{Fig.}{Figs.}
\crefname{table}{Tab.}{Tabs.}
\creflabelformat{equation}{#2#1#3}
\crefrangelabelformat{equation}{#3#1#4 to #5#2#6}
\providecommand{\keywords}[1]{\textbf{\textit{Keywords:}} #1}

\begingroup
\newcommand*{\DotsAndPage}
{\nobreak\leaders\hbox{\bfseries\normalsize\hbox to .75ex {\hss.\hss}}%
\hfill\nobreak
\makebox[\rightskip][r]{\bfseries\normalsize\etocpage}\par}

\graphicspath{{figures/}}

\title{A polymer model for the quantitative reconstruction of chromosome architecture from Hi-C and GAM data}

\date{}

\author{Guillaume Le Treut \\ {\normalsize \textit{Department of Physics, University of California San Diego, La Jolla, California 92093, USA.}} \and Fran\c{c}ois K\'ep\`es \thanks{Currently at Synovance, \'Evry, France.} \\ {\normalsize \textit{institute of Systems and Synthetic Biology, Genopole, CNRS,}} \\ {\normalsize \indent \textit{UEVE, Universit\'e Paris-Saclay, F-91030 \'Evry, France.}} \and Henri Orland \\ {\normalsize \textit{Institut de Physique Th\'eorique, CEA, CNRS-URA 2306, F-91191, Gif-sur-Yvette, France.}} \\[0.5em] {\normalsize \textit{Beijing Computational Science Research Center, No. 10 East Xibeiwang Road,}} \\{\normalsize \textit{Beijing 100193, China.}}}

\begin{document}
\maketitle

%abstract
\begin{abstract}
It is widely believed that the folding of the chromosome in the nucleus has a major effect on genetic expression. For example co-regulated genes in several species have been shown to colocalize in space despite being far away on the DNA sequence. In this manuscript, we present a new method to model the three-dimensional structure of the chromosome in live cells, based on DNA-DNA interactions measured in high-throughput chromosome conformation capture experiments (Hi-C) and genome architecture mapping experiments (GAM). Our approach incorporates a polymer model, and directly uses the contact probabilities measured in Hi-C and GAM experiments rather than estimates of average distances between genomic loci. Specifically, we model the chromosome as a Gaussian polymer with harmonic interactions and extract the coupling coefficients best reproducing the experimental contact probabilities. In contrast to existing methods, we give an exact expression of the contact probabilities at thermodynamic equilibrium. The Gaussian effective model (GEM) reconstructed with our method reproduces experimental contacts with high accuracy. We also show how Brownian Dynamics simulations of our reconstructed GEM can be used to study chromatin organization, and possibly give some clue about its dynamics.
\end{abstract}

.

% main text
\clearpage
\newpage
\renewcommand\thesection{\Roman{section}}
\newrefsegment %starting a segment for biblatex
%\newrefsection
\etocdepthtag.toc{tmain}
\section{Introduction}
While the chromosome has been classically seen as the carrier of the genetic information, there has been increasing evidence that its folding is a determinant of genetic regulation \cite{kepes2003transcriptionbased,junier2010spatial}. In particular, co-expressed genes were found to be more often in contact than unrelated genes \cite{Spilianakis2005,Llopis2010,Schoenfelder2010}, and the epigenetic state of the chromatin was shown to be related to its folding \cite{boettiger2016super}.
The advent of chromosome conformation capture (3C) experiments has provided unprecedented insights on chromosome architecture in live cells \cite{dekker2013exploring}, and the combination of 3C techniques with high-throughput sequencing methods has enabled the measurement of contacts between thousands of loci on the chromosome. Extensive high-throughput chromosome conformation capture experiments (Hi-C) data have now been generated for several eukaryotic cells including human \cite{liebermanaiden2009comprehensive,rao2014a}, yeast \cite{duan2010a}, fly \cite{sexton2012threedimensional}, but also bacteria \cite{umbarger2011the,cagliero2013genome,marbouty2015condensin}. In eukaryotes, the patterns observed in contact matrices generated from Hi-C experiments have revealed a high-level organization in sub-megabase-pair topologically associated domains (TADs) \cite{dixon2012topological,olivareschauvet2016capturing}. This organization displays significant changes throughout the cell cycle \cite{nagano2017cellcycle}, but also during cell differentiation \cite{fraser2015hierarchical} and in the context of cell pluripotency \cite{sexton2013the} or cell senescence \cite{chandra2015global}. More recently, the genome architecture mapping (GAM) technique was developed, representing an alternative way to measure interactions between chromosomal loci \cite{beagrie2017complex}. Its application to mouse embryonic stem cells confirmed that actively transcribed genes sometimes separated by large genomic distances were more often in contact. Based on these experimental findings several studies have suggested that chromosome architecture and genetic expression are intimately connected \cite{cavalli2007chromosome,ba2010the,nora2012spatial,stefano2013colocalization,jost2014modeling,stefano2016hicconstrained,soleroliva2017analysis}.

Several methods have been proposed to reconstruct the chromosome folding from Hi-C data (see \cref{sec:review_methods} in the Supplementary Information for a short review). A first class of models aimed at reconstructing chromosome configurations such that the distances $d_{ij}$ between chromosomal loci take prescribed values, inferred from the Hi-C contacts probabilities $c_{ij}$ \cite{duan2010a,umbarger2011the,ba2012genome,lesne20143d,wang2015inferential}. Those studies generally assumed that these average distances would scale like $d_{ij} \sim 1 / c_{ij}$. Yet a scaling analysis tells us that $d_{ij} \sim c_{ij}^{-\gamma}$, with $\gamma = 0.3$ for a self-avoiding chain (see \cref{sec:scaling_relation_cij_dij} in the Supplementary Information).
Another class of models aimed at finding an ensemble of chromosome configurations which reproduces the experimental contact probabilities, $c_{ij}^{exp}$ \cite{varoquaux2014a,tjong2016populationbased}. Yet most of these methods did not incorporate a realistic polymer model of the chromosome. Thus the configurations obtained may violate topological constraints imposed by the chain structure of the chromosome.

Here, we model the chromosome as a Gaussian polymer and introduce harmonic interactions to constrain its folding (see \cref{fig:gem}). The rigidity of these interactions will be determined by the cross-linking frequency between pairs of genomic loci obtained from the Hi-C protocol. This defines our Gaussian effective model (GEM). The inverse problem to solve consists in finding the effective couplings such that the contact probabilities of the model, $c_{ij}$, reproduce the contact probabilities obtained from a Hi-C experiment, $c_{ij}^{exp}$, similarly to previous studies \cite{giorgetti2014predictive,meluzzi2013recovering,chiariello2016polymer}. Yet in those methods, the contact probabilities of the model could only be computed through Monte-Carlo or Brownian Dynamics (BD) simulations. In contrast, we provide an exact relation between the contact probabilities and the harmonic couplings of our model. Based on this relation, we propose a minimization scheme to find a physical GEM with contact probabilities as close as possible to the experimental ones. We then apply our method to Hi-C and GAM data, thus demonstrating that experimental contact probability matrices can be quantitatively reproduced by our effective polymer model.

We suggest that our reconstructed GEM can be used to study chromatin organization. Typically, coarse-grained models of the chromosome are simulated by BD \cite{brackley2016predicting,michieletto2016polymer}. Due to the complexity of the DNA-DNA and DNA-protein interactions, practical implementations generally require some dimensional reduction or arbitrary choices for unknown parameters such as binding energies or protein binding sites. In contrast, BD simulations of the reconstructed GEM offer a simple alternative which reproduces faithfully the contacts observed in Hi-C or GAM experiments.

\section*{Model}

\subsection*{Gaussian effective model}

We model the chromosome as a beads-on-string polymer comprising $N+1$ monomers with coordinates $\lbrace \vec{r}_ i \rbrace_{i=0\ldots N}$, each monomer corresponding to a genomic bin with size $b$ which, depending on the resolution, may represent from \SI{5}{\kilo bp} to \SI{1}{\mega bp}. Despite some controversy \cite{fussner2011living} euchromatin is generally regarded as a fiber of diameter \SI{30}{\nm} and persistence length $l_p = \SI{60}{\nm} \approx \SI{6}{\kilo bp}$ \cite{Langowski2412006}. Thus we choose to neglect the bending rigidity of the chromosome, and consider the Gaussian chain potential for the chromosome backbone:
\begin{align}
  \beta U_{0}\left[ \left\{ \vec{r}_i \right\} \right] = \frac{3}{2 b^2} \sum \limits_{i=1}^N \left( \vec{r}_i - \vec{r}_{i-1} \right)^2,
  \label{eq:gem_gaussian_energy}
\end{align}
\vspace{0.6em}
where $\beta = (k_\mathrm{B} T)^{-1}$ is the inverse temperature.

\begin{figure}[hbt!]
  \centering
  \includegraphics[width = 0.6 \linewidth]{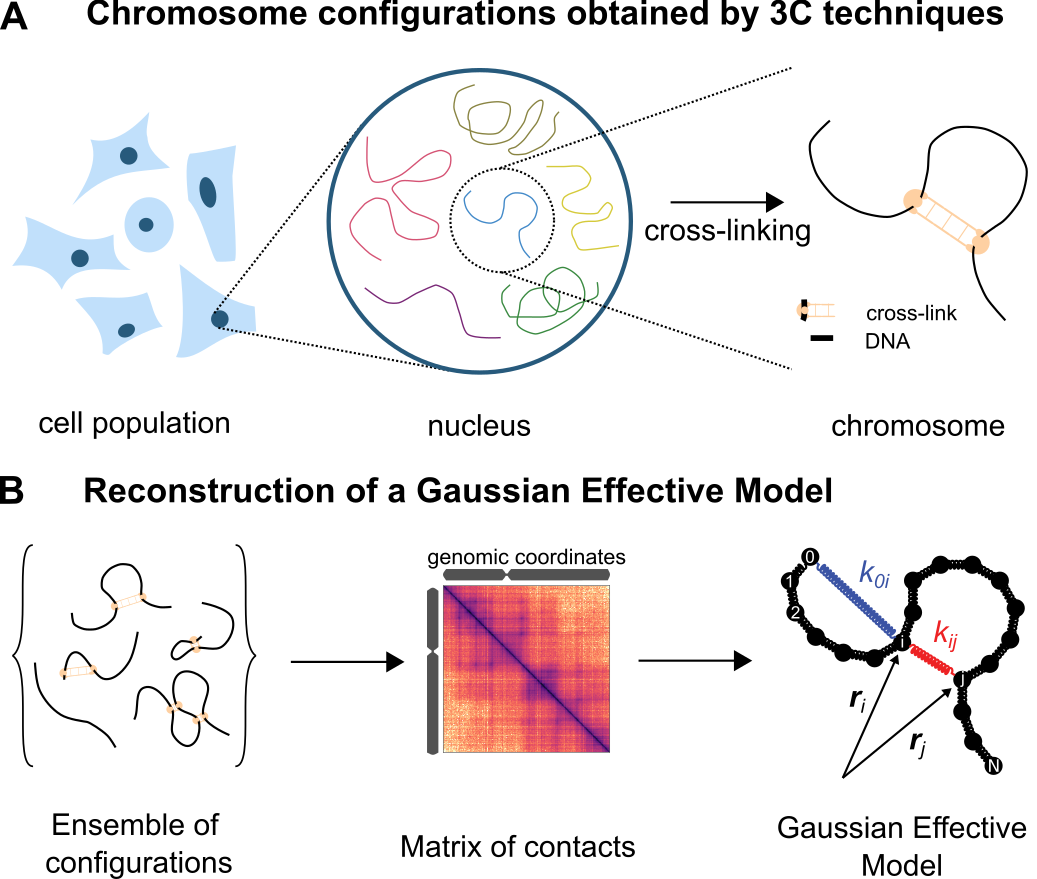}
  \caption{\myref{a} Configurations adopted by a chromosome in a cell population are retrieved using chromosome conformation capture (3C) techniques. \myref{b} We use the count matrix generated by the Hi-C protocol, containing information on the ensemble of chromosome configurations, to reconstruct a Gaussian effective model. Harmonic interactions with elastic coefficients $k_{ij}$ are added on top of a Gaussian polymer model and adjusted to reproduce the experimental contacts.}
  \label{fig:gem}
\end{figure}

The Hi-C protocol uses a cross-linking agent to induce proximity ligations between DNA fragments that are close to each other in the nucleus (\cref{fig:gem}A). The matrix of contacts generated subsequently encodes information on the ensemble of configurations adopted by the chromosome (\cref{fig:gem}B). We represent the underlying interactions which constrain its folding  as harmonic springs with rigidity $3 k_{ij} / b^2$, leading to the interaction potential:
\begin{equation}
  \beta U_I \left[ \left\{ \vec{r}_i \right\}  \right] = \frac{3}{2 b^2} \sum \limits_{0 \le i<j \le N} k_{ij} \left( \vec{r}_i -\vec{r}_j \right)^2.
  \label{eq:gem_interaction_energy}
\end{equation}

The probability of a particular configuration at equilibrium is given by a Boltzmann weight. Namely, if we denote the total energy as $U = U_0 + U_I$, we have:
\begin{equation}
  \proba{ \left\{ \vec{r}_i \right\}} = \frac{1}{Z} e^{-\beta U\left[ \left\{ \vec{r}_i \right\} \right]}.
  \label{eq:proba_boltzmann}
\end{equation}

Actually, the total energy is quadratic in the ${\vec{r}_i}$ variables and may be written:
\begin{align}
  \begin{aligned}
    \beta U\left[ \left\{ \vec{r}_i \right\} \right] &= \frac{3}{2 b^2} \sum \limits_{i,j} \sigma_{ij}^{-1} \vec{r}_i \cdot \vec{r}_j.
  \end{aligned}
  \label{eq:gem_energy}
\end{align}

As a result, the probability distribution in \cref{eq:proba_boltzmann} is Gaussian, hence the name of Gaussian effective model. The GEM is completely determined by its covariance matrix $\Sigma=[\sigma_{ij}]_{i,j=1 \ldots N}$ or equivalently its two-points correlation functions. In particular we have $\langle \vec{r}_i \cdot \vec{r}_j \rangle = \sigma_{ij} b^2$ and $\langle \vec{r}_i^2 \rangle = \sigma_{ii}$, where the brackets denote an average taken over the Gaussian distribution in \cref{eq:proba_boltzmann}. Its inverse is expressed as:
\begin{equation}
  \Sigma^{-1} = T + W,
  \label{eq:gem_inverse_correlation}
\end{equation}
where $T$ is a tridiagonal matrix enforcing the chain structure from \cref{eq:gem_gaussian_energy} and $W$ is a matrix of reduced couplings enforcing the interactions from \cref{eq:gem_interaction_energy}. The matrix $W$ has the structure of a Kirchhoff (or valency-adjacency) matrix as defined in graph theory \cite{kasteleyn1967graph}. These matrices read:
\begin{align}
  \begin{aligned}
    T &=
    \begin{pmatrix}
      2 & -1 & \ldots & 0 & 0 \\
      -1 & 2 & \ldots & 0 & 0 \\
      \vdots & \vdots & \ddots & \vdots & \vdots \\
      0 & 0 & \ldots & 2 & -1 \\
      0 & 0 & \ldots & -1 & 1
    \end{pmatrix},
    \\
    W &=
    \begin{pmatrix}
      \sum \limits_{\substack{j=0 \\ j \neq 1}} k_{1j} & -k_{12} & \ldots & -k_{1 N-1} & -k_{1N} \\
      -k_{21} &  \sum \limits_{\substack{j=0 \\ j \neq 2}} k_{2j}& \ldots & -k_{2 N-1} & -k_{2N} \\
      \vdots & \vdots & \ddots & \vdots & \vdots \\
      -k_{N-1 1} & -k_{N-1 2} & \ldots & \sum \limits_{\substack{j=0 \\ j \neq N-1}} k_{N-1 j} & -k_{N-1 N} \\
      -k_{N 1} & -k_{N 2} & \ldots & - k_{N N-1} & \sum \limits_{\substack{j=0 \\ j \neq N}} k_{N j}
    \end{pmatrix}.
  \end{aligned}
\label{eq:gem_trid_redk}
\end{align}

As an essential feature of the GEM, the pair distances have Gaussian distributions:
\begin{equation}
  \proba{\vec{r}_{ij} = \vec{r}} = \left( \frac{2 \pi \langle r_{ij}^2 \rangle}{3} \right)^{-3/2} \exp{\left( - \frac{3}{2} \frac{r^2}{\langle r_{ij}^2 \rangle} \right)},
  \label{gem:pair_pdf}
\end{equation}
where the mean-square distance $\langle r_{ij}^2 \rangle$ is related to the covariance matrix through the classical identities $\langle r_{ij}^2 \rangle = \langle \vec{r}_{i}^2 \rangle + \langle \vec{r}_{j}^2 \rangle - 2 \langle \vec{r}_i \cdot \vec{r}_j \rangle $.

We now formally express the contact probability between monomers $i$ and $j$ as:
\begin{align}
  \begin{aligned}
    c_{ij} &= \langle \mu(r_{ij}) \rangle, \\
    & = \int \ud{^3 \vec{r}} \mu(r) \langle \delta(\vec{r}_{ij} - \vec{r}) \rangle,
  \end{aligned}
  \label{eq:cij_thermo}
\end{align}

In \cref{eq:cij_thermo}, $\mu(r_{ij})$ is the probability that a cross-link is formed between monomers $i$ and $j$ that are separated by a distance $r_{ij}$. The cross-linking agent used in Hi-C experiments, namely formaldehyde, is known to polymerize in solution, resulting in cross-links of variable lengths \cite{Jackson1251999}. Therefore, in this work, we have considered a Gaussian form factor:
\begin{equation}
  \mu_{\xi}(r) = \exp{\left( - \frac{3}{2} \frac{r^2}{\xi^2} \right)},
  \label{eq:form_factor_gaussian_main}
\end{equation}
where the threshold $\xi$ represents the typical distance under which two monomers can be cross-linked. With this definition, we can compute the thermodynamic average in \cref{eq:cij_thermo} and obtain (see \cref{sec:gem_si} in the Supplementary Information):
\begin{equation}
  c_{ij} = \left( 1 + \frac{\langle r_{ij}^2 \rangle }{\xi^2} \right)^{-3/2}.
  \label{eq:cij_rij2}
\end{equation}

We have thus expressed explicitly the contact probability between monomers $i$ and $j$ as a function of their mean square distance. As might be expected, the contact probability $c_{ij}$ is a decreasing function of $\langle r_{ij}^2 \rangle$. Similar expressions can be obtained for other choices of form factors (see \cref{sec:gem_si} in the Supplementary Information).

In summary, \cref{eq:gem_inverse_correlation} and \cref{eq:cij_rij2} define a unique correspondence between the coupling matrix $[k_{ij}]_{i,j = 0 \ldots N}$ and the contact probability matrix $[c_{ij}]_{i,j = 0 \ldots N}$. The only free parameter is the threshold $\xi$. We can therefore reconstruct the GEM reproducing a given contact probability matrix. For example, we have successfully applied this method to contact probabilities obtained by sampling configurations of a predefined GEM through BD simulations (see \cref{sec:gem_si} in the Supplementary Information). We note that our model does not take into account excluded volume effects.

\subsection*{Reconstruction of an admissible GEM}

We realized that the presence of noise in the contact probabilities could lead to an unstable GEM, having a covariance matrix with negative eigenvalues and therefore a non-finite free energy (see \cref{sec:gem_directinversion} in the Supplementary Information). To solve this issue we reasoned that although a GEM is unstable, there may exist a stable GEM with very close contact probabilities. We therefore introduce the least-square estimator (LSE) between some experimental contact probability matrix and the one of a candidate (stable) GEM:
\begin{equation}
  \mathrm{LSE} = \frac{1}{(N+1)^2} \sum \limits_{i,j} (c_{ij} - c_{ij}^{exp})^2.
  \label{eq:cij_lse}
\end{equation}

In \cref{eq:cij_lse} the LSE is a function of the $k_{ij}$ variables since the $c_{ij}$ are computed from the coupling matrix using the GEM mapping introduced above. Our goal is then to minimize the LSE under the constraint that the GEM is stable. A rigorous enforcement of this principle would be to ensure that its covariance matrix $\Sigma$ has strictly positive eigenvalues, which is difficult to implement in practice. Instead we consider the more restrictive condition:
\begin{equation}
  k_{ij} \ge 0,
  \label{eq:positivity_couplings}
\end{equation}
which is a sufficient condition of stability of the GEM.

\subsection*{Implementation}

We use a steepest descent algorithm with projection to minimize \cref{eq:cij_lse} under the constraint in \cref{eq:positivity_couplings} (see \cref{sec:lse_minimization} in the Supplementary Information). We thus obtain the positive couplings $k_{ij}^{*}$ minimizing the LSE. As seen earlier, computing the $c_{ij}$ as a function of the $k_{ij}$ relies on the choice of a threshold $\xi$. Therefore, we repeat the above minimization procedure for several values of $\xi$, and choose the one with the smallest LSE. \textit{In fine}, the reconstructed couplings $k_{ij}^{opt}$ define the best physically admissible GEM with contact probabilities $c_{ij}^{opt}$ reproducing the experimental values of the contact probabilities.

\section*{Results}

\begin{figure}[hbt!]
  \centering
  \includegraphics[width = 0.5 \linewidth]{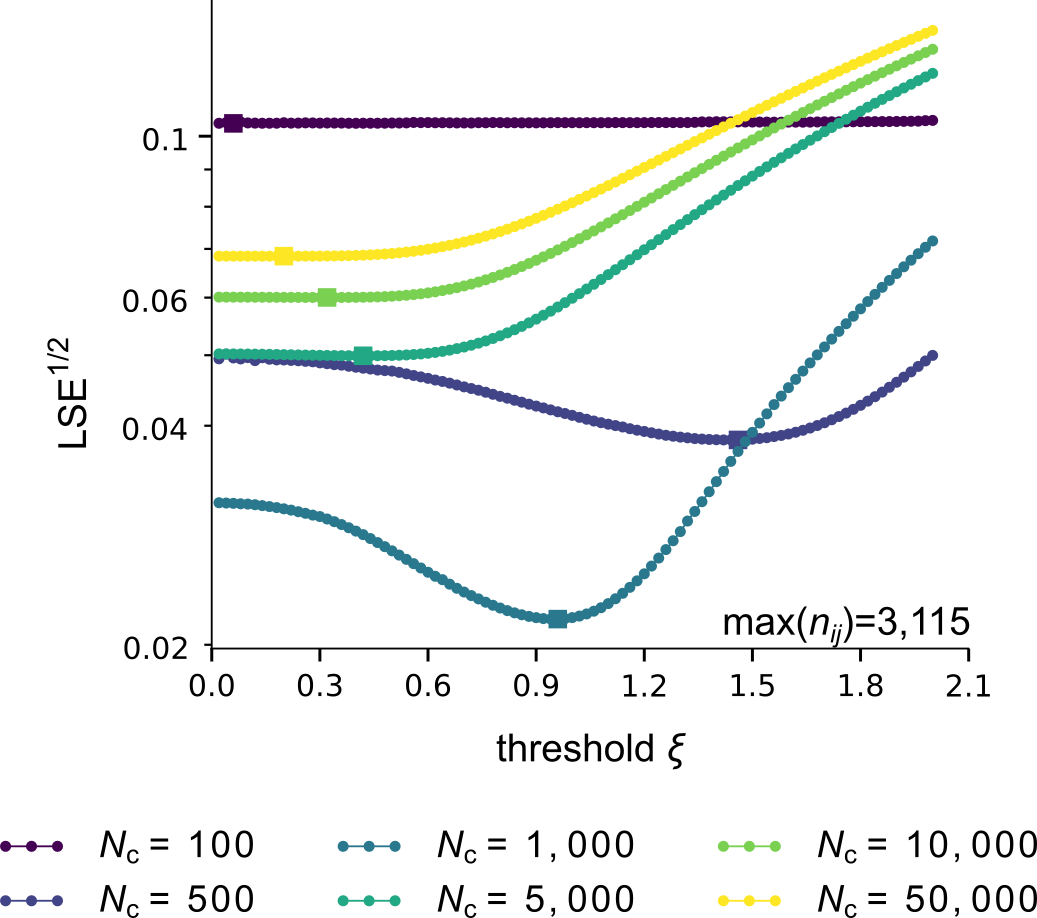}
  \caption{Application of the GEM reconstruction method to Hi-C data from \cite{rao2014a} for chromosome 8 at bin resolution \SI{5}{\kilo bp}. The best GEM is obtained for values of $\xi$ and $N_c$ that minimize the LSE between experimental and GEM contact probabilities. The maximum number of contacts detected among $(i,j)$ bin pairs is denoted as $\mathrm{max}{(n_{ij})}$.}
  \label{fig:rao_chr8_5Kbp_normalizations}
\end{figure}

We have applied our reconstruction method to Hi-C data generated from human lymphoblastoid cells (type GM12878) \cite{rao2014a}. For a given chromosome, this data comes under the form of count matrices, in which each entry $n_{ij}$ corresponds to the number of contacts detected between bins $i$ and $j$ on the chromosome. To compute the contact probability matrix, we applied a global normalization factor $N_c$ to the Hi-C count matrices, $c_{ij} = n_{ij} / N_c$ (see \cref{sec:contact_proba_exp} in the Supplementary Information). One may picture $N_c$ as the number of cells in the experimental sample. Since this normalization is not known, we adjusted both free parameters $\xi$ and $N_c$ when applying our reconstruction method, so as to minimize the least-square estimator (LSE) between experimental and GEM contact probabilities. For data of the chromosome 8 at a bin resolution of \SI{5}{\kilo bp}, the best reconstructed GEM was obtained for $N_c = \num{e3}$ and $\xi = \num{0.96}$ (see \cref{fig:rao_chr8_5Kbp_normalizations}).

\begin{figure}[hbt!]
  \centering
  \includegraphics[width = \linewidth]{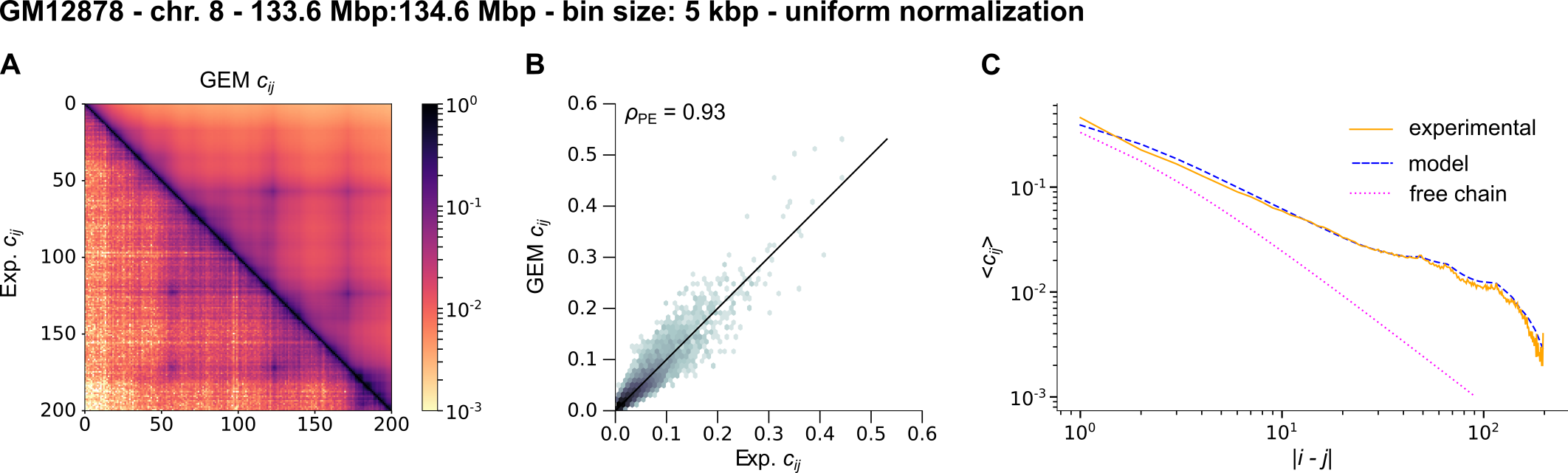}
  \caption{Best reconstructed GEM for Hi-C data of human chromosome 8 at \SI{5}{\kilo bp} resolution \cite{rao2014a}. \myref{a} Comparison between experimental (lower left) and GEM (upper right) contact probabilities. \myref{b} Comparison of experimental and GEM contact probabilities (2d-histogram). We give the Pearson correlation coefficient. \myref{c} Comparison of the average contact probability as a function of the contour length.}
  \label{fig:rao_chr8_5Kbp_contacts}
\end{figure}

The typical discrepancy between experimental and GEM contact probabilities was small, $\mathrm{LSE}^{1/2} = \num{0.022}$, suggesting that this chromosome region can be well represented by a GEM. Much of the structure found in the experimental contact probability matrix was indeed well captured in the reconstructed model (\cref{fig:rao_chr8_5Kbp_contacts}A). This agreement was also readily seen when considering the average contact probability $\langle c_{ij} \rangle$ at a given contour length (\cref{fig:rao_chr8_5Kbp_contacts}C).

Other methods, more sophisticated than the one used above, have been proposed to estimate contact probabilities from Hi-C count matrices \cite{imakaev2012iterative,yaffe2011probabilistic,cournac2012normalization,rao2014a}. For completeness, we have also applied our reconstruction procedure to contact probabilities generated from the same Hi-C data, but using the matrix balancing normalization, which produces a stochastic matrix of contact probabilities (see \cref{sec:contact_proba_exp} in the Supplementary Information). In this case, the only free parameter to adjust was the threshold $\xi$. We found that the reconstructed GEM also reproduced well the experimental contact probabilities (see \cref{fig:rao_chr8_5Kbp_stochastic} in the Supplementary Information). Yet the LSE was larger than for the previous normalization. A possible explanation for this increased value may be that a stochastic contact probability matrix is a poor representation of a cross-linked polymer.

\begin{figure}[hbt!]
  \centering
  \includegraphics[width = \linewidth]{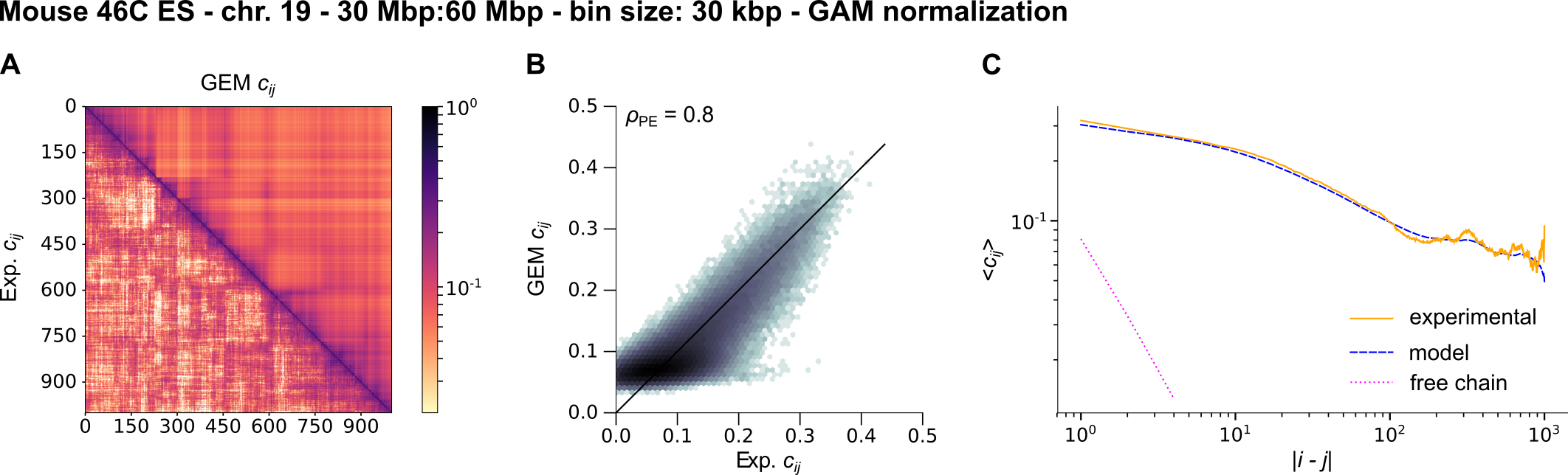}
  \caption{Best reconstructed GEM for GAM data of mouse chromosome 19 at \SI{30}{\kilo bp} resolution \cite{beagrie2017complex}. \myref{a} Comparison between experimental (lower left) and GEM (upper right) contact probabilities. \myref{b} Comparison of experimental and GEM contact probabilities (2d-histogram). We give the Pearson correlation coefficient. \myref{c} Comparison of the average contact probability as a function of the contour length.}
  \label{fig:beagrie_chr19_30Kbp_contacts}
\end{figure}

To demonstrate that the effectiveness of our method is not limited to Hi-C data only, we have also applied our reconstruction procedure to GAM experimental data of mouse embryonic stem cells \cite{beagrie2017complex}. Briefly, with this technique, slices of cell nuclei are obtained by making cryosections, and their DNA content is sequenced. The main output is an array of co-segregation frequencies, representing the probability for two genomic bins to be present in the same slice. We developed a normalization scheme to convert these co-segregation frequencies into contact probabilities (see \cref{sec:contact_proba_exp} in the Supplementary Information). This does not introduce additional parameters, so when applying our reconstruction procedure, we only had to adjust the threshold $\xi$. For example, we applied our method to GAM data generated from mouse embryonic stems cells, for the chromosome 19 with a bin resolution of \SI{30}{\kilo bp} (\cref{fig:beagrie_chr19_30Kbp_contacts}). Again, the reconstructed model well reproduced the experimental contact probabilities, with a typical discrepancy $\mathrm{LSE}^{1/2} = \num{0.032}$. Although this value is slightly greater than in the Hi-C case presented above, the size of the corresponding polymer is larger, with $N=1000$. Therefore the quantitative agreement between experiment and reconstructed model remains very good. Note that the optimal threshold of the reconstruction was quite small, $\xi^{opt} = \num{0.48}$. Eventually, it appears that the precise value of the threshold is not critical. Indeed, below $\xi \lesssim 1.0$, the relative variations of the LSE became very small (see \cref{fig:beagrie_chr19_30Kbp}). Hence, the threshold may actually be seen as a regularization parameter for the reconstructed contact probability matrix.

We have applied our reconstruction procedure to various chromosomes and bin resolutions from either Hi-C or GAM data sets (see \cref{tab:supp_table} together with \cref{fig:rao_chr7_5Kbp_global,fig:rao_chr7_10Kbp_global,fig:rao_chr8_5Kbp_global,fig:rao_chr10_5Kbp_global,fig:rao_chr14_10Kbp_N200_global,fig:rao_chr14_10Kbp_N1000_global,fig:rao_chr14_100Kbp_global,fig:rao_chr16_5Kbp_global,fig:rao_chr7_5Kbp_stochastic,fig:rao_chr7_10Kbp_stochastic,fig:rao_chr8_5Kbp_stochastic,fig:rao_chr10_5Kbp_stochastic,fig:rao_chr14_10Kbp_N200_stochastic,fig:rao_chr14_10Kbp_N1000_stochastic,fig:rao_chr14_100Kbp_stochastic,fig:rao_chr16_5Kbp_stochastic,fig:beagrie_chr19_30Kbp,fig:beagrie_chr19_100Kbp,fig:beagrie_chr19_1Mbp,fig:beagrie_chr12_30Kbp,fig:beagrie_chr12_100Kbp,fig:beagrie_chr12_1Mbp,fig:beagrie_chr1_30Kbp,fig:beagrie_chr1_100Kbp,fig:beagrie_chr1_1Mbp}). Overall, the contact probabilities of the reconstructed GEMs quantitatively reproduced the experimental ones. We found in general that the typical distance between experimental and reconstructed model contact probabilities was $\mathrm{LSE}^{1/2} \sim \numrange{0.01}{0.05}$. Thus we conclude that our method allows to represent to a quantifiable accuracy the ensemble of configurations adopted by the chromosome.

\begin{figure}[hbt!]
  \centering
  \includegraphics[width = 0.6 \linewidth]{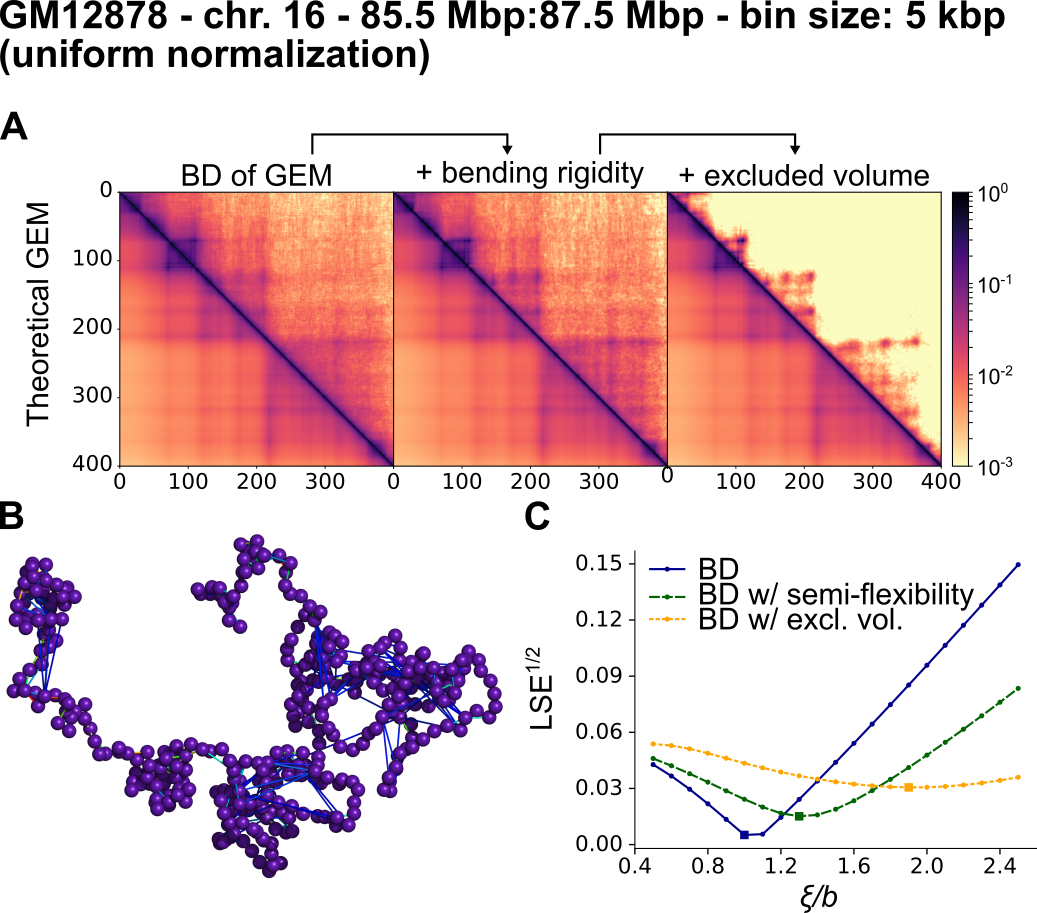}
  \caption{Brownian dynamics (BD) of the reconstructed GEM for Hi-C data of human chromosome 16 \cite{rao2014a} (\SI{5}{\kilo bp} resolution). \textbf{(A)} Contact probability matrices obtained through BD simulation of: (i) the GEM, (ii) the GEM with bending rigidity, and (iii) the GEM with bending rigidity and with excluded volume. The contact probabilities were computed from BD trajectories and are compared with the theoretical values for the GEM. \textbf{(B)} Snapshot of a configuration obtained by BD of the reconstructed GEM with bending rigidity and excluded volume. The couplings are represented by tie lines, from weak couplings (in blue) to strong couplings (in red). \textbf{(C)} LSE as a function of the threshold $\xi$ between contact probabilities computed from the BD trajectory and the theoretical values.}
  \label{fig:rao_chr16_5Kbp_brownian_dynamics_main}
\end{figure}

In order to illustrate possible applications of our method to study chromosome organization, we used the reconstructed coupling matrices to perform BD simulations of the chromosome (see \cref{sec:brownian_dynamics} in the Supplementary Information). To do so, we replaced the Gaussian chain potential in \cref{eq:gem_gaussian_energy} by a finitely-extensible non-linear elastic bond potential, we took into account the polymer bending rigidity and we introduced excluded volume interactions. We then performed BD simulations and used the sampled configurations to compute the equilibrium contact probabilities, which we compared to the ones of the GEM (see \cref{fig:rao_chr16_5Kbp_brownian_dynamics_main}A, \cref{fig:rao_chr8_5Kbp_brownian_dynamics,fig:beagrie_chr19_30Kbp_brownian_dynamics}). In the presence of excluded volume and semi-flexibility, the obtained contact probabilities were not as close to the GEM ones. Yet the essential structure of the contact probability matrix remained. In \cref{fig:rao_chr16_5Kbp_brownian_dynamics_main}B we show a typical configuration for the human chromosome 16.

\section{Discussion}
In this article, we have proposed a polymer model constrained by Hi-C or GAM experimental measurements to represent the chromosome. We modeled the DNA as a flexible polymer (since the resolution is much larger than the persistence length of the DNA), with harmonic interactions between chromosomal loci encoding the contact frequency in Hi-C and GAM experiments. The spring constants are chosen so as to best reproduce the experimentally measured contact probabilities. We computed the explicit mapping defined in \cref{eq:gem_inverse_correlation,eq:cij_rij2} which relates the harmonic couplings to the contact probabilities between monomers. We then used this property to reconstruct a physically admissible GEM of the chromosome by minimizing the distance between experimental and model contact probabilities. We applied this method to many chromosomes and data sets. Overall, the quantitative agreement obtained suggested that the GEM offers a good representation of the chromosome. In order to illustrate potential applications of our method, we then used the reconstructed GEM to perform BD simulation of the chromosome. While it is not a substitute to first principles molecular dynamics simulations, this approach is valuable because the trajectories simulated by BD reproduce the experimental contact probabilities.

\subsection*{Models for cross-linked polymer}

Properties of cross-linked polymers have been extensively studied \cite{Solf66551995,Kantor52631996,Bryngelson5421996}. However, in those studies the rigidities of the harmonic interactions were uniform, \textit{i.e.} $k_{ij} = k$ in \cref{eq:gem_energy}. A similar model was also re-introduced to account for the particular scaling of the radius of gyration of the chromosome in the interphase nucleus, in which the $k_{ij}$ were distributed as Bernoulli variables and hence defined random loops \cite{Bohn0518052007,Mateos-Langerak38122009}. Recently, an other model with quadratic interactions was proposed to obtain polymer states with arbitrary fractal dimension \cite{polovnikov2018effective}, in which the harmonic couplings followed a power law of the contour distances. Yet these studies did not attempt to compute Hi-C contact probabilities or to predict chromatin conformations. Our model also presents some similarities with the Gaussian Elastic Network model used in the context of protein folding \cite{bahar1997direct,haliloglu1997gaussian}.

\subsection*{Do the reconstructed couplings represent biological interactions?}

Hi-C data are often generated from a population of cells. Thus if a pair of chromosomal loci has a number of contacts which is statistically significant, it means that specific interactions should favor their co-localization. Therefore the couplings $k_{ij}$ can be seen as defining coarse-grained potentials representing the superimposition of many microscopical interactions, such as the bridging by divalent proteins, and used as effective interactions in coarse-grained models of the chromosome. Yet the mean pair potentials $e_{ij} = 3/2 k_{ij} \langle r_{ij}^2 \rangle$, expressed in $k_\mathrm{B} T$, provide a more physical interpretation of the reconstructed interactions. Yet the effective model obtained can give clues about where the major constraints that determine the folding of the chromosome are applied.

\subsection*{Fractal globule scaling of the contact probabilities}

It is believed that the so-called fractal globule model (or crumpled polymer) provides a more realistic framework to describe the chromosome than classical polymer models \cite{Grosbert1993373,mirny2011the}. In short, the presence of excluded volume and confinement results in high energy barriers from one configuration to the other, leading to a behavior different from an ideal polymer. In particular, the fractal globule was shown to reproduce the scaling for the mean contact probability as a function of the contour length, $c_{ij} \propto |i-j|^{-1}$, observed in Hi-C experiments \cite{liebermanaiden2009comprehensive}. We note that although our GEM does not incorporate excluded volume, it reproduces the experimental scaling because the couplings are reconstructed from the experimental contacts.

\subsection*{Robustness of the method}
\begin{figure}[hbt!]
  \centering
  \includegraphics[width = 0.6 \linewidth]{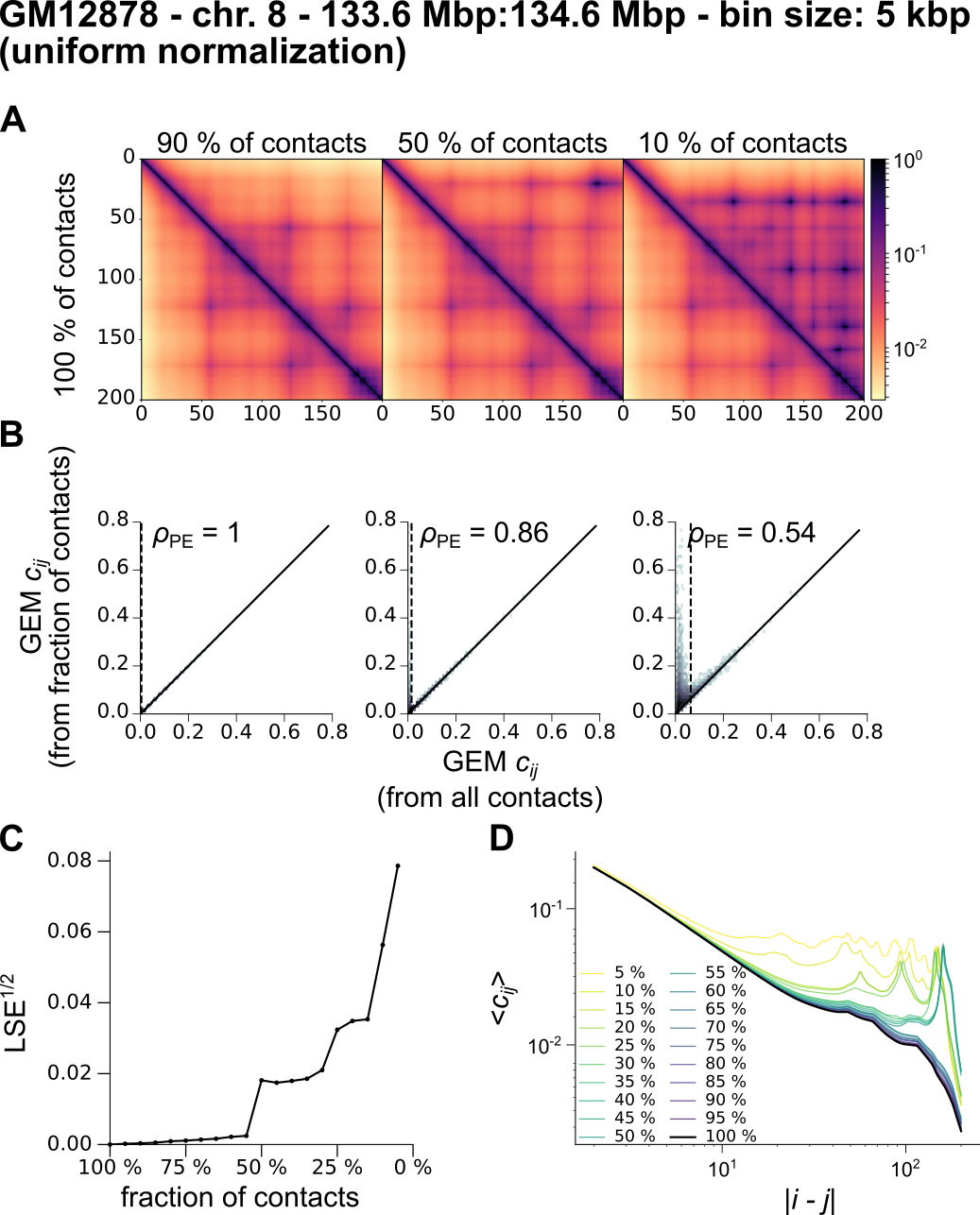}
  \caption{%
    Robustness of GEM reconstruction for Hi-C data of human chromosome 8 \cite{rao2014a} (\SI{5}{\kilo bp} resolution). For all GEM reconstructions we used a threshold $\xi = \num{1}$ and a normalization factor $N_c = \num{e3}$. \textbf{(A)} Comparison of the contact probabilities of the reconstructed GEM with those of a GEM obtained by performing the minimization only on the top \SI{90}{\percent}, \SI{50}{\percent} and \SI{10}{\percent} experimental contacts. \textbf{(B)} 2d-histograms corresponding to the matrices shown in \textbf{(A)}. We give the Pearson correlation coefficients. The thresholding quantiles are represented by vertical dashed lines. \textbf{(C)} Comparison of the GEMs reconstructed from a decreasing fraction of the experimental contacts with the original GEM. $\mathrm{LSE}^{1/2}$ is the Euclidean distance between contact probabilities divided by ($N+1$). \textbf{(D)} Average contact probability as a function of the contour length for GEMs reconstructed from a decreasing fraction of the experimental contacts.%
  }%
  \label{fig:rao_chr8_5Kbp_significant_contacts}
\end{figure}

In order to investigate the robustness of the reconstructed GEM, we repeated the minimization procedure but considered only a subset of the experimental contacts in the sum from \cref{eq:cij_lse}. Specifically, we retained only the top fraction of the experimental contact probabilities. In \cref{fig:rao_chr8_5Kbp_significant_contacts}A, we compared the contact probabilities of the original reconstructed GEM for the human chromosome 8 with the contact probabilities of the GEMs reconstructed by considering only the top \SI{90}{\percent}, \SI{50}{\percent} and \SI{10}{\percent}. Starting from \SI{50}{\percent}, we noticed that some artifacts appear in the reconstructed GEM for long-range contacts. These are located in regions that are sparse in contacts in the experimental contact probability matrix. As a result, very few significant contacts are retained in those regions for the minimization procedure. In fact, contacts below the thresholding quantile, that were discarded from the reconstruction, tend to be overestimated in the newly reconstructed GEM (\cref{fig:rao_chr8_5Kbp_significant_contacts}B). This suggests that regions of the contact probability matrix that contain little meaningful information (significant contacts in our case) will be poorly reconstructed. Overall, \cref{fig:rao_chr8_5Kbp_significant_contacts}C shows that the distance to the original reconstructed GEM increases as the fraction of contacts retained shrinks, and \cref{fig:rao_chr8_5Kbp_significant_contacts}D illustrates that long-range contacts are indeed the first to suffer from reconstruction artifacts. The same analysis for other data sets are given in \cref{fig:rao_chr16_5Kbp_significant_contacts,fig:beagrie_chr19_30Kbp_significant_contacts}.

\subsection*{Future improvements}

A first improvement to our model would be to explicitly include semi-flexibility in the polymer structure. This can be done by adding harmonic interactions extending to second nearest neighbors in \cref{eq:gem_gaussian_energy}. However, this refinement might appear superfluous as long as we consider bin resolutions beyond $\sim \SI{5}{\kilo bp}$. A second improvement would be to extend the method to several chromosomes, by adjusting the matrix $T$ which defines the chain structure.

\small
\section*{Author Contributions}

F.K. and H.O. designed the research. G.L.T. and H.O. performed the research. G.L.T. wrote the code and analyzed the data. All authors contributed to the writing of the article.

\section*{Acknowledgments}

This work was supported by the "IDI 2013" project funded by the IDEX Paris-Saclay, ANR-11-IDEX-0003-02. G.L.T. is grateful to the iSSB and the IPhT for giving him access to their computing facilities.

\normalsize

% vim: set sw=2 expandtab tabstop=2 foldcolumn=4:
% vim: set spell spelllang=en_us,en_gb:

\printbibliography[title=References, segment=1]
%\printbibliography[title=References, section=1]

% supplementary figures
\renewcommand\thesection{\arabic{section}}
\clearpage
\title{Supplementary Information}
\date{}
\author{}
\emptythanks
\maketitle
\renewcommand\thefigure{S\arabic{figure}}
\renewcommand\thetable{S\arabic{table}}
\setcounter{figure}{0}
\etocdepthtag.toc{tappendix}
\etocsettagdepth{tmain}{none}
\etocsettagdepth{tfigures}{section}
\etocsettagdepth{tappendix}{section}
\tableofcontents
\newpage
\etocdepthtag.toc{tfigures}
\setcounter{section}{0}
%%%%%%%%%%%%%%%%%%%%%%%%%%%%%%%%%%%%%%%%%%%%%%%%%%%%%%%%%%%%%%%%%%%%%%%%%%%%
% GEM reconstruction
%%%%%%%%%%%%%%%%%%%%%%%%%%%%%%%%%%%%%%%%%%%%%%%%%%%%%%%%%%%%%%%%%%%%%%%%%%%%
% master table
\sisetup{range-phrase=:}
\footnotesize
\begin{landscape}% Landscape page
  \centering % Center table
  \begin{longtabu} to \linewidth {|X[1,l]|X[2.8,l]|X[1,l]|X[5,l]|X[4,l]|X[1,l]|X[1,l]|X[3.0,l]|X[1,l]|}
    \caption{Application of the GEM reconstruction method to several experimental data sets.} \label{tab:summary}\label{tab:supp_table} \\
    \hline
    \rowfont\bfseries Figure & Reference & Data & Cell type & Genomic range & Resolution & N & Normalization & LSE$^{\mathbf{1/2}}$ \\
    \hline
%  % Rao 2014
% uniform normalization
  %% chr7
    \cref{fig:rao_chr7_5Kbp_global} & Rao \textit{et al.} (2014) & Hi-C & GM12878 (human) & Chr. 7 \SIrange{137}{138}{\mega bp} & \SI{5}{\kilo bp} & \num{200} & uniform & \num{0.023} \\ \hline
    \cref{fig:rao_chr7_10Kbp_global} & Rao \textit{et al.} (2014) & Hi-C & GM12878 (human) & Chr. 7 \SIrange{130}{140}{\mega bp} & \SI{10}{\kilo bp} & \num{1000} & uniform & \num{0.013} \\ \hline
  %% chr8
    \cref{fig:rao_chr8_5Kbp_global} & Rao \textit{et al.} (2014) & Hi-C & GM12878 (human) & Chr. 8 \SIrange{133.6}{134.6}{\mega bp} & \SI{5}{\kilo bp} & \num{200} & uniform & \num{0.022} \\ \hline
  %% chr10
    \cref{fig:rao_chr10_5Kbp_global} & Rao \textit{et al.} (2014) & Hi-C & GM12878 (human) & Chr. 10 \SIrange{90.5}{91.5}{\mega bp} & \SI{5}{\kilo bp} & \num{200} & uniform & \num{0.023} \\ \hline
  %% chr14
    \cref{fig:rao_chr14_10Kbp_N200_global} & Rao \textit{et al.} (2014) & Hi-C & GM12878 (human) & Chr. 14 \SIrange{94}{96}{\mega bp} & \SI{10}{\kilo bp} & \num{200} & uniform &  \num{0.022} \\ \hline
    \cref{fig:rao_chr14_10Kbp_N1000_global} & Rao \textit{et al.} (2014) & Hi-C & GM12878 (human) & Chr. 14 \SIrange{86}{96}{\mega bp} & \SI{10}{\kilo bp} & \num{1000} & uniform & \num{0.014} \\ \hline
    \cref{fig:rao_chr14_100Kbp_global} & Rao \textit{et al.} (2014) & Hi-C & GM12878 (human) & Chr. 14 \SIrange{19}{107.2}{\mega bp} & \SI{100}{\kilo bp} & \num{882} & uniform & \num{0.013} \\ \hline
  %% chr16
    \cref{fig:rao_chr16_5Kbp_global} & Rao \textit{et al.} (2014) & Hi-C & GM12878 (human) & Chr. 16 \SIrange{85.5}{87.5}{\mega bp} & \SI{5}{\kilo bp} & \num{400} & uniform &  \num{0.019} \\ \hline
% Matrix balancing normalization
  %% chr7
    \cref{fig:rao_chr7_5Kbp_stochastic} & Rao \textit{et al.} (2014) & Hi-C & GM12878 (human) & Chr. 7 \SIrange{137}{138}{\mega bp} & \SI{5}{\kilo bp} & \num{200} & matrix balancing &  \num{0.057} \\ \hline
    \cref{fig:rao_chr7_10Kbp_stochastic} & Rao \textit{et al.} (2014) & Hi-C & GM12878 (human) & Chr. 7 \SIrange{130}{140}{\mega bp} & \SI{10}{\kilo bp} & \num{1000} & matrix balancing &  \num{0.026} \\ \hline
  %% chr8
    \cref{fig:rao_chr8_5Kbp_stochastic} & Rao \textit{et al.} (2014) & Hi-C & GM12878 (human) & Chr. 8 \SIrange{133.6}{134.6}{\mega bp} & \SI{5}{\kilo bp} & \num{200} & matrix balancing & \num{0.056} \\ \hline
  %% chr10
    \cref{fig:rao_chr10_5Kbp_stochastic} & Rao \textit{et al.} (2014) & Hi-C & GM12878 (human) & Chr. 10 \SIrange{90.5}{91.5}{\mega bp} & \SI{5}{\kilo bp} & \num{200} & matrix balancing &  \num{0.059}\\ \hline
  %% chr14
    \cref{fig:rao_chr14_10Kbp_N200_stochastic} & Rao \textit{et al.} (2014) & Hi-C & GM12878 (human) & Chr. 14 \SIrange{94}{96}{\mega bp} & \SI{10}{\kilo bp} & \num{200} & matrix balancing & \num{0.056} \\ \hline
    \cref{fig:rao_chr14_10Kbp_N1000_stochastic} & Rao \textit{et al.} (2014) & Hi-C & GM12878 (human) & Chr. 14 \SIrange{86}{96}{\mega bp} & \SI{10}{\kilo bp} & \num{1000} & matrix balancing &  \num{0.026}\\ \hline
    \cref{fig:rao_chr14_100Kbp_stochastic} & Rao \textit{et al.} (2014) & Hi-C & GM12878 (human) & Chr. 14 \SIrange{19}{107.2}{\mega bp} & \SI{100}{\kilo bp} & \num{882} & matrix balancing &  \num{0.026} \\ \hline
  %% chr16
    \cref{fig:rao_chr16_5Kbp_stochastic} & Rao \textit{et al.} (2014) & Hi-C & GM12878 (human) & Chr. 16 \SIrange{85.5}{87.5}{\mega bp} & \SI{5}{\kilo bp} & \num{400} & matrix balancing &  \num{0.042}\\ \hline

  % Beagrie 2017
  %% chr19
    \cref{fig:beagrie_chr19_30Kbp} & Beagrie \textit{et al.} (2017) & GAM  & mouse 46C line embryonic stem cells & Chr. 19 \SIrange{30}{60}{\mega bp} & \SI{30}{\kilo bp} & \num{1000} & GAM & \num{0.032}\\ \hline
    \cref{fig:beagrie_chr19_100Kbp} & Beagrie \textit{et al.} (2017) & GAM  & mouse 46C line embryonic stem cells & Chr. 19 \SIrange{3}{61.2}{\mega bp} & \SI{100}{\kilo bp} & \num{582} & GAM & \num{0.028} \\ \hline
    \cref{fig:beagrie_chr19_1Mbp} & Beagrie \textit{et al.} (2017) & GAM & mouse 46C line embryonic stem cells & Chr. 19 \SIrange{3}{60}{\mega bp} & \SI{1}{\mega bp} & \num{57} & GAM &  \num{0.021}\\ \hline
  %% chr12
    \cref{fig:beagrie_chr12_30Kbp} & Beagrie \textit{et al.} (2017) & GAM  & mouse 46C line embryonic stem cells & Chr. 12 \SIrange{40}{70}{\mega bp} & \SI{30}{\kilo bp} & \num{1000} & GAM & \num{0.033} \\ \hline
    \cref{fig:beagrie_chr12_100Kbp} & Beagrie \textit{et al.} (2017) & GAM  & mouse 46C line embryonic stem cells & Chr. 12 \SIrange{30}{120}{\mega bp} & \SI{100}{\kilo bp} & \num{900} & GAM & \num{0.029} \\ \hline
    \cref{fig:beagrie_chr12_1Mbp} & Beagrie \textit{et al.} (2017) & GAM  & mouse 46C line embryonic stem cells & Chr. 12 \SIrange{3}{120}{\mega bp} & \SI{1}{\mega bp} & \num{117} & GAM &  \num{0.025}\\ \hline
  %% chr1
    \cref{fig:beagrie_chr1_30Kbp} & Beagrie \textit{et al.} (2017) & GAM  & mouse 46C line embryonic stem cells & Chr. 1 \SIrange{135}{165}{\mega bp} & \SI{30}{\kilo bp} & \num{1000} & GAM & \num{0.032} \\ \hline
    \cref{fig:beagrie_chr1_100Kbp} & Beagrie \textit{et al.} (2017) & GAM  & mouse 46C line embryonic stem cells & Chr. 1 \SIrange{90}{190}{\mega bp} & \SI{100}{\kilo bp} & \num{1000} & GAM & \num{0.029} \\ \hline
    \cref{fig:beagrie_chr1_1Mbp} & Beagrie \textit{et al.} (2017) & GAM  & mouse 46C line embryonic stem cells & Chr. 1 \SIrange{3}{196}{\mega bp} & \SI{1}{\mega bp} & \num{193} & GAM & \num{0.026} \\
    \hline
  \end{longtabu}
\end{landscape}
\normalsize
\sisetup{range-phrase=-}

% one figure per page
\makeatletter
\@fpsep\textheight
\makeatother

\begin{figure}%
  \centering%
  \includegraphics[width = \linewidth]{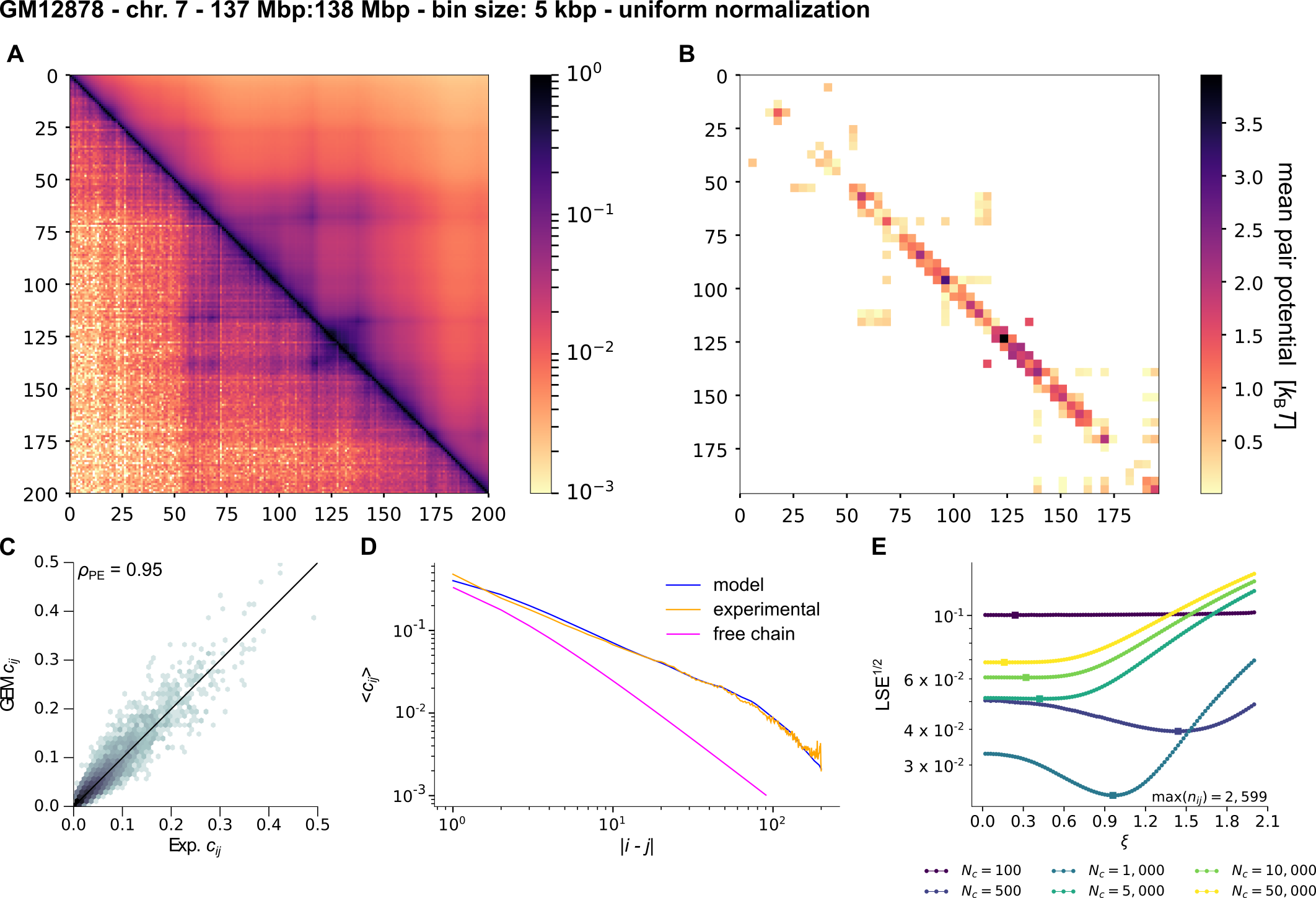}
  \caption{%
    GEM reconstruction for Hi-C data of human chromosome 7 \cite{rao2014a} (\SI{5}{\kilo bp} resolution), normalized by applying a global factor. \textbf{(A)} Comparison between experimental (lower left) and GEM (upper right) contact probabilities. \textbf{(B)} Matrix of mean pair potentials, binned with a ratio 1:4. \textbf{(C)} Comparison of experimental and GEM contact probabilities (2d-histogram). We give the Pearson correlation coefficient. \textbf{(D)} Average contact probability as a function of the contour length. \textbf{(E)} LSE as a function of the threshold $\xi$ used for the GEM mapping, and for different normalizations $N_c$ of the Hi-C counts.%
  }%
  \label{fig:rao_chr7_5Kbp_global}%
\end{figure}

\begin{figure}%
  \centering%
  \includegraphics[width = \linewidth]{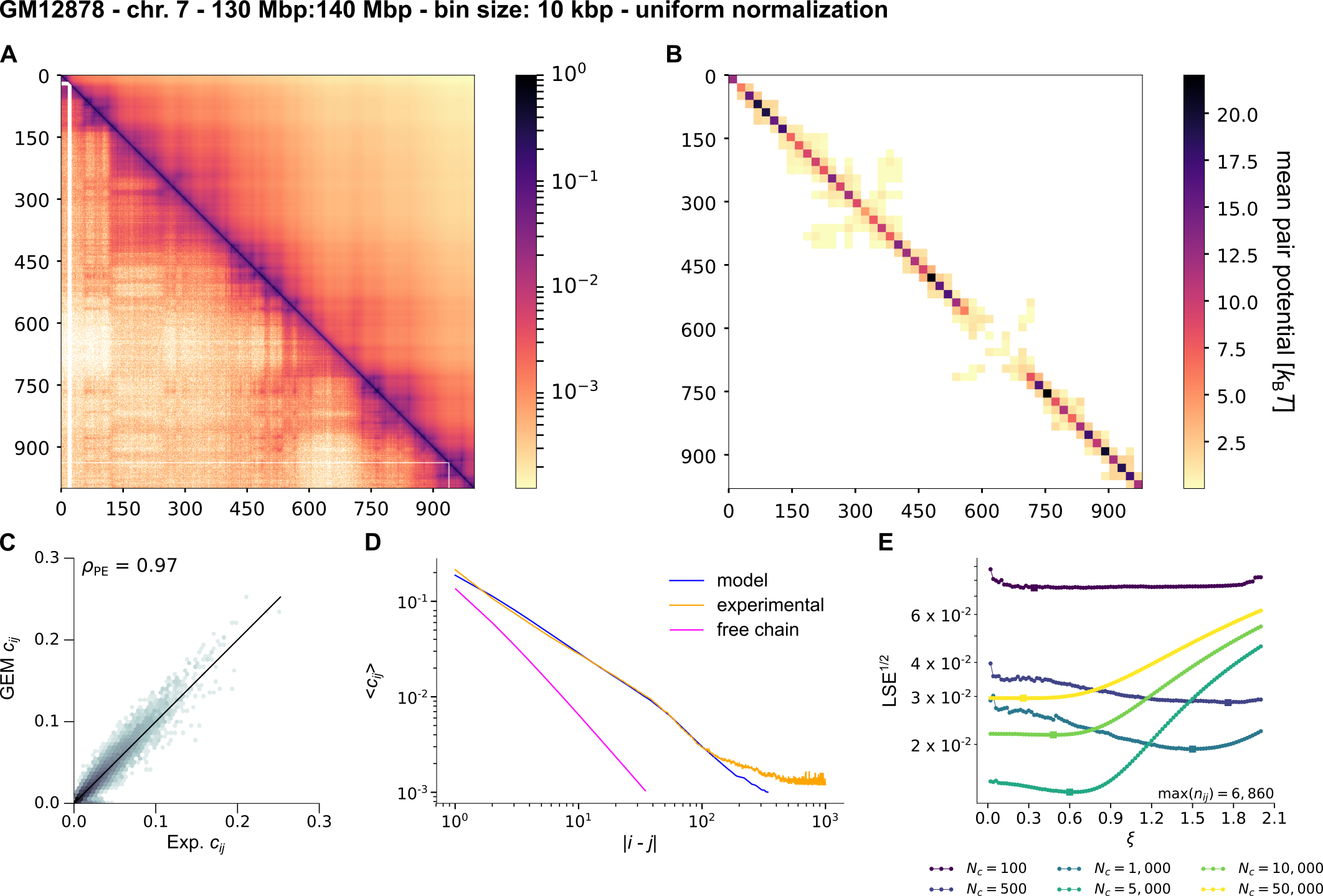}
  \caption{%
    GEM reconstruction for Hi-C data of human chromosome 7 \cite{rao2014a} (\SI{10}{\kilo bp} resolution), normalized by applying a global factor. \textbf{(A)} Comparison between experimental (lower left) and GEM (upper right) contact probabilities. \textbf{(B)} Matrix of mean pair potentials, binned with a ratio 1:20. \textbf{(C)} Comparison of experimental and GEM contact probabilities (2d-histogram). We give the Pearson correlation coefficient. \textbf{(D)} Average contact probability as a function of the contour length. \textbf{(E)} LSE as a function of the threshold $\xi$ used for the GEM mapping, and for different normalizations $N_c$ of the Hi-C counts.%
  }%
  \label{fig:rao_chr7_10Kbp_global}%
\end{figure}

\begin{figure}%
  \centering%
  \includegraphics[width = \linewidth]{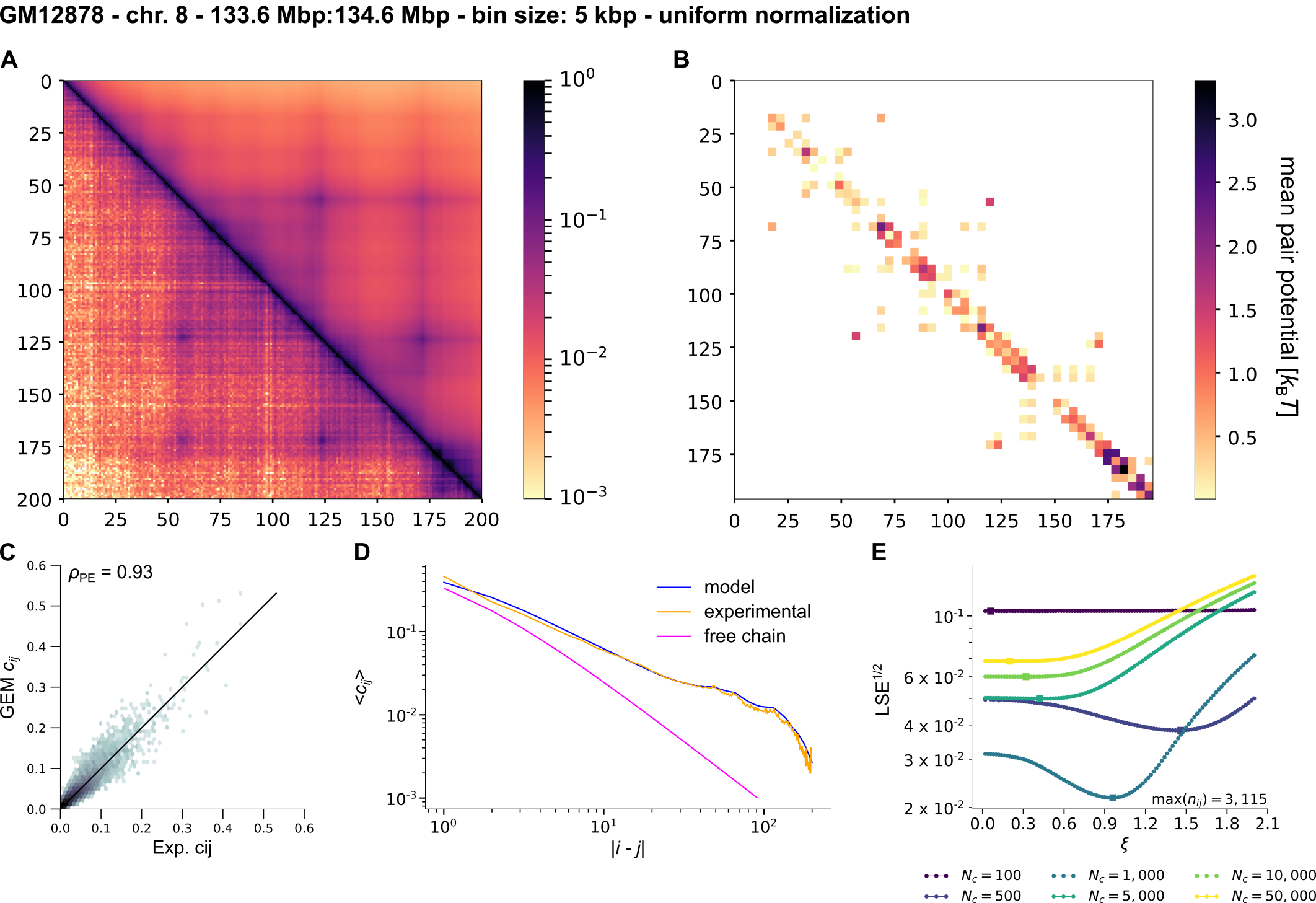}
  \caption{%
    GEM reconstruction for Hi-C data of human chromosome 8 \cite{rao2014a} (\SI{5}{\kilo bp} resolution), normalized by applying a global factor. \textbf{(A)} Comparison between experimental (lower left) and GEM (upper right) contact probabilities. \textbf{(B)} Matrix of mean pair potentials, binned with a ratio 1:4. \textbf{(C)} Comparison of experimental and GEM contact probabilities (2d-histogram). We give the Pearson correlation coefficient. \textbf{(D)} Average contact probability as a function of the contour length. \textbf{(E)} LSE as a function of the threshold $\xi$ used for the GEM mapping, and for different normalizations $N_c$ of the Hi-C counts.%
  }%
  \label{fig:rao_chr8_5Kbp_global}%
\end{figure}

\begin{figure}%
  \centering%
  \includegraphics[width = \linewidth]{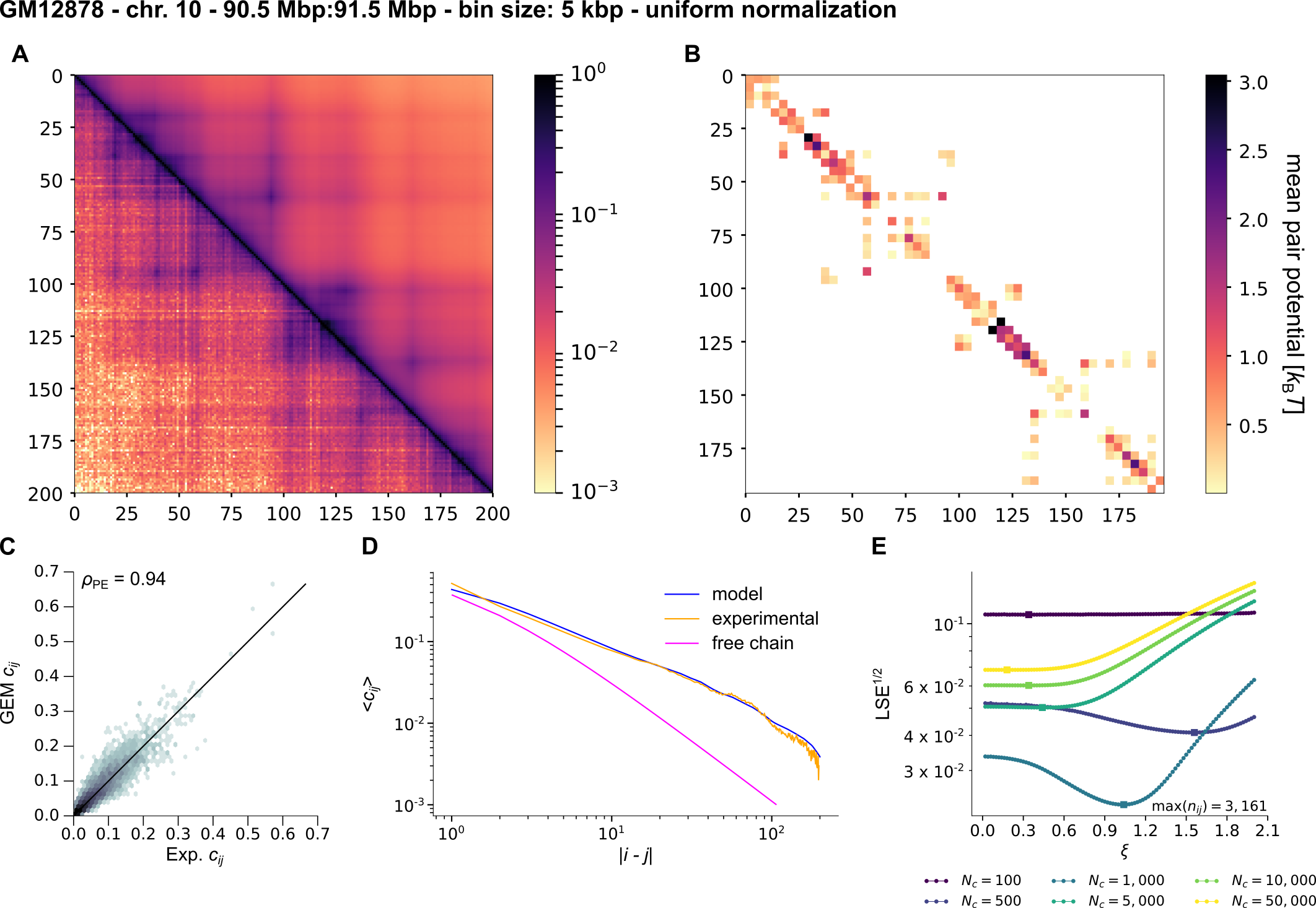}
  \caption{%
    GEM reconstruction for Hi-C data of human chromosome 10 \cite{rao2014a} (\SI{5}{\kilo bp} resolution), normalized by applying a global factor. \textbf{(A)} Comparison between experimental (lower left) and GEM (upper right) contact probabilities. \textbf{(B)} Matrix of mean pair potentials, binned with a ratio 1:4. \textbf{(C)} Comparison of experimental and GEM contact probabilities (2d-histogram). We give the Pearson correlation coefficient. \textbf{(D)} Average contact probability as a function of the contour length. \textbf{(E)} LSE as a function of the threshold $\xi$ used for the GEM mapping, and for different normalizations $N_c$ of the Hi-C counts.%
  }%
  \label{fig:rao_chr10_5Kbp_global}%
\end{figure}

\begin{figure}%
  \centering%
  \includegraphics[width = \linewidth]{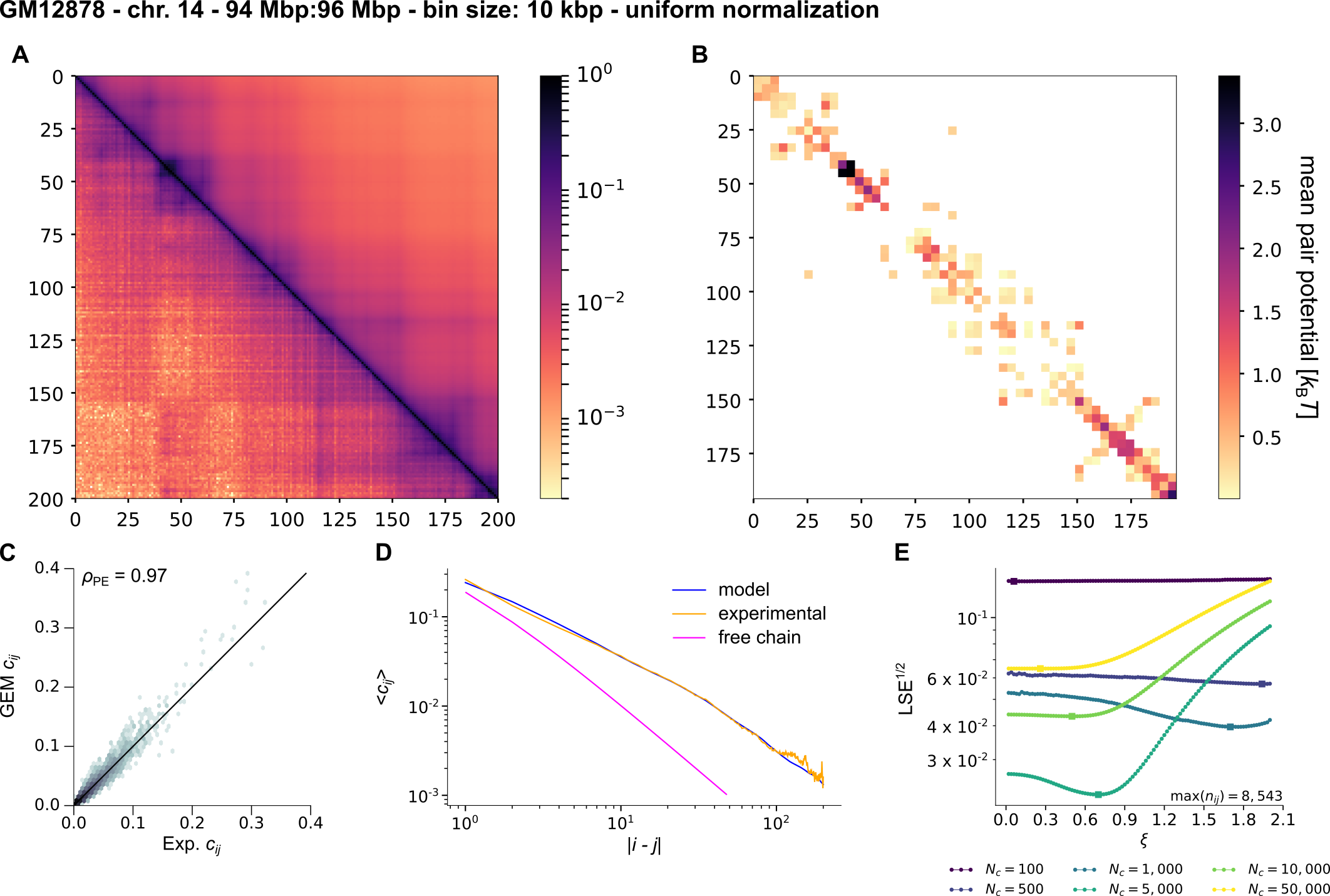}
  \caption{%
    GEM reconstruction for Hi-C data of human chromosome 14 \cite{rao2014a} (\SI{10}{\kilo bp} resolution), normalized by applying a global factor. \textbf{(A)} Comparison between experimental (lower left) and GEM (upper right) contact probabilities. \textbf{(B)} Matrix of mean pair potentials, binned with a ratio 1:4. \textbf{(C)} Comparison of experimental and GEM contact probabilities (2d-histogram). We give the Pearson correlation coefficient. \textbf{(D)} Average contact probability as a function of the contour length. \textbf{(E)} LSE as a function of the threshold $\xi$ used for the GEM mapping, and for different normalizations $N_c$ of the Hi-C counts.%
  }%
  \label{fig:rao_chr14_10Kbp_N200_global}%
\end{figure}

\begin{figure}%
  \centering%
  \includegraphics[width = \linewidth]{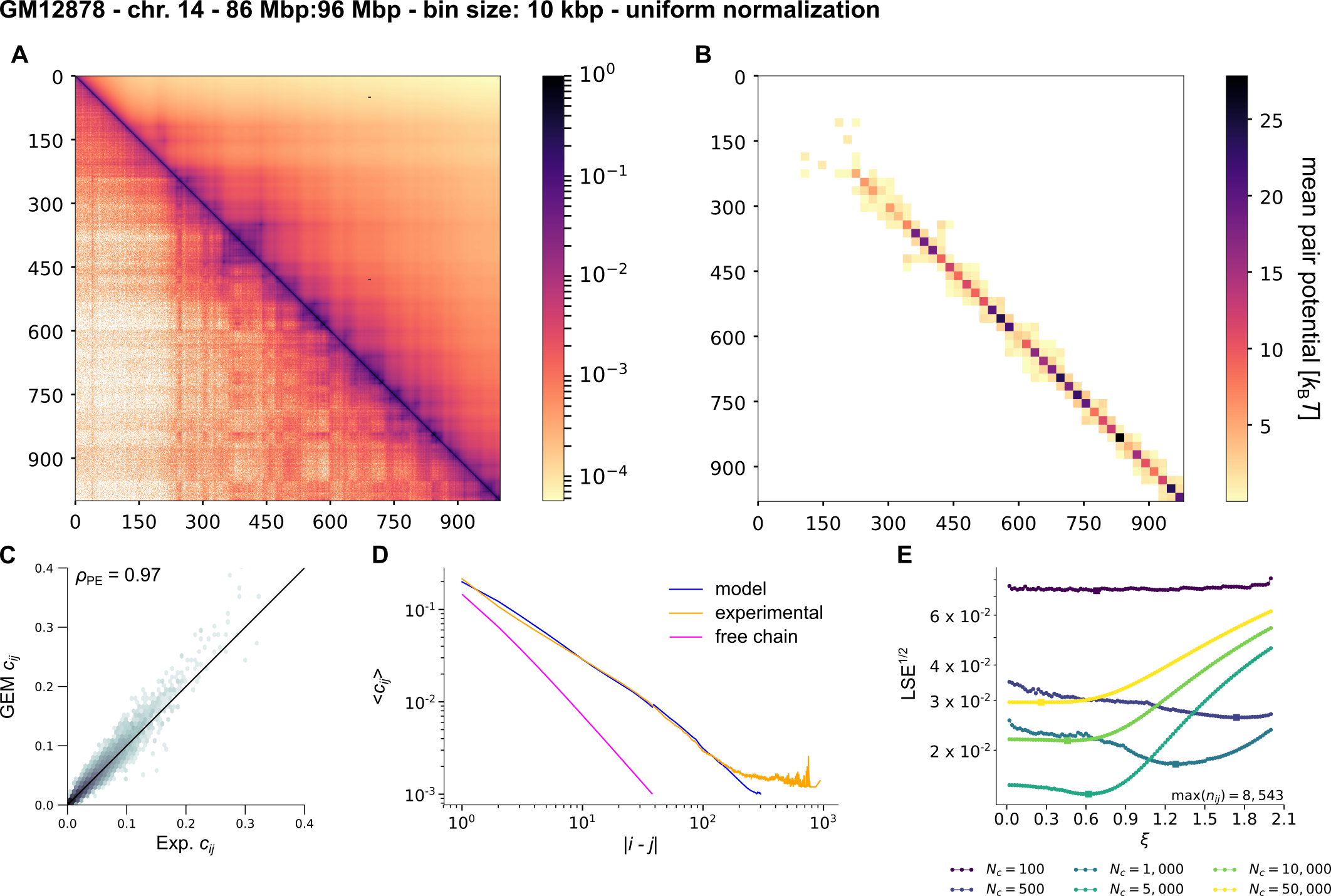}
  \caption{%
    GEM reconstruction for Hi-C data of human chromosome 14 \cite{rao2014a} (\SI{10}{\kilo bp} resolution), normalized by applying a global factor. \textbf{(A)} Comparison between experimental (lower left) and GEM (upper right) contact probabilities. \textbf{(B)} Matrix of mean pair potentials, binned with a ratio 1:20. \textbf{(C)} Comparison of experimental and GEM contact probabilities (2d-histogram). We give the Pearson correlation coefficient. \textbf{(D)} Average contact probability as a function of the contour length. \textbf{(E)} LSE as a function of the threshold $\xi$ used for the GEM mapping, and for different normalizations $N_c$ of the Hi-C counts.%
  }%
  \label{fig:rao_chr14_10Kbp_N1000_global}%
\end{figure}

\begin{figure}%
  \centering%
  \includegraphics[width = \linewidth]{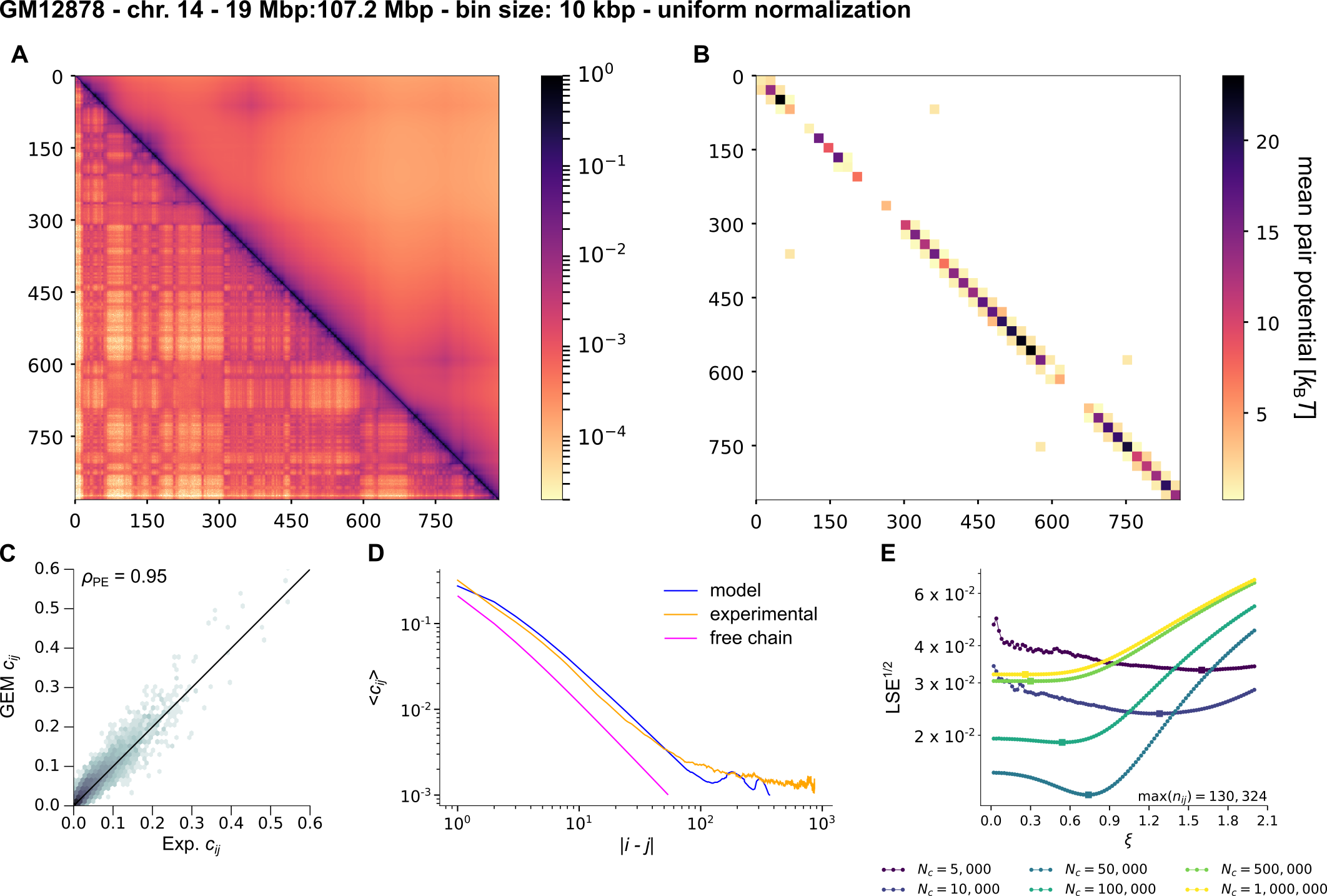}
  \caption{%
    GEM reconstruction for Hi-C data of human chromosome 14 \cite{rao2014a} (\SI{100}{\kilo bp} resolution), normalized by applying a global factor. \textbf{(A)} Comparison between experimental (lower left) and GEM (upper right) contact probabilities. \textbf{(B)} Matrix of mean pair potentials, binned with a ratio 1:20. \textbf{(C)} Comparison of experimental and GEM contact probabilities (2d-histogram). We give the Pearson correlation coefficient. \textbf{(D)} Average contact probability as a function of the contour length. \textbf{(E)} LSE as a function of the threshold $\xi$ used for the GEM mapping, and for different normalizations $N_c$ of the Hi-C counts.%
  }%
  \label{fig:rao_chr14_100Kbp_global}%
\end{figure}

\begin{figure}%
  \centering%
  \includegraphics[width = \linewidth]{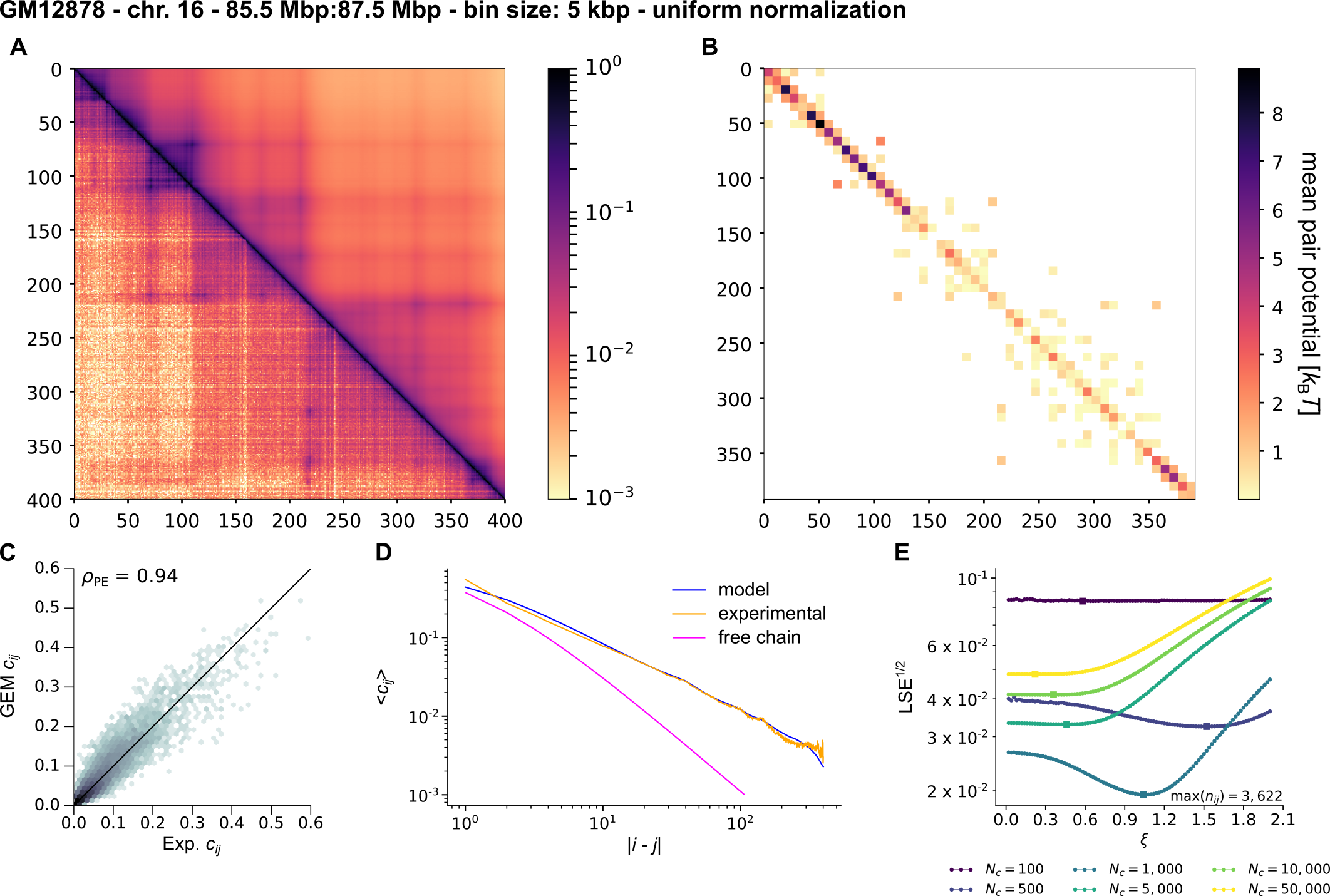}
  \caption{%
    GEM reconstruction for Hi-C data of human chromosome 16 \cite{rao2014a} (\SI{5}{\kilo bp} resolution), normalized by applying a global factor. \textbf{(A)} Comparison between experimental (lower left) and GEM (upper right) contact probabilities. \textbf{(B)} Matrix of mean pair potentials, binned with a ratio 1:8. \textbf{(C)} Comparison of experimental and GEM contact probabilities (2d-histogram). We give the Pearson correlation coefficient. \textbf{(D)} Average contact probability as a function of the contour length. \textbf{(E)} LSE as a function of the threshold $\xi$ used for the GEM mapping, and for different normalizations $N_c$ of the Hi-C counts.%
  }%
  \label{fig:rao_chr16_5Kbp_global}%
\end{figure}

\begin{figure}%
  \centering%
  \includegraphics[width = \linewidth]{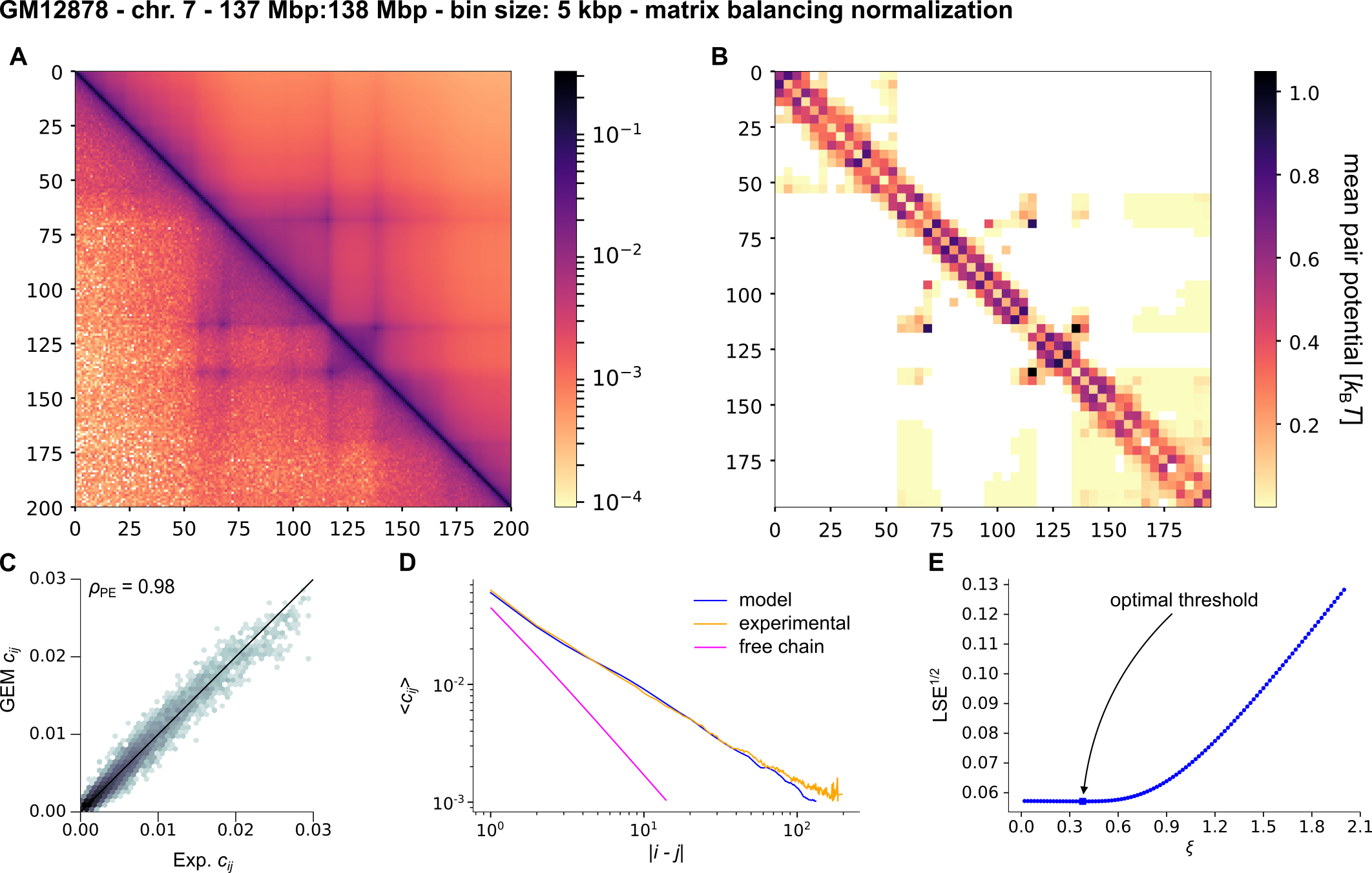}
  \caption{%
    GEM reconstruction for Hi-C data of human chromosome 7 \cite{rao2014a} (\SI{5}{\kilo bp} resolution), normalized by matrix balancing. \textbf{(A)} Comparison between experimental (lower left) and GEM (upper right) contact probabilities. \textbf{(B)} Matrix of mean pair potentials, binned with a ratio 1:4. \textbf{(C)} Comparison of experimental and GEM contact probabilities (2d-histogram). We give the Pearson correlation coefficient. \textbf{(D)} Average contact probability as a function of the contour length. \textbf{(E)} LSE as a function of the threshold $\xi$ used for the GEM mapping.%
  }%
  \label{fig:rao_chr7_5Kbp_stochastic}%
\end{figure}

\begin{figure}%
  \centering%
  \includegraphics[width = \linewidth]{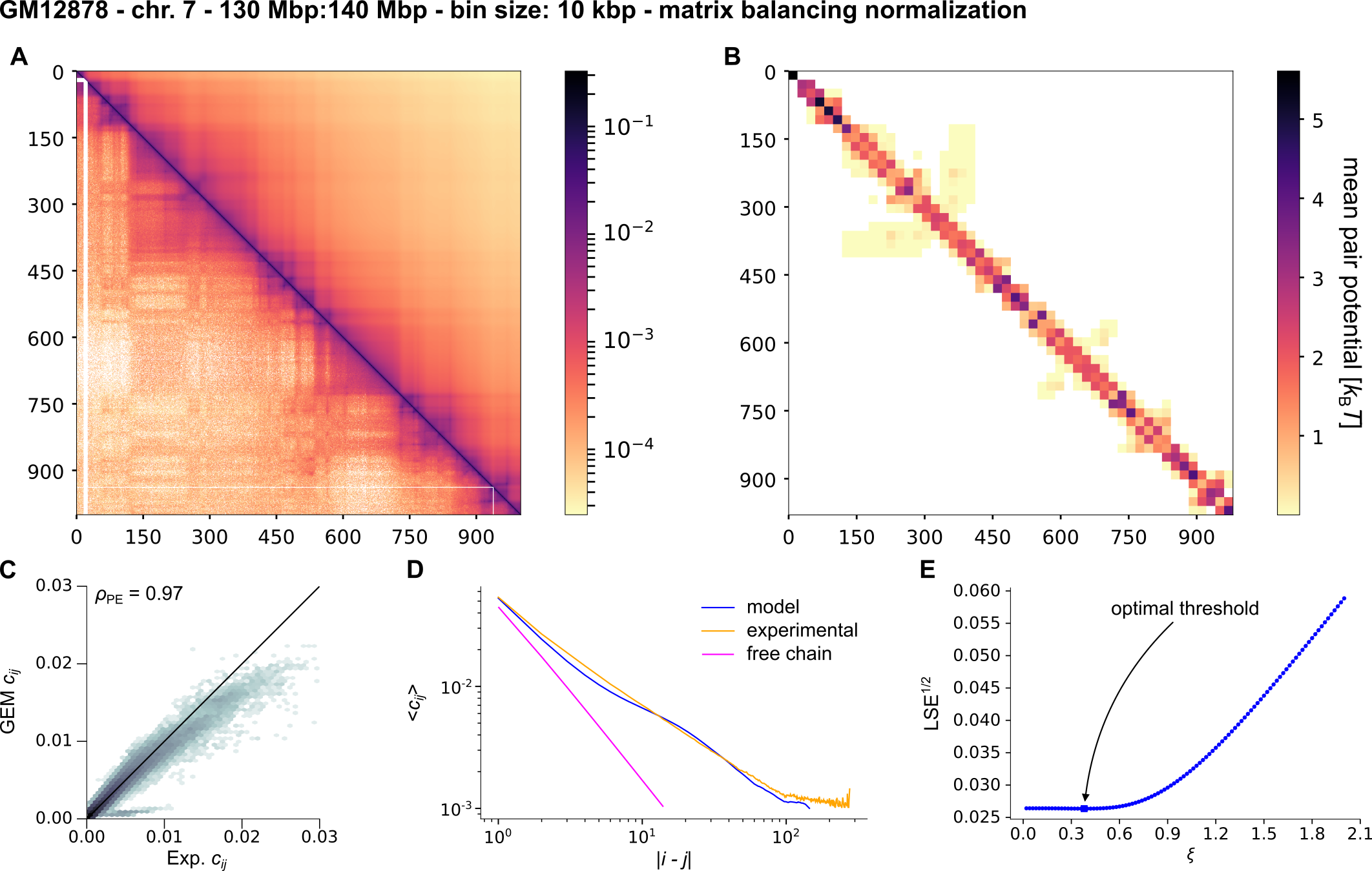}
  \caption{%
    GEM reconstruction for Hi-C data of human chromosome 7 \cite{rao2014a} (\SI{10}{\kilo bp} resolution), normalized by matrix balancing. \textbf{(A)} Comparison between experimental (lower left) and GEM (upper right) contact probabilities. \textbf{(B)} Matrix of mean pair potentials, binned with a ratio 1:20. \textbf{(C)} Comparison of experimental and GEM contact probabilities (2d-histogram). We give the Pearson correlation coefficient. \textbf{(D)} Average contact probability as a function of the contour length. \textbf{(E)} LSE as a function of the threshold $\xi$ used for the GEM mapping.%
  }%
  \label{fig:rao_chr7_10Kbp_stochastic}%
\end{figure}

\begin{figure}%
  \centering%
  \includegraphics[width = \linewidth]{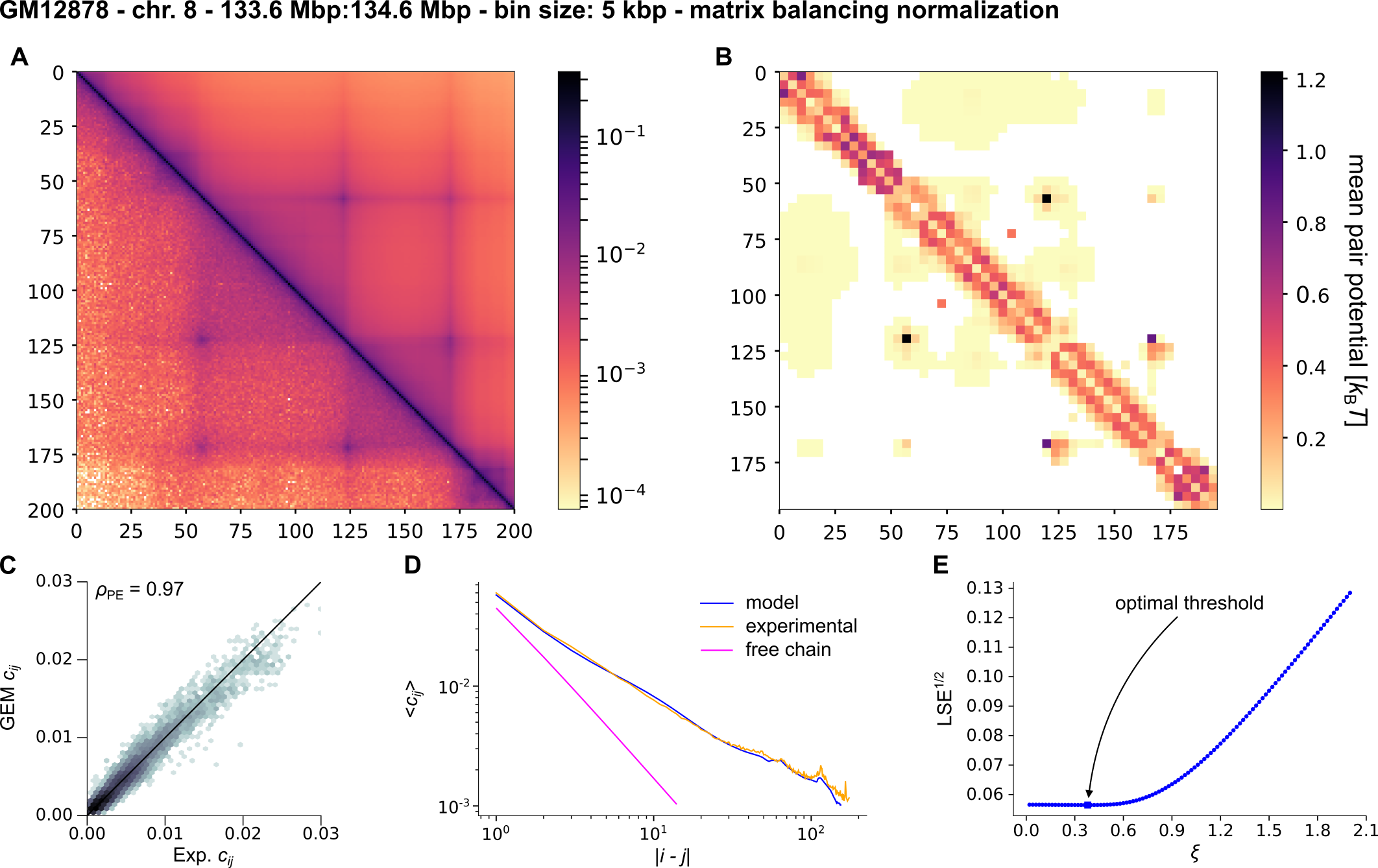}
  \caption{%
    GEM reconstruction for Hi-C data of human chromosome 8 \cite{rao2014a} (\SI{5}{\kilo bp} resolution), normalized by matrix balancing. \textbf{(A)} Comparison between experimental (lower left) and GEM (upper right) contact probabilities. \textbf{(B)} Matrix of mean pair potentials, binned with a ratio 1:4. \textbf{(C)} Comparison of experimental and GEM contact probabilities (2d-histogram). We give the Pearson correlation coefficient. \textbf{(D)} Average contact probability as a function of the contour length. \textbf{(E)} LSE as a function of the threshold $\xi$ used for the GEM mapping.%
  }%
  \label{fig:rao_chr8_5Kbp_stochastic}%
\end{figure}

\begin{figure}%
  \centering%
  \includegraphics[width = \linewidth]{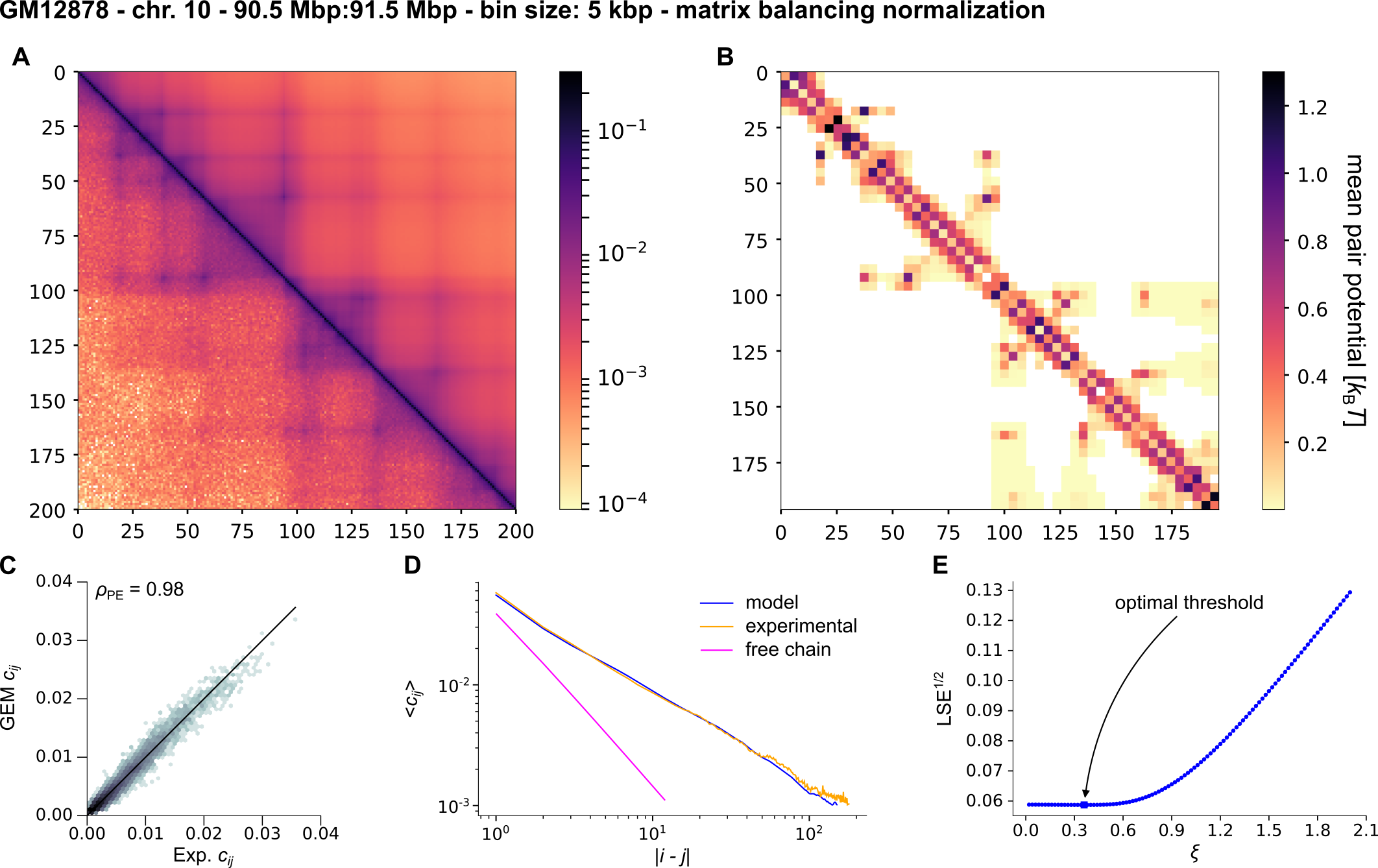}
  \caption{%
    GEM reconstruction for Hi-C data of human chromosome 10 \cite{rao2014a} (\SI{5}{\kilo bp} resolution), normalized by matrix balancing. \textbf{(A)} Comparison between experimental (lower left) and GEM (upper right) contact probabilities. \textbf{(B)} Matrix of mean pair potentials, binned with a ratio 1:4. \textbf{(C)} Comparison of experimental and GEM contact probabilities (2d-histogram). We give the Pearson correlation coefficient. \textbf{(D)} Average contact probability as a function of the contour length. \textbf{(E)} LSE as a function of the threshold $\xi$ used for the GEM mapping.%
  }%
  \label{fig:rao_chr10_5Kbp_stochastic}%
\end{figure}

\begin{figure}%
  \centering%
  \includegraphics[width = \linewidth]{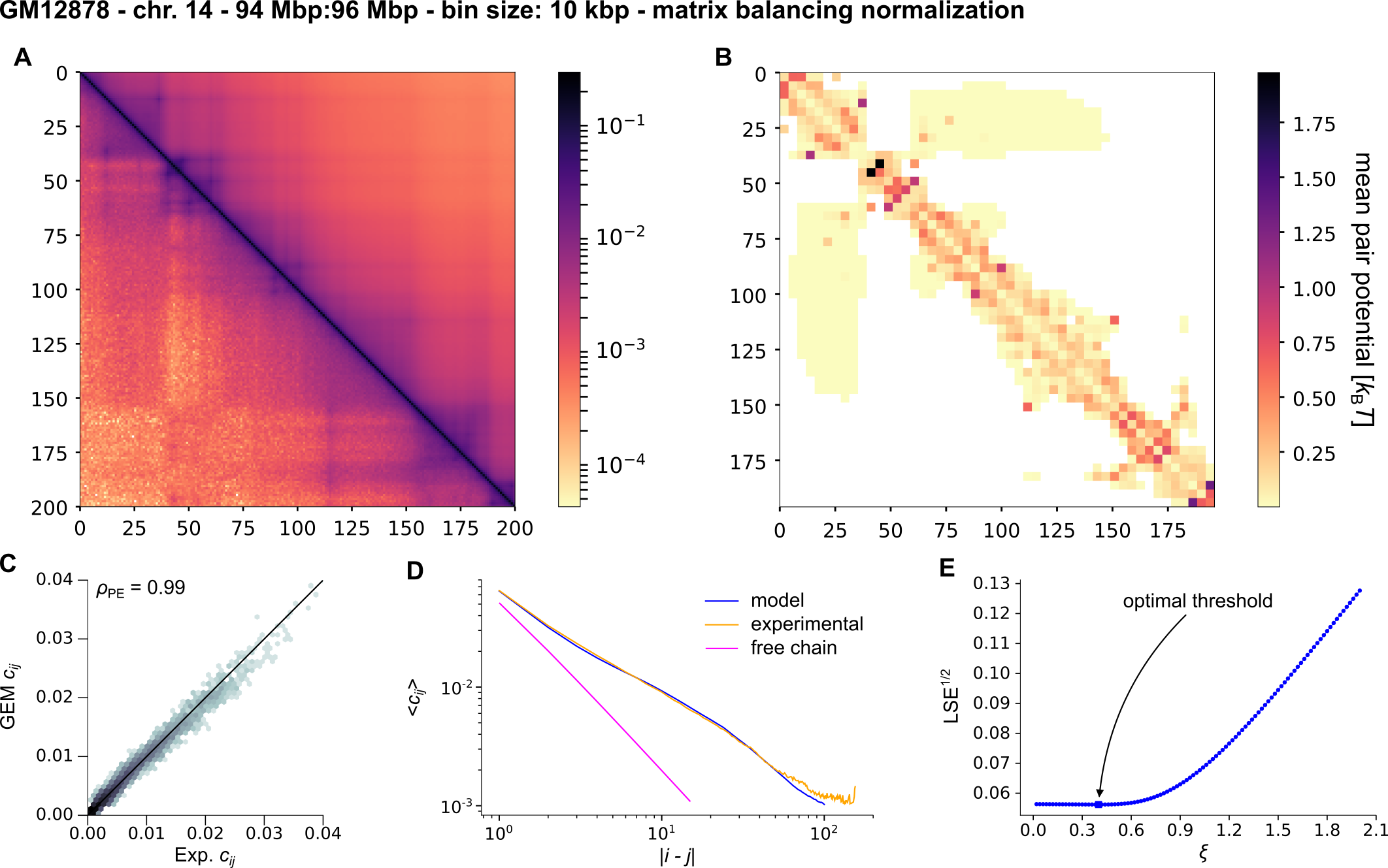}
  \caption{%
    GEM reconstruction for Hi-C data of human chromosome 14 \cite{rao2014a} (\SI{10}{\kilo bp} resolution), normalized by matrix balancing. \textbf{(A)} Comparison between experimental (lower left) and GEM (upper right) contact probabilities. \textbf{(B)} Matrix of mean pair potentials, binned with a ratio 1:4. \textbf{(C)} Comparison of experimental and GEM contact probabilities (2d-histogram). We give the Pearson correlation coefficient. \textbf{(D)} Average contact probability as a function of the contour length. \textbf{(E)} LSE as a function of the threshold $\xi$ used for the GEM mapping.%
  }%
  \label{fig:rao_chr14_10Kbp_N200_stochastic}%
\end{figure}

\begin{figure}%
  \centering%
  \includegraphics[width = \linewidth]{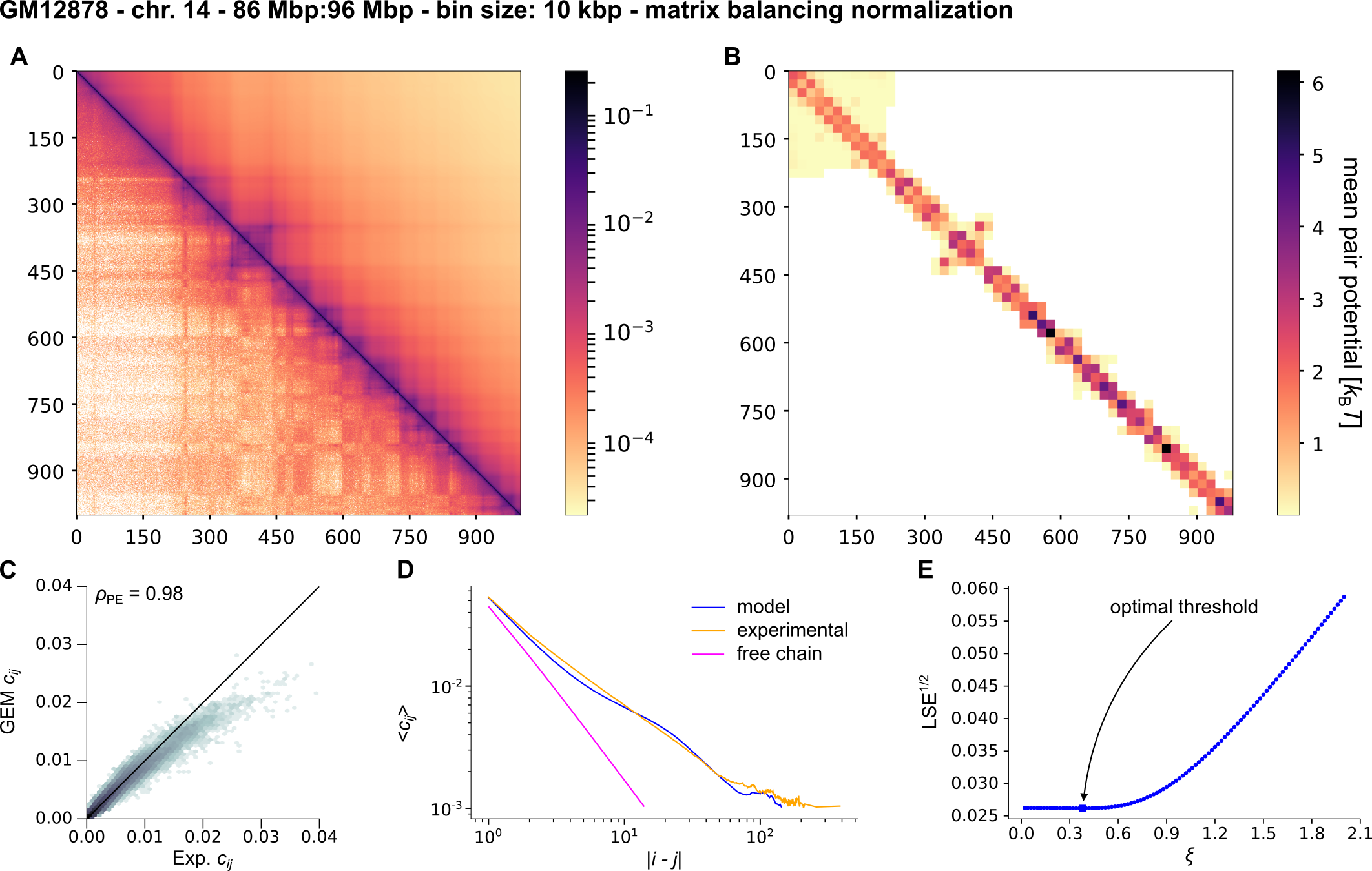}
  \caption{%
    GEM reconstruction for Hi-C data of human chromosome 14 \cite{rao2014a} (\SI{10}{\kilo bp} resolution), normalized by matrix balancing. \textbf{(A)} Comparison between experimental (lower left) and GEM (upper right) contact probabilities. \textbf{(B)} Matrix of mean pair potentials, binned with a ratio 1:20. \textbf{(C)} Comparison of experimental and GEM contact probabilities (2d-histogram). We give the Pearson correlation coefficient. \textbf{(D)} Average contact probability as a function of the contour length. \textbf{(E)} LSE as a function of the threshold $\xi$ used for the GEM mapping.%
  }%
  \label{fig:rao_chr14_10Kbp_N1000_stochastic}%
\end{figure}

\begin{figure}%
  \centering%
  \includegraphics[width = \linewidth]{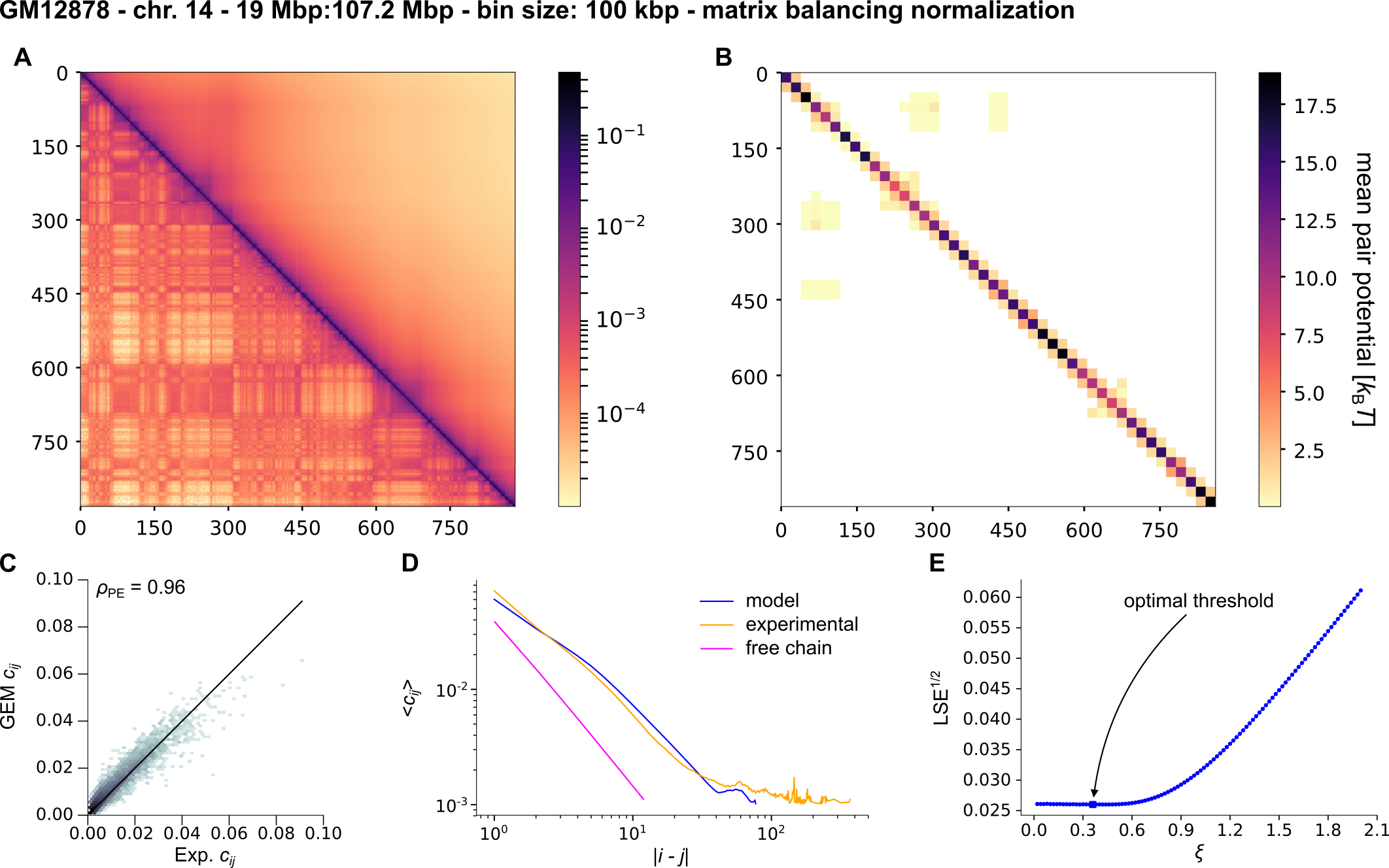}
  \caption{%
    GEM reconstruction for Hi-C data of human chromosome 14 \cite{rao2014a} (\SI{100}{\kilo bp} resolution), normalized by matrix balancing. \textbf{(A)} Comparison between experimental (lower left) and GEM (upper right) contact probabilities. \textbf{(B)} Matrix of mean pair potentials, binned with a ratio 1:20. \textbf{(C)} Comparison of experimental and GEM contact probabilities (2d-histogram). We give the Pearson correlation coefficient. \textbf{(D)} Average contact probability as a function of the contour length. \textbf{(E)} LSE as a function of the threshold $\xi$ used for the GEM mapping.%
  }%
  \label{fig:rao_chr14_100Kbp_stochastic}%
\end{figure}

\begin{figure}%
  \centering%
  \includegraphics[width = \linewidth]{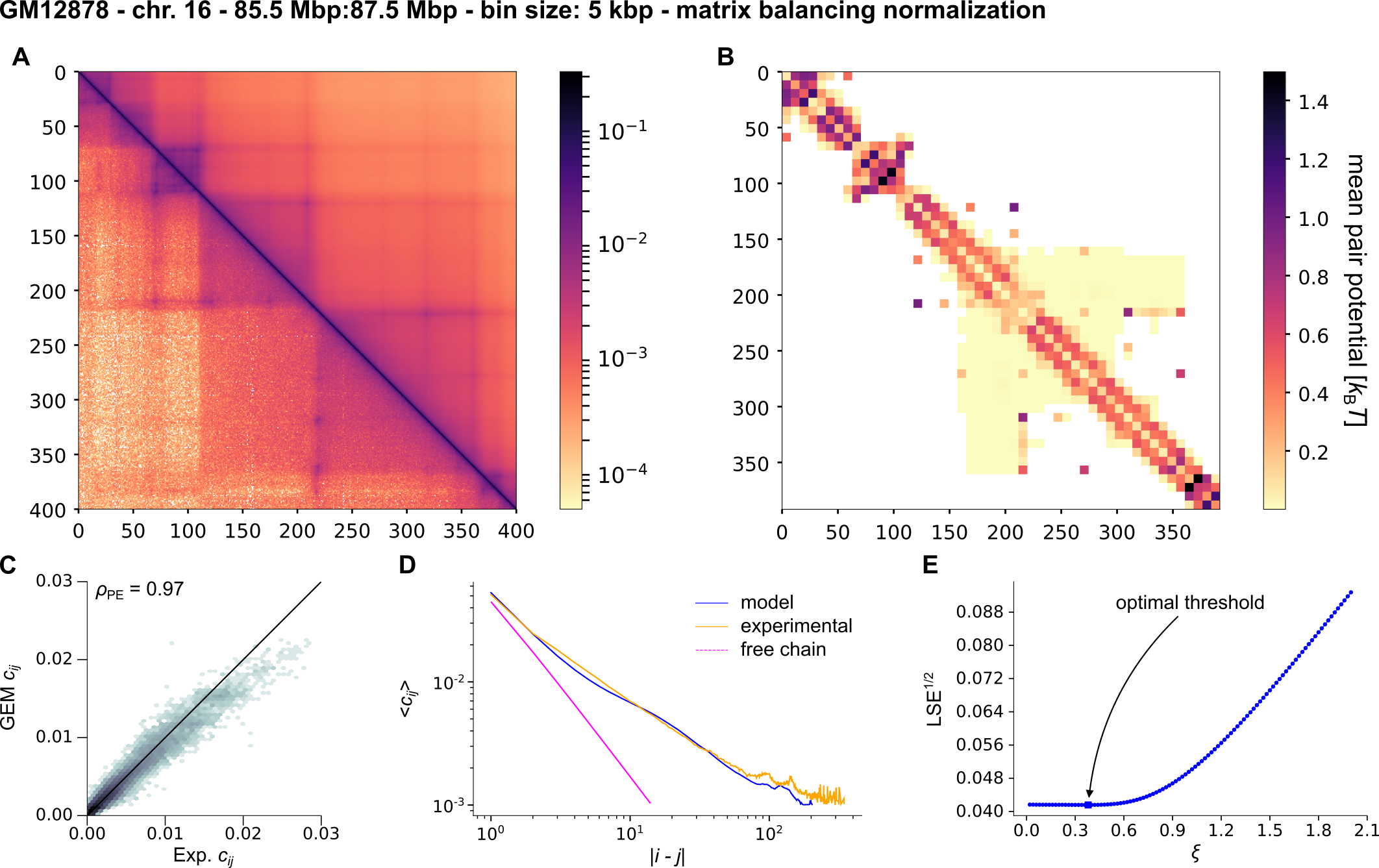}
  \caption{%
    GEM reconstruction for Hi-C data of human chromosome 16 \cite{rao2014a} (\SI{5}{\kilo bp} resolution), normalized by matrix balancing. \textbf{(A)} Comparison between experimental (lower left) and GEM (upper right) contact probabilities. \textbf{(B)} Matrix of mean pair potentials, binned with a ratio 1:8. \textbf{(C)} Comparison of experimental and GEM contact probabilities (2d-histogram). We give the Pearson correlation coefficient. \textbf{(D)} Average contact probability as a function of the contour length. \textbf{(E)} LSE as a function of the threshold $\xi$ used for the GEM mapping.%
  }%
  \label{fig:rao_chr16_5Kbp_stochastic}%
\end{figure}

\begin{figure}%
  \centering%
  \includegraphics[width = \linewidth]{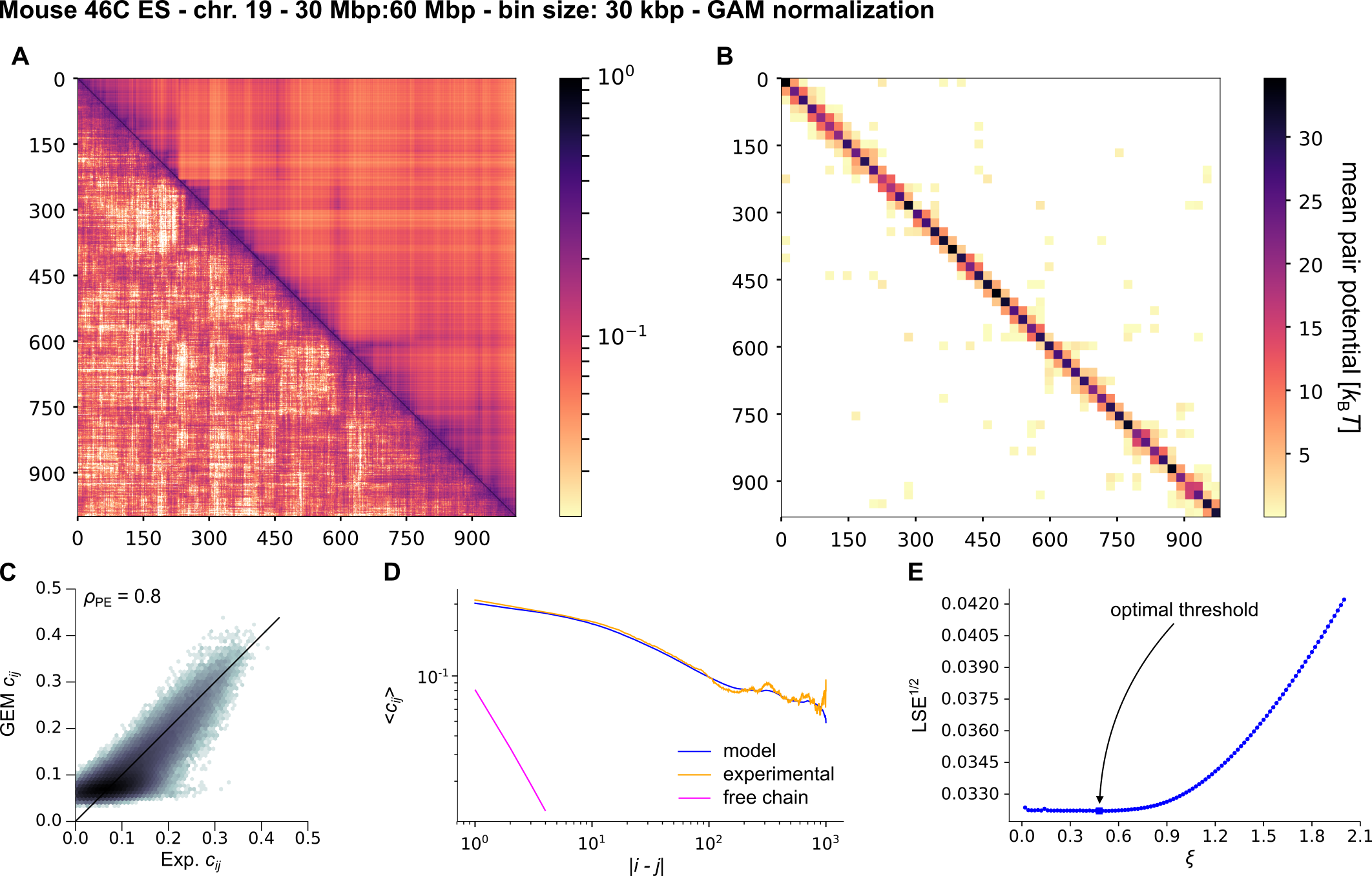}
  \caption{%
    GEM reconstruction for GAM data of mouse chromosome 19 \cite{beagrie2017complex} (\SI{30}{\kilo bp} resolution). \textbf{(A)} Comparison between experimental (lower left) and GEM (upper right) contact probabilities. \textbf{(B)} Matrix of mean pair potentials, binned with a ratio 1:20. \textbf{(C)} Comparison of experimental and GEM contact probabilities (2d-histogram). We give the Pearson correlation coefficient. \textbf{(D)} Average contact probability as a function of the contour length. \textbf{(E)} LSE as a function of the threshold $\xi$ used for the GEM mapping.%
  }%
  \label{fig:beagrie_chr19_30Kbp}%
\end{figure}

\begin{figure}%
  \centering%
  \includegraphics[width = \linewidth]{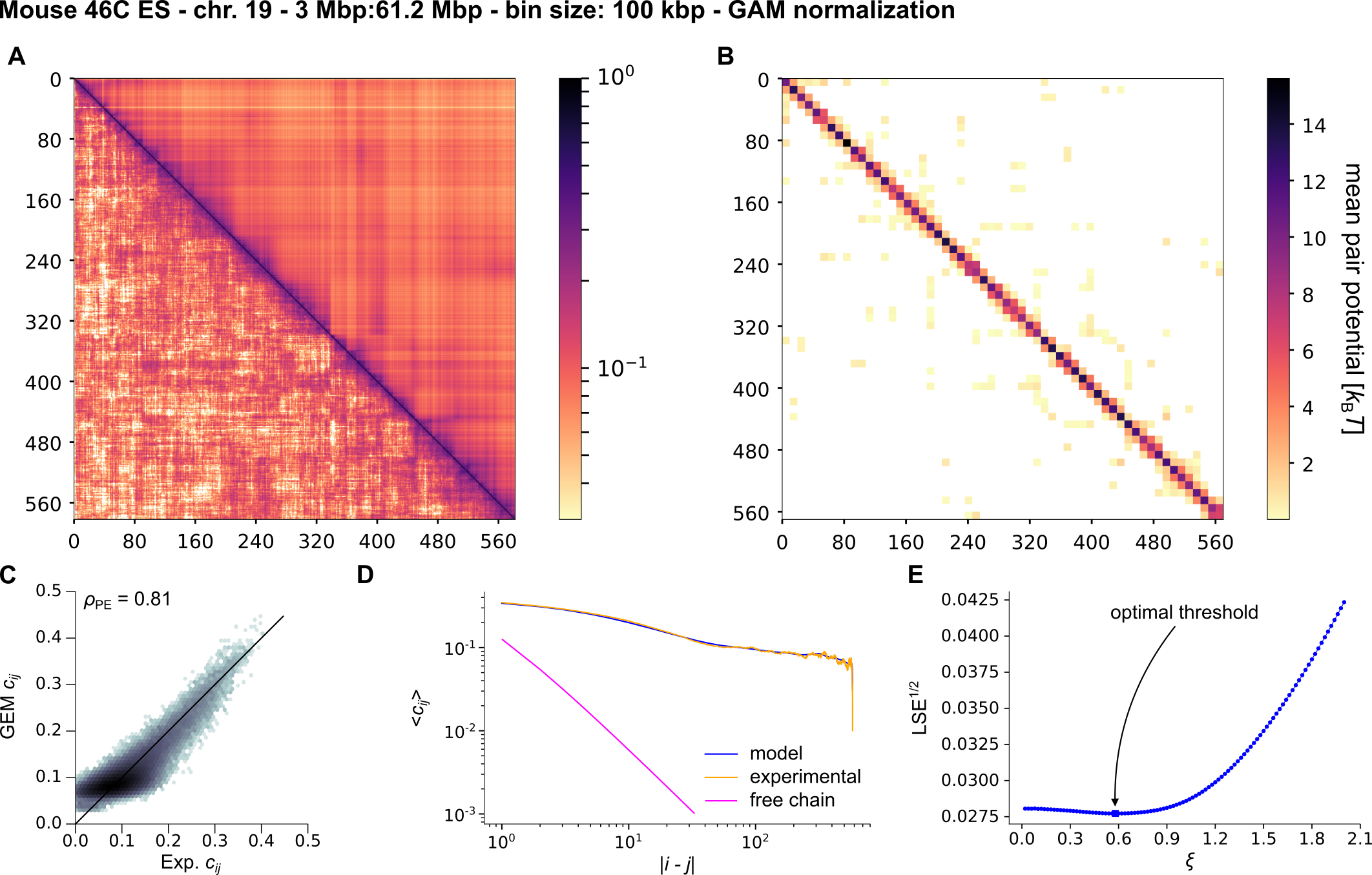}
  \caption{%
    GEM reconstruction for GAM data of mouse chromosome 19 \cite{beagrie2017complex} (\SI{100}{\kilo bp} resolution). \textbf{(A)} Comparison between experimental (lower left) and GEM (upper right) contact probabilities. \textbf{(B)} Matrix of mean pair potentials, binned with a ratio 1:10. \textbf{(C)} Comparison of experimental and GEM contact probabilities (2d-histogram). We give the Pearson correlation coefficient. \textbf{(D)} Average contact probability as a function of the contour length. \textbf{(E)} LSE as a function of the threshold $\xi$ used for the GEM mapping.%
  }%
  \label{fig:beagrie_chr19_100Kbp}%
\end{figure}

\begin{figure}%
  \centering%
  \includegraphics[width = \linewidth]{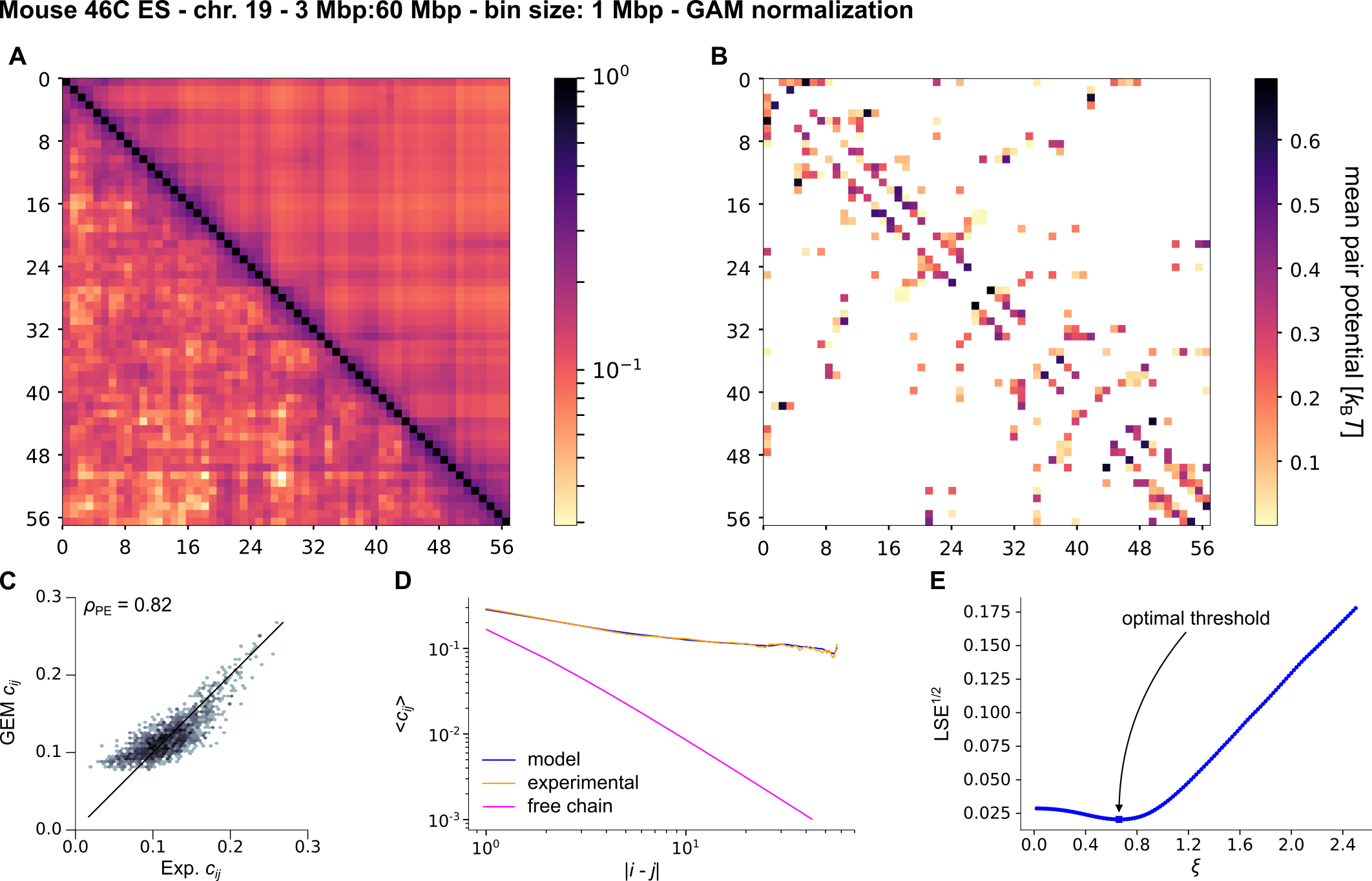}
  \caption{%
    GEM reconstruction for GAM data of mouse chromosome 19 \cite{beagrie2017complex} (\SI{1}{\mega bp} resolution). \textbf{(A)} Comparison between experimental (lower left) and GEM (upper right) contact probabilities. \textbf{(B)} Matrix of mean pair potentials. \textbf{(C)} Comparison of experimental and GEM contact probabilities (2d-histogram). We give the Pearson correlation coefficient. \textbf{(D)} Average contact probability as a function of the contour length. \textbf{(E)} LSE as a function of the threshold $\xi$ used for the GEM mapping.%
  }%
  \label{fig:beagrie_chr19_1Mbp}%
\end{figure}

\begin{figure}%
  \centering%
  \includegraphics[width = \linewidth]{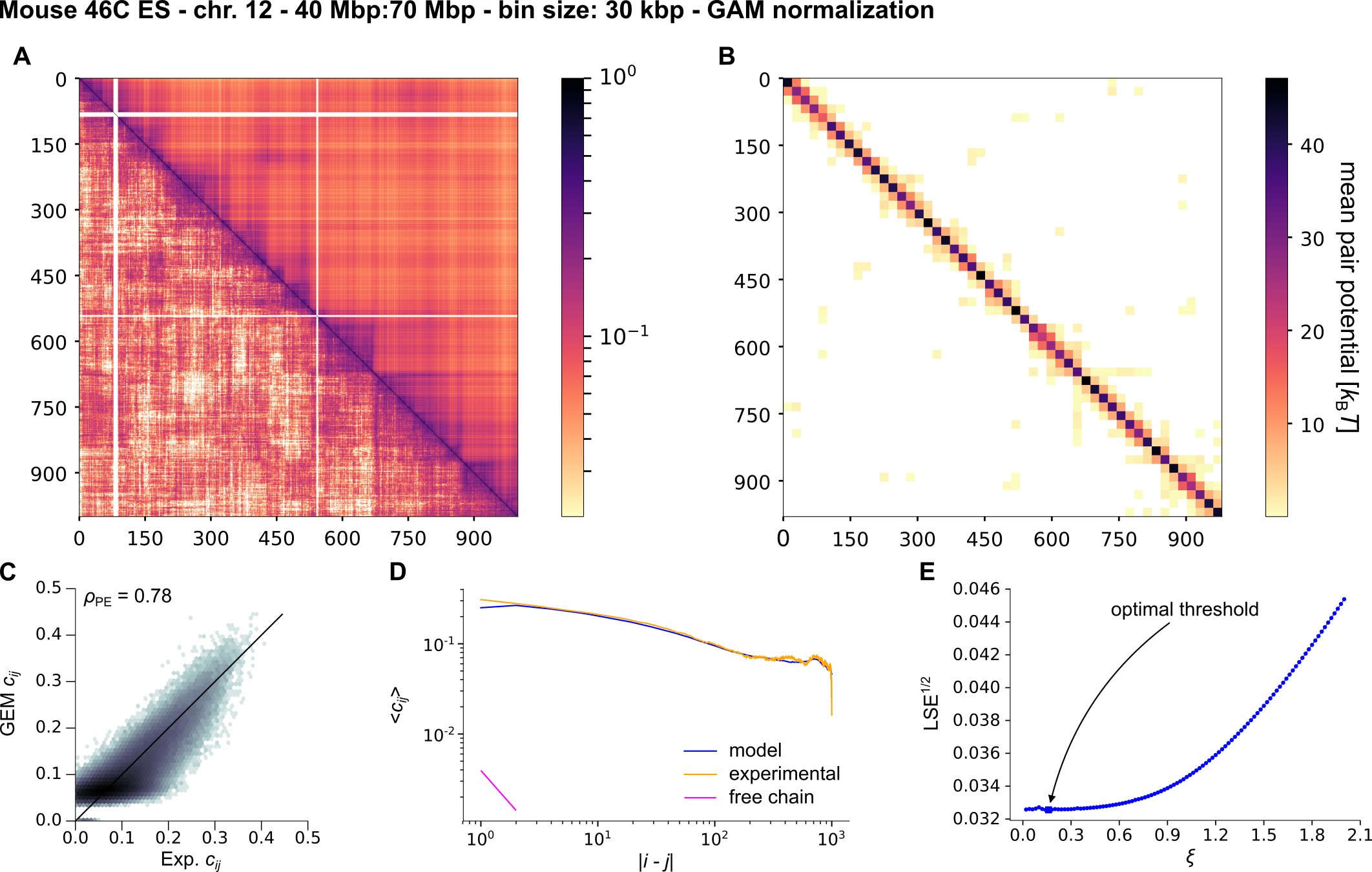}
  \caption{%
    GEM reconstruction for GAM data of mouse chromosome 12 \cite{beagrie2017complex} (\SI{30}{\kilo bp} resolution). \textbf{(A)} Comparison between experimental (lower left) and GEM (upper right) contact probabilities. \textbf{(B)} Matrix of mean pair potentials, binned with a ratio 1:20. \textbf{(C)} Comparison of experimental and GEM contact probabilities (2d-histogram). We give the Pearson correlation coefficient. \textbf{(D)} Average contact probability as a function of the contour length. \textbf{(E)} LSE as a function of the threshold $\xi$ used for the GEM mapping.%
  }%
  \label{fig:beagrie_chr12_30Kbp}%
\end{figure}

\begin{figure}%
  \centering%
  \includegraphics[width = \linewidth]{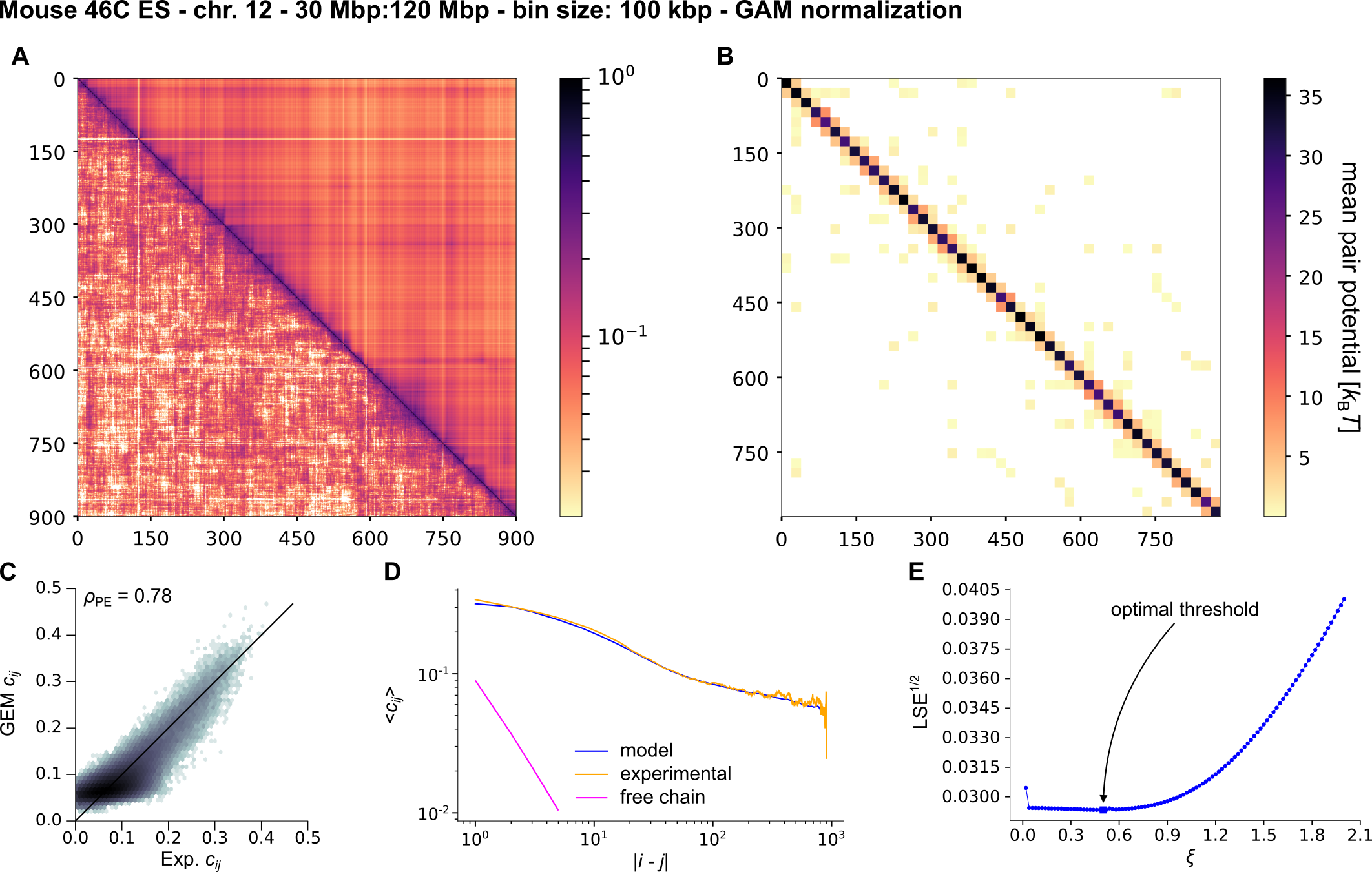}
  \caption{%
    GEM reconstruction for GAM data of mouse chromosome 12 \cite{beagrie2017complex} (\SI{100}{\kilo bp} resolution). \textbf{(A)} Comparison between experimental (lower left) and GEM (upper right) contact probabilities. \textbf{(B)} Matrix of mean pair potentials, binned with a ratio 1:20. \textbf{(C)} Comparison of experimental and GEM contact probabilities (2d-histogram). We give the Pearson correlation coefficient. \textbf{(D)} Average contact probability as a function of the contour length. \textbf{(E)} LSE as a function of the threshold $\xi$ used for the GEM mapping.%
  }%
  \label{fig:beagrie_chr12_100Kbp}%
\end{figure}

\begin{figure}%
  \centering%
  \includegraphics[width = \linewidth]{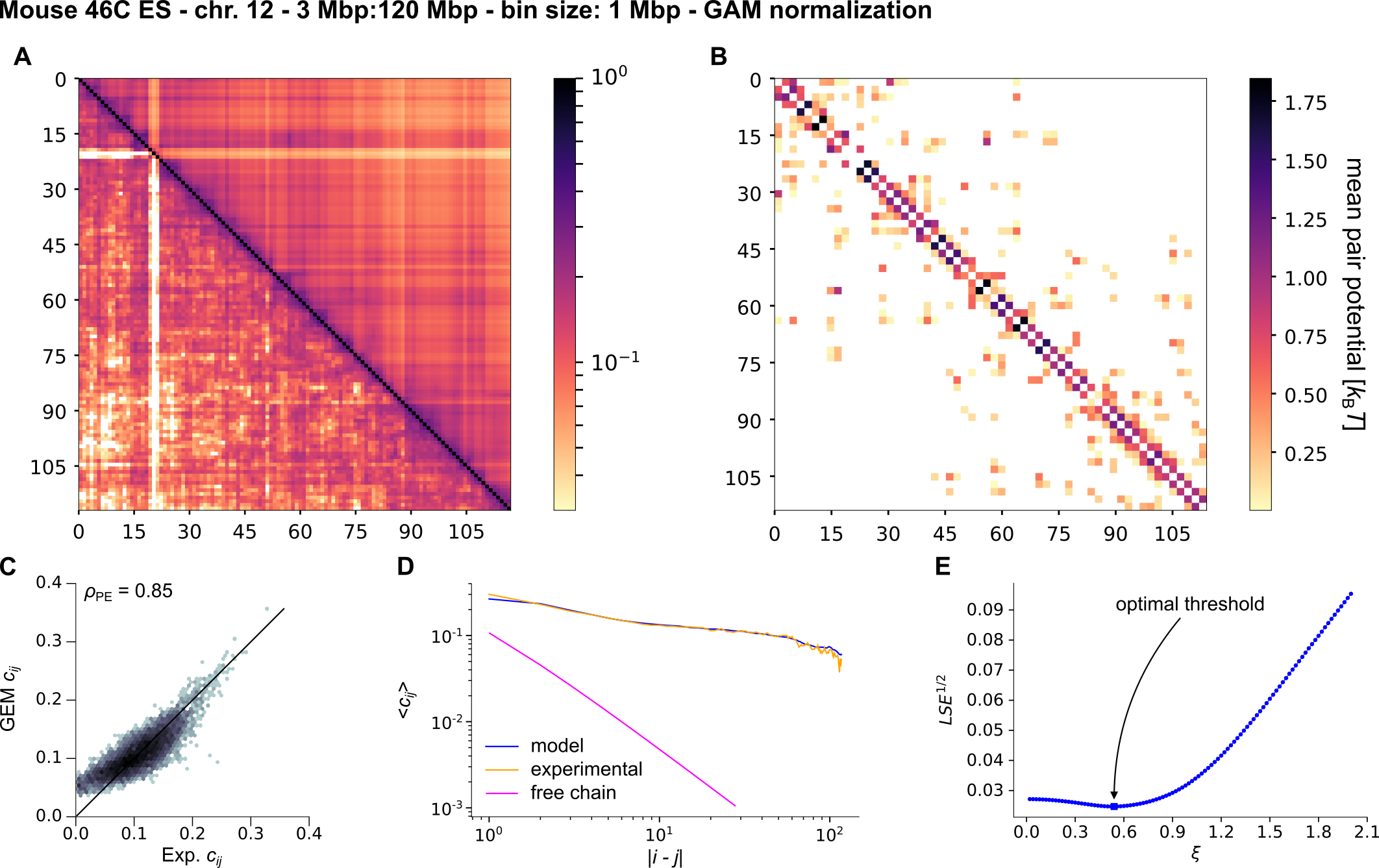}
  \caption{%
    GEM reconstruction for GAM data of mouse chromosome 12 \cite{beagrie2017complex} (\SI{1}{\mega bp} resolution). \textbf{(A)} Comparison between experimental (lower left) and GEM (upper right) contact probabilities. \textbf{(B)} Matrix of mean pair potentials. \textbf{(C)} Comparison of experimental and GEM contact probabilities (2d-histogram). We give the Pearson correlation coefficient. \textbf{(D)} Average contact probability as a function of the contour length. \textbf{(E)} LSE as a function of the threshold $\xi$ used for the GEM mapping.%
  }%
  \label{fig:beagrie_chr12_1Mbp}%
\end{figure}

\begin{figure}%
  \centering%
  \includegraphics[width = \linewidth]{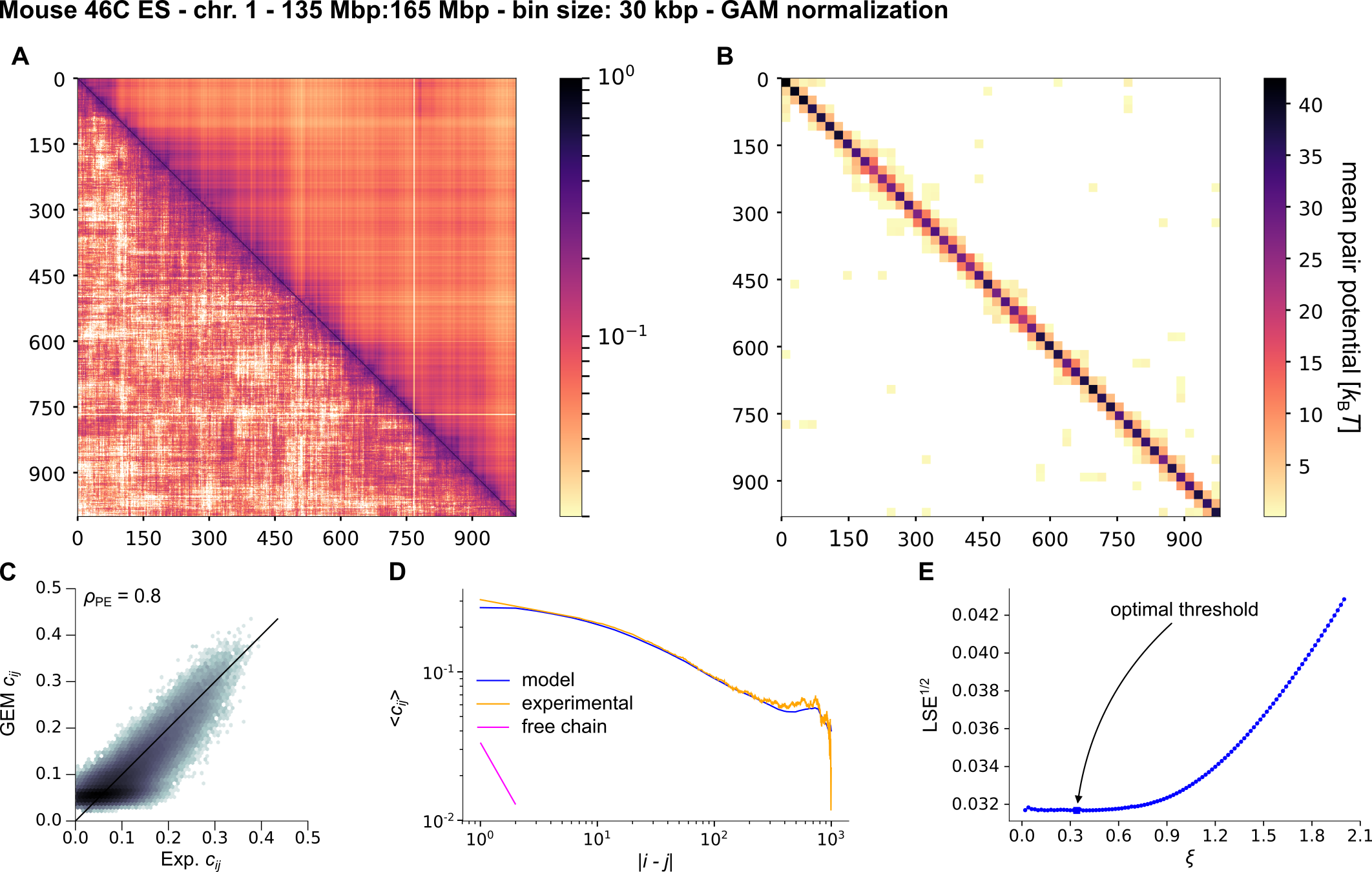}
  \caption{%
    GEM reconstruction for GAM data of mouse chromosome 1 \cite{beagrie2017complex} (\SI{30}{\kilo bp} resolution). \textbf{(A)} Comparison between experimental (lower left) and GEM (upper right) contact probabilities. \textbf{(B)} Matrix of mean pair potentials, binned with a ratio 1:20. \textbf{(C)} Comparison of experimental and GEM contact probabilities (2d-histogram). We give the Pearson correlation coefficient. \textbf{(D)} Average contact probability as a function of the contour length. \textbf{(E)} LSE as a function of the threshold $\xi$ used for the GEM mapping.%
  }%
  \label{fig:beagrie_chr1_30Kbp}%
\end{figure}

\begin{figure}%
  \centering%
  \includegraphics[width = \linewidth]{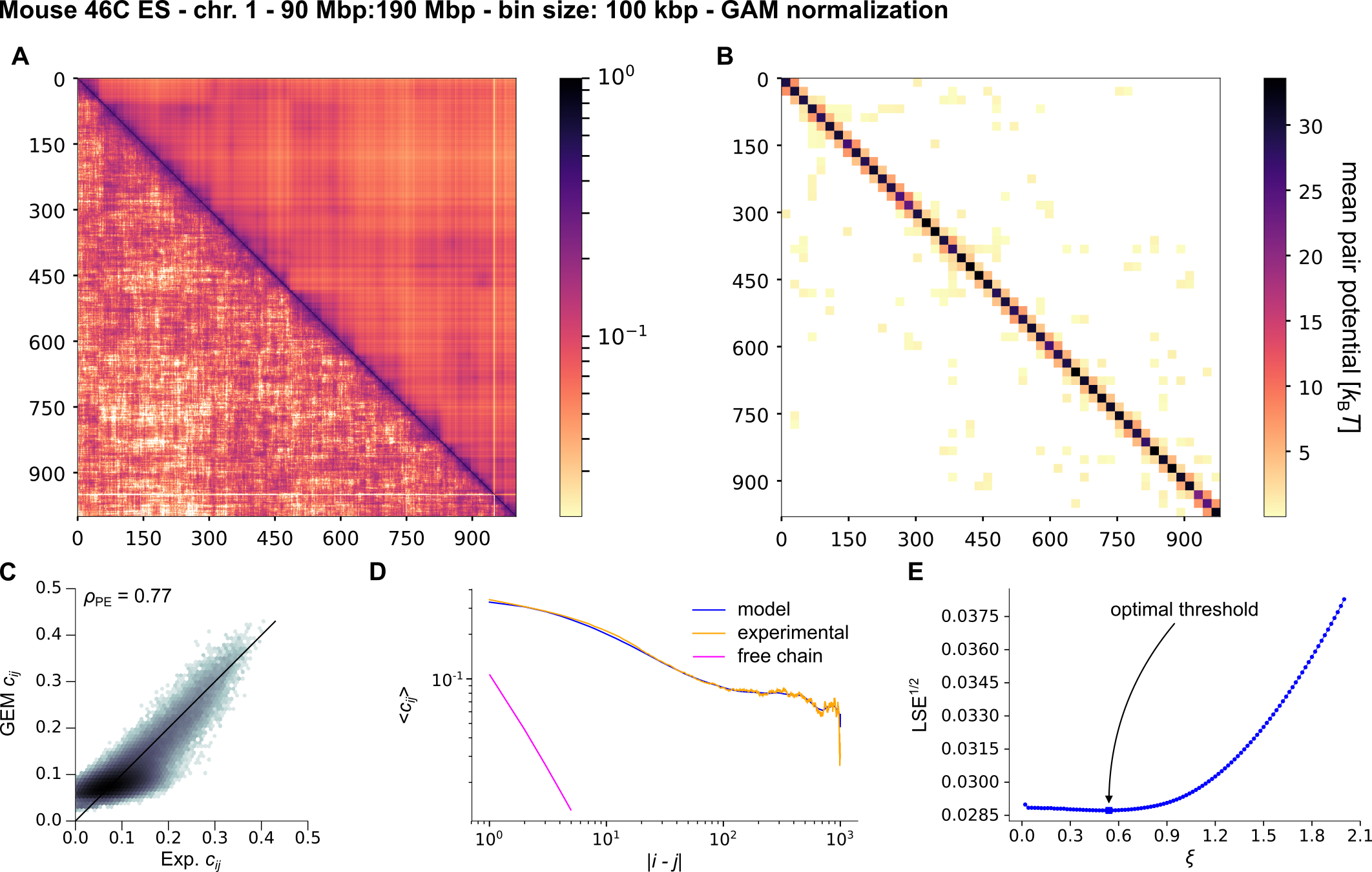}
  \caption{%
    GEM reconstruction for GAM data of mouse chromosome 1 \cite{beagrie2017complex} (\SI{100}{\kilo bp} resolution). \textbf{(A)} Comparison between experimental (lower left) and GEM (upper right) contact probabilities. \textbf{(B)} Matrix of mean pair potentials, binned with a ratio 1:20. \textbf{(C)} Comparison of experimental and GEM contact probabilities (2d-histogram). We give the Pearson correlation coefficient. \textbf{(D)} Average contact probability as a function of the contour length. \textbf{(E)} LSE as a function of the threshold $\xi$ used for the GEM mapping.%
  }%
  \label{fig:beagrie_chr1_100Kbp}%
\end{figure}

\begin{figure}%
  \centering%
  \includegraphics[width = \linewidth]{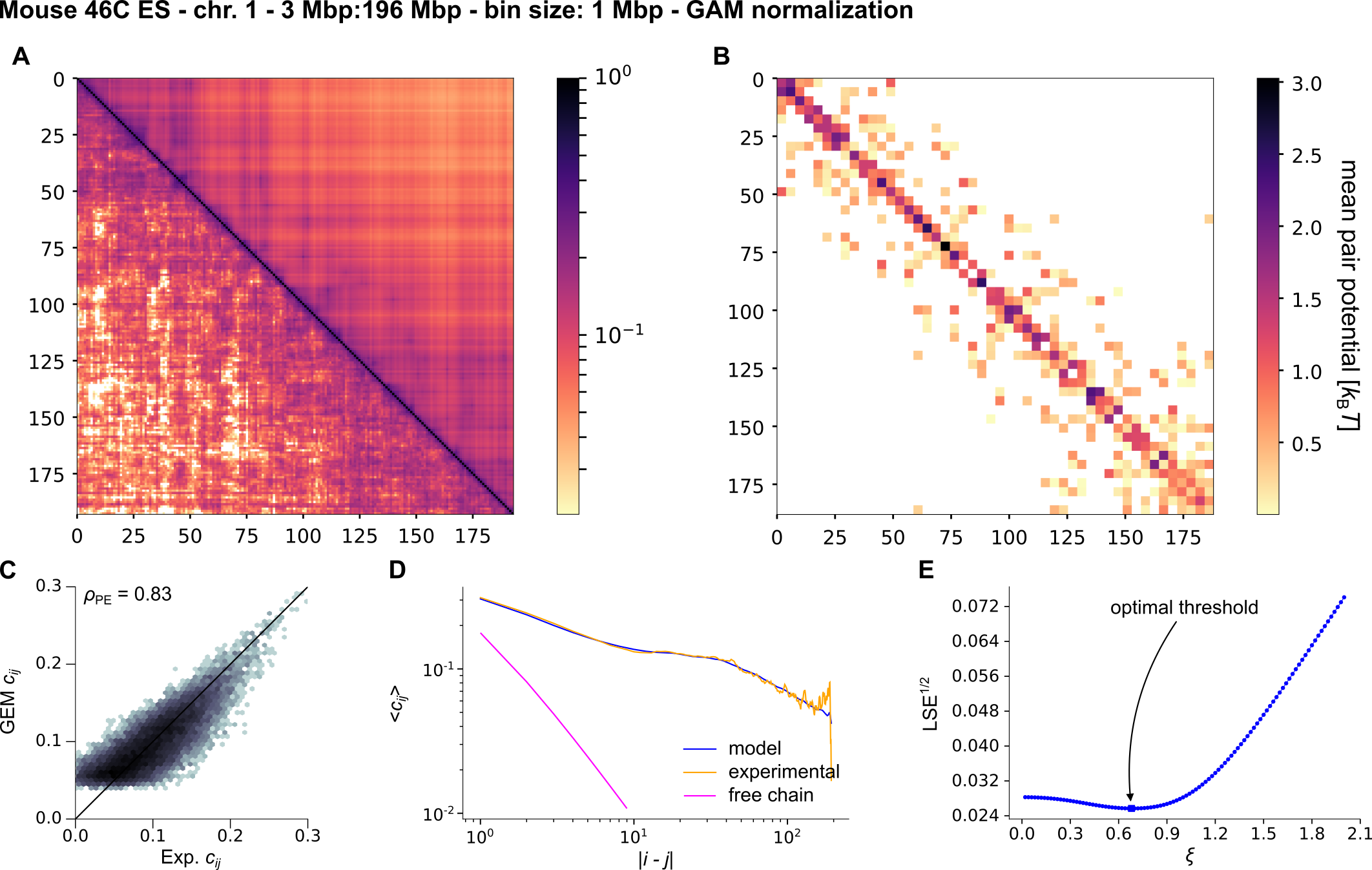}
  \caption{%
    GEM reconstruction for GAM data of mouse chromosome 1 \cite{beagrie2017complex} (\SI{1}{\mega bp} resolution). \textbf{(A)} Comparison between experimental (lower left) and GEM (upper right) contact probabilities. \textbf{(B)} Matrix of mean pair potentials. \textbf{(C)} Comparison of experimental and GEM contact probabilities (2d-histogram). We give the Pearson correlation coefficient. \textbf{(D)} Average contact probability as a function of the contour length. \textbf{(E)} LSE as a function of the threshold $\xi$ used for the GEM mapping.%
  }%
  \label{fig:beagrie_chr1_1Mbp}%
\end{figure}

\begin{figure}%
  \centering%
  \includegraphics[width = 0.6 \linewidth]{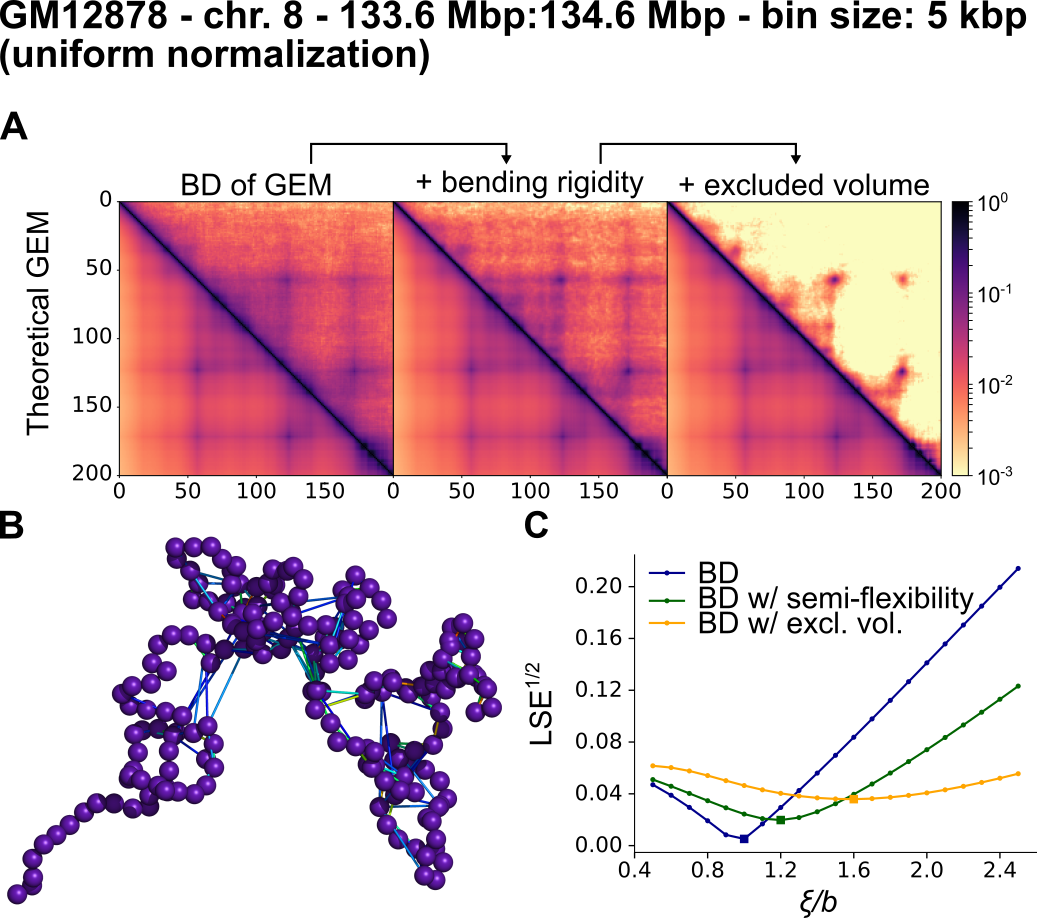}
  \caption{%
    Brownian dynamics (BD) of the reconstructed GEM for Hi-C data of human chromosome 8 \cite{rao2014a} (\SI{5}{\kilo bp} resolution). \textbf{(A)} Contact probability matrices obtained through BD simulation of: (i) the GEM, (ii) the GEM with bending rigidity, and (iii) the GEM with bending rigidity and with excluded volume. The contact probabilities were computed from BD trajectories and are compared with the theoretical values for the GEM. \textbf{(B)} Snapshot of a configuration obtained by BD of the reconstructed GEM with bending rigidity and excluded volume. The couplings are represented by tie lines, from weak couplings (in blue) to strong couplings (in red). \textbf{(C)} LSE as a function of the threshold $\xi$ between contact probabilities computed from the BD trajectory and the theoretical values.%
  }%
  \label{fig:rao_chr8_5Kbp_brownian_dynamics}%
\end{figure}

\begin{figure}%
  \centering%
  \includegraphics[width = 0.6 \linewidth]{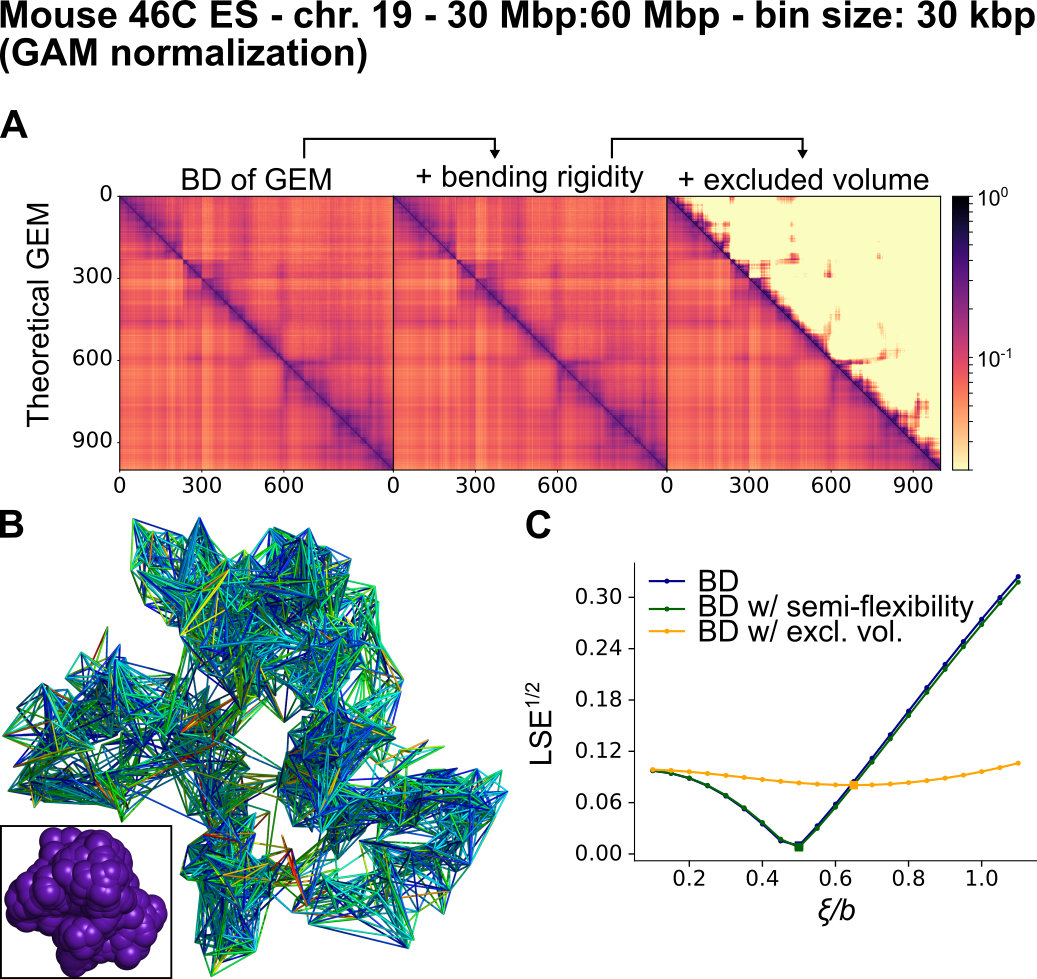}
  \caption{%
    Brownian dynamics (BD) of the reconstructed GEM for GAM data of mouse chromosome 19 \cite{beagrie2017complex} (\SI{30}{\kilo bp} resolution). \textbf{(A)} Contact probability matrices obtained through BD simulation of: (i) the GEM, (ii) the GEM with bending rigidity, and (iii) the GEM with bending rigidity and with excluded volume. The contact probabilities were computed from BD trajectories and are compared with the theoretical values for the GEM. \textbf{(B)} Snapshot of a configuration obtained by BD of the reconstructed GEM with bending rigidity and excluded volume. The couplings are represented by tie lines, from weak couplings (in blue) to strong couplings (in red). The inset shows the same configuration with the monomers. Note that the hard-core distance is $\sigma=1$ whereas the bond length is $b=8$. \textbf{(C)} LSE as a function of the threshold $\xi$ between contact probabilities computed from the BD trajectory and the theoretical values.%
  }%
  \label{fig:beagrie_chr19_30Kbp_brownian_dynamics}%
\end{figure}

\begin{figure}%
  \centering%
  \includegraphics[width = 0.6 \linewidth]{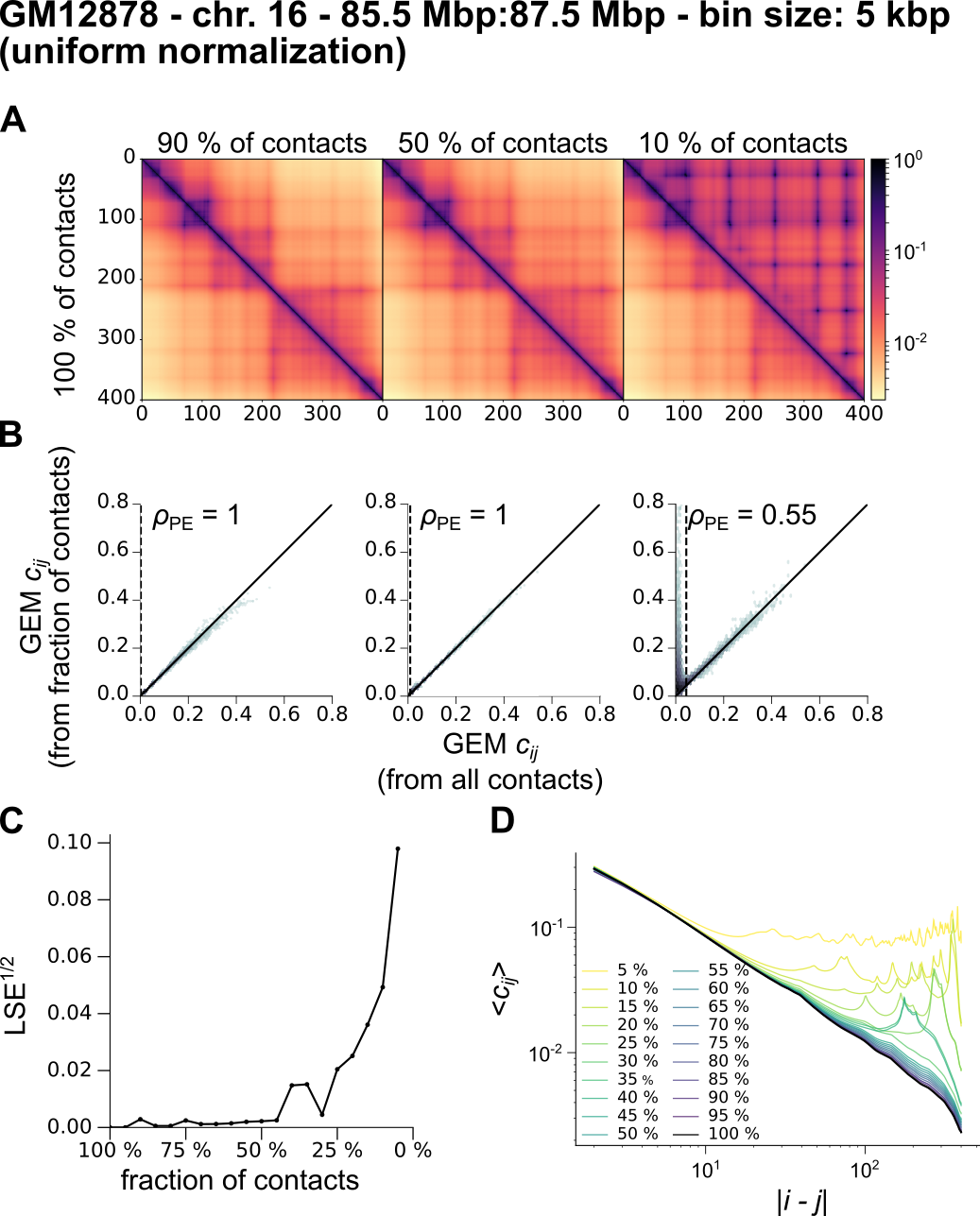}
  \caption{%
    Robustness of GEM reconstruction for Hi-C data of human chromosome 16 \cite{rao2014a} (\SI{5}{\kilo bp} resolution). For all GEM reconstructions we used a threshold $\xi = \num{1}$ and a normalization factor $N_c = \num{e3}$. \textbf{(A)} Comparison of the contact probabilities of the reconstructed GEM with those of a GEM obtained by performing the minimization only on the top \SI{90}{\percent}, \SI{50}{\percent} and \SI{10}{\percent} experimental contacts. \textbf{(B)} 2d-histograms corresponding to the matrices shown in \textbf{(A)}. We give the Pearson correlation coefficients. The thresholding quantiles are represented by vertical dashed lines. \textbf{(C)} Comparison of the GEMs reconstructed from a decreasing fraction of the experimental contacts with the original GEM. $\mathrm{LSE}^{1/2}$ is the Euclidean distance between contact probabilities divided by ($N+1$). \textbf{(D)} Average contact probability as a function of the contour length for GEMs reconstructed from a decreasing fraction of the experimental contacts.%
}%
  \label{fig:rao_chr16_5Kbp_significant_contacts}%
\end{figure}

\begin{figure}%
  \centering%
  \includegraphics[width = 0.6 \linewidth]{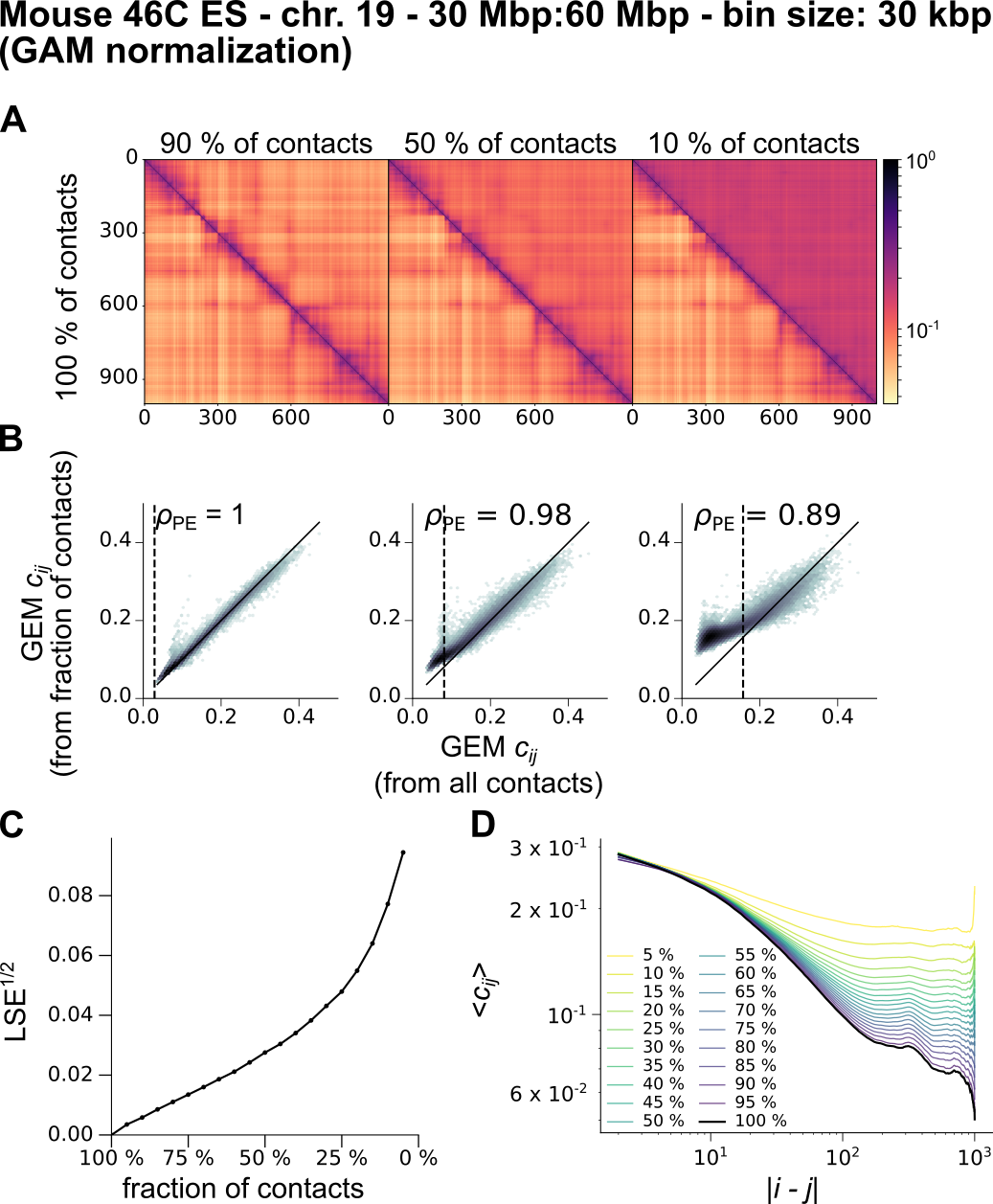}
  \caption{%
    Robustness of GEM reconstruction for GAM data of mouse chromosome 19 \cite{beagrie2017complex} (\SI{30}{\kilo bp} resolution). For all GEM reconstructions we used a threshold $\xi = \num{0.5}$. \textbf{(A)} Comparison of the contact probabilities of the reconstructed GEM with those of a GEM obtained by performing the minimization only on the top \SI{90}{\percent}, \SI{50}{\percent} and \SI{10}{\percent} experimental contacts. \textbf{(B)} 2d-histograms corresponding to the matrices shown in \textbf{(A)}. We give the Pearson correlation coefficients. The thresholding quantiles are represented by vertical dashed lines. \textbf{(C)} Comparison of the GEMs reconstructed from a decreasing fraction of the experimental contacts with the original GEM. $\mathrm{LSE}^{1/2}$ is the Euclidean distance between contact probabilities divided by ($N+1$). \textbf{(D)} Average contact probability as a function of the contour length for GEMs reconstructed from a decreasing fraction of the experimental contacts.%
}%
  \label{fig:beagrie_chr19_30Kbp_significant_contacts}%
\end{figure}

% to cancel one float per page
\clearpage
\makeatletter
\afterpage{\global\setlength\@fpsep{8\p@ \@plus 2fil}}
\makeatother

% vim: set sw=2 expandtab tabstop=2 foldcolumn=4:
% vim: set spell spelllang=en_us,en_gb:

% appendix
\clearpage
\newpage
%starting a segment for biblatex
\newrefsegment
%\newrefsection
%\renewcommand\thefigure{A\arabic{figure}}
%\setcounter{figure}{0}
\etocdepthtag.toc{tappendix}
\section{Existing methods to reconstruct chromosome architecture}
\label{sec:review_methods}
Let us review some of the models which have been proposed in the past to address the reconstruction of chromosome architecture from 3C data. Our aim is not to review thoroughly the available methods, but rather to emphasize essential differences with our own approach. For a more detailed review of the existing methods for reconstructing chromosome architecture we refer the interested reader to \cite{serra2015restraintbased}.

\subsection{Non-polymer models}
\label{subsec:review_nonpolymer_models}
\subsubsection{Harmonic model}
A numerical procedure relying on the introduction of harmonic potentials has been proposed to reconstruct the equilibrium configurations of the chromosome from the experimental contact probabilities \cite{umbarger2011the,ba2012genome}. Harmonic interactions are introduced between each chromosomal bin pair $(i,j)$, such that the contribution to the internal energy is:
\begin{equation}
  U(\lbrace \vec{r}_{i} \rbrace) = \sum \limits_{i<j} \frac{k}{2} \left( r_{ij} - r_{ij}^0 \right)^2,
  \label{eq:previous_models:umbarger}
\end{equation}
in which $r_{ij} = | \vec{r}_{j} - \vec{r}_i |$ is the distance between loci $i$ and $j$, $k$ is an arbitrarily chosen elastic constant and $r_{ij}^0$ is the length of the isolated spring. A Monte-Carlo simulation is then performed to sample equilibrium configurations of the system defined in \cref{eq:previous_models:umbarger}. These configurations are used to represent the chromosome configurations.

In this method, the elastic constant was assigned arbitrarily to $k= 5 \, k_\mathrm{B} T$. The fact that this elastic constant is the same for all $(i,j)$ is a first limitation in this approach. The spring lengths are taken such that $r_{ij}^0 =  d_{ij}$, where $d_{ij}$ is the distance desired between beads $i$ and $j$. The authors assumed that the equilibrium distance between two chromosomal loci is inversely proportional to the contact probability, $d_{ij} = 1 / c_{ij}$. We will come back to this assumption.

\subsubsection{Constraint satisfaction}
Another approach is to cast the problem of reconstituting chromosome architecture into a constraint satisfaction problem \cite{duan2010a}. The reformulated problem then consists in finding the coordinates $\lbrace \vec{r}_{i} \rbrace$ such that the distances between any pair of chromosomal bins $(i,j)$ is bounded from below and from above:
\begin{equation}
  a_{ij} < r_{ij} < b_{ij}.
  \label{eq:previous_model:duan}
\end{equation}

In \cref{eq:previous_model:duan} the upper bound is taken inversely proportional to the experimental contact probability, $b_{ij} \propto 1 / c_{ij}$, and the proportionality coefficient is a parameter of the method. The lower bound $a_{ij}$ is introduced to take into account excluded volume between any pair of chromosomal loci, and to penalize contacts between adjacent loci due to the chromosome bending rigidity. This is a constraint satisfaction problem, which can be solved with the simplex method. The obtained solution is then used to represent a chromosome configuration.

The main limitation of this approach is clearly that the choice of the lower and upper bounds must be adjusted by the user and adapted to each data set. Beside, this is not a physical model of the chromosome architecture.

\subsubsection{Singular value decomposition of the spatial correlation matrix}
Let us consider the matrix $R$ of size $d \times N$, where $d=3$ is the space dimension and $N$ is the number of bins in the Hi-C contact matrix. The matrix element $r_{\alpha i}$ is therefore the spatial coordinate of loci $i$ along the $\alpha$-axis ($\alpha=x,y,z$). Next we consider the Singular Value Decomposition (SVD) of $R$:
\begin{equation}
  r_{\alpha i} = \sum \limits_{\gamma =1}^d \lambda_\gamma u_{\alpha \gamma} v_{ i \gamma},
  \label{eq:previous_models:mozziconacci:svd}
\end{equation}
where $U=[u_{\alpha \gamma}]$ and $V=[v_{\gamma i}]$ are two orthogonal matrices, and $\left\lbrace \lambda_\gamma \right\rbrace_{\gamma=1,\dots,d}$ are the singular values of $R$. Then $C=R^T R$ and $\tilde{C} = R R^T$ have the same non-zero eigenvalues, which are $\lambda_1^2$, $\lambda_2^2$ and $\lambda_3^2$ (if $d=3$). Finally we introduce the matrix of distances, $D$, with elements:
\begin{equation}
  d_{ij} = \sqrt{\sum \limits_{\alpha=1}^d \left( r_{\alpha i} - r_{\alpha j} \right)^2}.
  \label{eq:previous_models:mozziconacci:distances}
\end{equation}
It turns out that the correlation matrix $C$ can be obtained from the distance matrix $D$ \cite{lesne20143d,marbouty2015condensin}. Therefore, from the knowledge of the distances, one can infer the singular values of the coordinates matrix, and obtain an approximation for $R$.

\subsection{Polymer models}
Models presented in \cref{subsec:review_nonpolymer_models} lack a physical model of the chromosome. In clear, the Hi-C bins define a gas of particles with coordinates $\lbrace \vec{r}_i \rbrace$ and minimizing \cref{eq:previous_models:umbarger} (resp. solving \cref{eq:previous_model:duan,eq:previous_models:mozziconacci:distances}) can result in configurations that violate topological constraints of the polymer chain representing the chromosome. Therefore, subsequent improvements have consisted in incorporating a polymer model of the chromosome when attempting to reconstruct chromosome architecture.

\subsubsection{Random walk backbone with tethered loops}
Another way to look at Hi-C data is to consider that when the contact probability between loci $i$ and $j$ is high enough, it defines a DNA loop. This is the approach taken in \cite{jhunjhunwala2008the}. In short, whenever
\begin{equation}
  c_{ij} > c_{min},
  \label{eq:previous_models:jhunjhunwala}
\end{equation}
with an arbitrary lower bound $c_{min}$ on the contact probability, the authors considered that the DNA subchain in the interval $[i,j]$ constitutes a loop, with $\vec{r}_i=\vec{r}_j$. The chromosome is then represented by a backbone polymer with Gaussian statistics on which are tethered polymer loops with varying sizes. Numerical simulations are then performed on the basis of this polymer model of the chromosome.

\subsubsection{First-principle approach}
In \cite{jost2014modeling,brackley2016predicting}, the authors start from a polymer representation of the chromosome, and add interactions between different regions of the chromosome. However, due to the complexity of chromosome interactions with proteins, this kind of studies can only be made under strong simplifying assumptions. For example, a unique generic type of protein is included and/or the variety in the binding energies with different loci on the chromosome is replaced by a single binding energy (or just a few). For this reason comparisons with experimental contact matrices have been rather qualitative.

\subsubsection{Inverse approach}
As mentioned in the main text, chromosome architecture might be well described with an effective model in which microscopical details, such as proteins and sequence effects, are coarse-grained. In particular, the effect of structuring proteins can be taken into account implicitly by introducing an effective potential $V_{ij}(r)$ between each $(i,j)$ monomer pair. In other words, each location on the genome experiences an effective interaction with the other loci on the genome, which mimics the effect of multivalent proteins. This type of approach was used, in which such potentials are considered to be short-range square potentials \cite{giorgetti2014predictive}:
\begin{align}
  V_{ij}(r) =
  \begin{cases}
    +\infty & \text{ if } r < \sigma \\
    - \varepsilon_{ij} & \text{ if } \sigma < r < \xi \\
    0 & \text{ otherwise,}
  \end{cases}
  \label{eq:pair_potential_square}
\end{align}
where $\sigma$ is the hard-core distance and $\xi$ is a threshold which defines at the same time the range of the potential and the distance below which monomers $i$ and $j$ are said to be in contact. By performing MC simulations on a polymer model with the pair potentials in \cref{eq:pair_potential_square}, one can obtain equilibrium configurations and use them to compute contact probabilities between monomer pairs.

Let us note $c_{ij}^{exp}$ the experimental contact probability between restriction fragments $i$ and $j$ obtained from Hi-C experiments, and $c_{ij}$ the contact probability between monomers $i$ and $j$ obtained from MC simulations of a polymer model with potentials as in \cref{eq:pair_potential_square}. We define the least-square estimator between the experimental and the predicted contact matrices:
\begin{align}
  d(c_{ij},c_{ij}^{exp}) = \frac{2}{N(N+1)} \sum \limits_{i<j} \left( c_{ij} - c_{ij}^{exp} \right)^2,
   \label{eq:chisq_contacts}
\end{align}

Finding a good model for chromosome architecture now consists in finding a collection of potentials $V_{ij}(r)$ that minimize $d(c_{ij},c_{ij}^{exp})$. The solution is achieved at the optimal values for $\sigma$, $\xi$ and the matrix of binding energy $\varepsilon_{ij}$. In \cite{giorgetti2014predictive}, a MC simulation was performed at each step of the minimization procedure, in order to re-sample equilibrium configurations of the chromosome and compute the $c_{ij}$ values. Therefore the computational burden is high.

\section{Scaling of contact probabilities of a polymer}
\label{sec:scaling_relation_cij_dij}
Several of the methods we have presented \cite{umbarger2011the,duan2010a,lesne20143d} have the inconvenience to rely on an estimate of the average distances between loci on the chromosome taken to be inversely proportional to the contact probabilities:
\begin{equation}
  d_{ij} \propto 1 / c_{ij}.
  \label{eq:dij_cij_prop}
\end{equation}

While \cref{eq:dij_cij_prop} may appear to be a reasonable assumption, there is no fundamental reason to support it. As pointed out in \cite{serra2015restraintbased},  a more general functional dependence would be $d_{ij} \sim c_{ij}^{-\gamma}$. For instance, if we model the chromosome as a polymer with scaling exponent $\nu$, we have \cite{deGennes1979}:
\begin{align}
  \begin{aligned}
    & \proba{\vec{r}_{ij}} \simeq \frac{1}{\langle r_{ij} \rangle^d} f_p \left( \frac{r_{ij}}{\langle r_{ij} \rangle} \right), \qquad  f_p(x) \underset{x\sim 0}{\sim} x^g \\
  & \langle r_{ij} \rangle \simeq b \mid i -j \mid^\nu.
  \end{aligned}
\end{align}

Let us consider that the contact probabilities are given by $c_{ij} = \proba{r_{ij}=b}$, and write $d_{ij}=\langle r_{ij} \rangle$. Then, we obtain the relation:
\begin{equation}
  d_{ij} \sim 1 / c_{ij}^{1/(d+g)}.
  \label{eq:dij_cij_polymer}
\end{equation}

For a Gaussian chain, we have $g=0$, and for a self-avoiding chain, $g=1/3$. Hence we obtain ($d=3$), $d_{ij} \sim 1 / c_{ij}^{0.33}$ and $d_{ij} \sim 1 / c_{ij}^{0.3}$, in direct contradiction with \cref{eq:dij_cij_prop}.

Reducing chromosome architecture to a mere conformation characterized by the average pair distances $d_{ij}$ is probably unrealistic. Indeed, co-localization of loci on the chromosome results from the effect of divalent (or multivalent) proteins. We may estimate the strength of the binding by considering contributions of about one $k_\mathrm{B} T$ per significant contact \cite{Sheinman2012}. Thus, we may consider that structuring proteins have a binding energy with DNA in the range $\varepsilon = 3-20 \, k_\mathrm{B} T$. Consequently, the probability to form a DNA loop between monomers $i$ and $j$ should read:
\begin{equation}
  \proba{r_{ij} = b} \simeq \frac{1}{\mid i - j\mid^{\nu(d+g)} } e^{\beta \varepsilon} \qquad (b=1),
\end{equation}
where $\nu(d+g)=2$ for a self-avoiding polymer chain with scaling exponent $\nu=3/5$. For example, considering a relatively strong transcription factor, with $\varepsilon = 10 \, k_\mathrm{B} T$, the contact probability $c_{ij} \approx 1$ when $\mid i-j\mid = 150$ monomers and falls quickly to zero for larger contour distances. Here a monomer typically represents the diameter of the DNA fiber. In eukaryotes, a monomer typically represents \SI{3000}{bp}. Therefore, it is very unlikely that chromosome loops are stable for contour length beyond $\SI{500}{\kilo bp}$ approximatively. In other words, thermodynamic fluctuations may provide the chromosome folding with a non negligible conformational entropy.

\section{Conversion of Hi-C and GAM data into contact probabilities}
\label{sec:contact_proba_exp}
In this section, we present the methods that have been used in this article to estimate experimental contact probabilities from the experimental measurements.

\subsection{Hi-C}
\label{sec:normalization_hic}
After sequencing, the read-pairs obtained in Hi-C experiments are mapped to a reference genome. Provided that the genome is divided into bins of equal size, each read can then be associated to a unique bin, say $i$, on the genome. Therefore, each read-pair defines a contact between the corresponding bin-pair. \textit{In fine}, a contact count matrix $[n_{ij}]$ can be constructed, where each entry $n_{ij}$ represents the number of times bins $i$ and $j$ were found in contact in the experiment. From this count matrix, the matrix of contact probabilities can be estimated. In the sequel we present the two methods that have been used in this article to compute the contact probability matrix $[c_{ij}]$ from the count matrix $[n_{ij}]$.

\subsubsection{Uniform normalization}
In first approximation, it seems reasonable to consider that $n_{ij}$ represents the number of cells in which bins $i$ and $j$ were found in contact. Assuming that $N_c$ is the number of cells in the experiment sample, the contact probability between bins $i$ and $j$ is simply:
\begin{equation}
  c_{ij} = \frac{n_{ij}}{N_c}.
\end{equation}

The previous expression suggests that the matrix of contact probabilities can be obtained from the count matrix by applying a global normalization factor. In practice however, the number of cells in the sample is unknown. Therefore, when using this normalization method to reconstruct the optimal Gaussian effective model, we have tried several values for $N_c$ and chosen the value giving the smallest distance between contact probabilities of the model and of the experiment.

\subsubsection{Matrix balancing}
Although intuitive, the ``uniform normalization'' presented above suffers from several pitfalls inherent to the Hi-C protocol. Sources of bias in the $n_{ij}$ counts comprise: chromatin accessibility to the restriction enzyme, alignability (\textit{e.g.} one bin containing many repeats may result in very few detected contacts because reads cannot be aligned uniquely) and restriction site density on the chromosome. For example, if one bin $i$ suffers from a bias leading to undersampling, the entry $n_{ij}$ will underestimate the contact frequency between bis $i$ and $j$.

The problem of count matrix normalization has been thoroughly studied \cite{yaffe2011probabilistic,imakaev2012iterative,rao2014a}. In short, these methods apply a different normalization factor to each entry of the count matrix $[n_{ij}]$. Among them, matrix balancing can be used to construct a corrected count matrix $[\tilde{n}_{ij}]$ such that the number of interactions with other bins on the chromosome is the same for every bin. To be more accurate, matrix balancing yields two vectors $\vec{U}$ and $\vec{V}$ such that:
\begin{align}
  \begin{array}{c c c c c}
    n_{ij} &=& U_i \tilde{n}_{ij} V_i &&\\
    N_c & = & \sum \limits_{j} \tilde{n}_{ij} & = & \sum \limits_{i} \tilde{n}_{ij}
  \end{array}.
\end{align}

The matrix of contact probabilities is then computed as: $c_{ij} = \tilde{n}_{ij}/N_c$. The resulting matrix, $[c_{ij}]$, is bistochastic: each row and column sums to one.

As pointed out in \cite{rao2014a}, the problem of matrix balancing has been well studied. In particular, an efficient algorithm is available to balance any non-negative matrix with total support \cite{knight2013}. Other implementations of matrix balancing dedicated to Hi-C data sets are also available (see for instance \cite{mirny_cooler}).

In this article, we considered the contact probability matrix obtained by matrix balancing for the Hi-C data coming from \cite{rao2014a}. The normalized matrices, using the algorithm from \cite{knight2013}, were readily available.

\subsection{GAM}
\label{subsec:normalization_gam}
Genome Architecture Mapping (GAM) is a recent experimental technique which has been proposed as an alternative to the Hi-C technique to collect information on chromosome architecture \cite{beagrie2017complex}. The procedure may be summarized as follows:
\begin{enumerate}
  \item Collect slices of a cell population by cryosectioning.
  \item Sequence DNA contained in each slice.
  \item Map reads to genomic coordinates by aligning to a reference genome.
  \item Assign genomic coordinates to bins corresponding to a regular subdivision of the genome.
\end{enumerate}

\begin{figure}[!htbp]
  \centering
  \includegraphics[width = \linewidth]{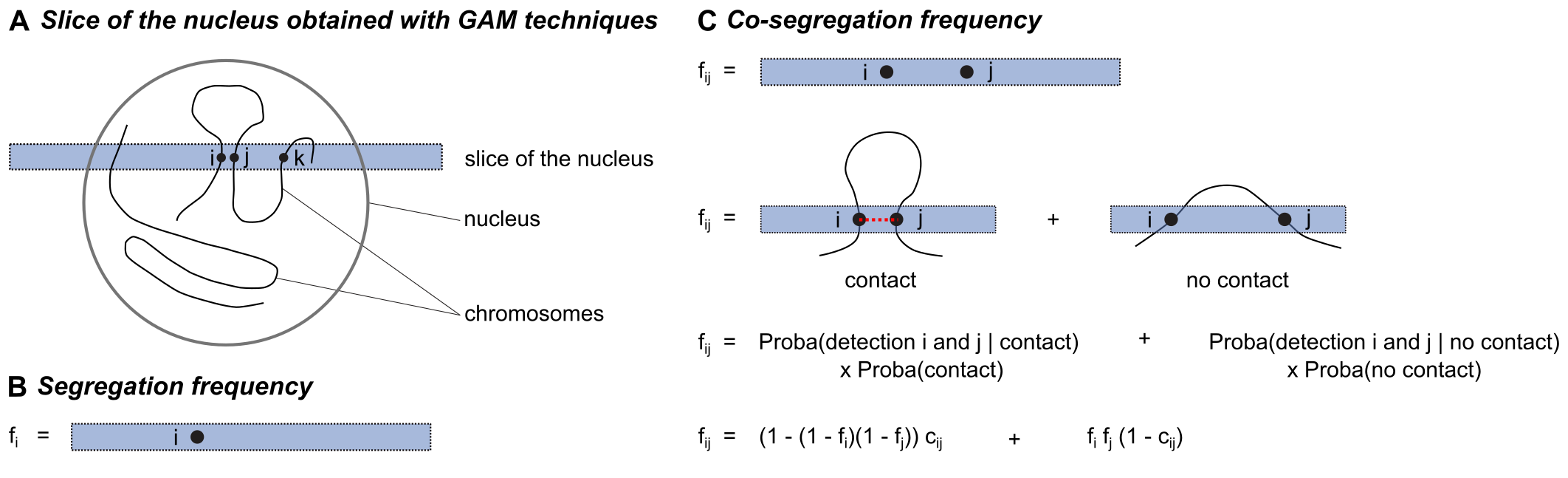}
  \caption{Estimation of the contact probability matrix from GAM data sets. \textbf{(A)} GAM experiments use cryosectioning to obtain thin slices of a cell population sample. Each slice (or nuclear profile) cuts many nuclei in random orientations. The genomic content of each slice is sequenced. \textbf{(B)} The segregation frequency $f_i$ is the fraction of nuclear profiles containing a specific genomic locus $i$. \textbf{(C)} The co-segregation frequency $f_{ij}$ is the fraction of nuclear profiles containing a pair of specific loci $i$ and $j$. The segregation and co-segregation frequencies can be used to estimate the contact probability of $c_{ij}$ (see \cref{eq:normalization_gam_cij}).}
  \label{fig:gam_normalization}
\end{figure}

Each slice collected contains thin layers of many nuclei with random orientations. Such a slice is represented in \cref{fig:gam_normalization}. Let us stress that a pair of DNA sequences detected in the same slice are not necessarily in contact. However, given that cells have been sliced in different orientations, if this pair is repeatedly found in the same slices, it means that these sequences belong to regions of the chromosome with a high contact probability. We now present the method used in this article to infer contact probabilities $c_{ij}$ from the GAM experimental data.

The main output of GAM experiments is a segregation matrix $[s_{ia}]$ in which: rows correspond to bins on the genome, columns correspond to slices collected and each entry $s_{ia}=1$ if bin $i$ was detected in slice $a$ and $s_{ia}=0$ otherwise. Assuming that there are $P$ slices, we define following reference \cite{beagrie2017complex}:
\begin{itemize}
  \item The segregation frequency for bin $i$:
    \begin{equation}
      f_i = \frac{1}{P}\sum \limits_{a=1}^P s_{ia}.
    \end{equation}
  \item The co-segregation frequency for bins $i$ and $j$:
    \begin{equation}
      f_{ij} = \frac{1}{P}\sum \limits_{a=1}^P s_{ia} s_{ja}.
    \end{equation}
\end{itemize}

We now relate the segregation and co-segregation frequencies to actual contact probabilities. The probability that bins $i$ and $j$ are detected in a slice $S_a$ (\textit{i.e.} $f_{ij}$) can be decomposed according to the law of total probability as:
\begin{align}
  \begin{array}{l c l}
    \proba{i \text{ and } j \text{ in } S_a} & = & \proba{\left. i \text{ and } j \text{ in } S_a \middle|i \text{ and } j \text{ in contact} \right.}\proba{i \text{ and } j \text{ in contact}} \\
    &+& \proba{\left. i \text{ and } j \text{ in } S_a \middle|i \text{ and } j \text{ not in contact} \right.}\proba{i \text{ and } j \text{ not in contact}}
  \end{array}
\end{align}

The probability that bins $i$ and $j$ are detected in a slice, conditioned to the fact that they are in contact (first term in the right hand side of the previous equation), is the probability that at least one of the bins is detected in the slice. Therefore, the previous expression is expressed in terms of the segregation frequencies, co-segregation frequencies and contact probabilities as:
\begin{equation}
  f_{ij} = (1 - (1- f_i) (1-f_j)) c_{ij} + f_i f_j (1 -c_{ij}).
\end{equation}

We finally obtain for the contact probability between bins $i$ and $j$:
\begin{equation}
  c_{ij} = \frac{f_i f_j - f_{ij}}{f_i + f_j - 2 f_i f_j}
  \label{eq:normalization_gam_cij}
\end{equation}

In this article, we have used the above equation to estimate the contact probability matrix from the experimental segregation matrix. Actually, \cref{eq:normalization_gam_cij} ensures that $c_{ij} < 1$. However, the nominator can be negative, in which case we set $c_{ij} \leftarrow \max{(c_{ij},0)}$.

\section{The Gaussian effective model}
\label{sec:gem_si}
\subsection{Partition function}
We consider the Gaussian effective model (GEM) with energy defined in the main text. To break the translational invariance, we attach the first monomer to the origin: $\vec{r}_0=0$. We can now write the GEM partition function as a Gaussian integral:
\begin{align}
  \begin{aligned}
    Z   &= \int \prod \limits_{i=1}^N \ud{^3 \vec{r}_i} \exp{\left( - \beta U[\lbrace \vec{r}_i \rbrace] \right)} \\
    &= \int \prod \limits_{i=1}^N \ud{^3 \vec{r}_i} \exp{\left( - \frac{3}{2 b^2} \sum_{i,j} \vec{r}_i \cdot \vec{r}_j \sigma^{-1}_{ij} \right)},
  \end{aligned}
  \label{eq:gem_partition_function}
\end{align}
where we have introduced the inverse covariance matrix $\Sigma^{-1}$ with elements $\sigma^{-1}_{ij}$ and formally expressed as:
\begin{equation}
  \Sigma^{-1}=T + W,
  \label{eq:gem_inverse_correlation_app}
\end{equation}
with:
\begin{align}
  T =
\begin{pmatrix}
  2 & -1 & \ldots & 0 & 0 \\
  -1 & 2 & \ldots & 0 & 0 \\
  \vdots & \vdots & \ddots & \vdots & \vdots \\
  0 & 0 & \ldots & 2 & -1 \\
  0 & 0 & \ldots & -1 & 1
\end{pmatrix}
,
\qquad
W =
\begin{pmatrix}
  \sum \limits_{\substack{j=0 \\ j \neq 1}} k_{1j} & -k_{12} & \ldots & -k_{1 N-1} & -k_{1N} \\
  -k_{21} &  \sum \limits_{\substack{j=0 \\ j \neq 2}} k_{2j}& \ldots & -k_{2 N-1} & -k_{2N} \\
  \vdots & \vdots & \ddots & \vdots & \vdots \\
  -k_{N-1 1} & -k_{N-1 2} & \ldots & \sum \limits_{\substack{j=0 \\ j \neq N-1}} k_{N-1 j} & -k_{N-1 N} \\
  -k_{N 1} & -k_{N 2} & \ldots & - k_{N N-1} & \sum \limits_{\substack{j=0 \\ j \neq N}} k_{N j}
\end{pmatrix}.
  \label{eq:gem_trid_redk_supp}
\end{align}

The partition function can be conveniently computed by separating the integration along each dimension:
\begin{align}
  \begin{aligned}
    Z &= \prod \limits_{a=x,y,z} \left[ \int \prod \limits_{i=1}^N \ud{r_i^a} \exp{\left( - \frac{3}{2 b^2} \sum_{i,j} r_i^a r_j^a \sigma^{-1}_{ij} \right)} \right] \\
      & = \left[ \int \prod \limits_{i=1}^N \ud{x_i} \exp{\left( - \frac{3}{2 b^2} \sum_{i,j} x_i x_j \sigma^{-1}_{ij} \right)} \right]^3 \\
      &= z^3, \qquad \text{ with } z = \left( \frac{2 \pi b^2}{3} \right)^{N/2} \det{\Sigma}^{1/2}.
  \end{aligned}
\end{align}

For any function of the monomer coordinates, $A(\lbrace \vec{r}_i \rbrace)$, we can therefore define the thermodynamical average:
\begin{equation}
  \langle A\left( \left\{ \vec{r}_i \right\} \right) \rangle = \frac{1}{Z} \int \prod_{i=1}^N \ud{^3 \vec{r}_i} A\left( \left\{ \vec{r}_i \right\} \right) \exp{\left( - \beta U \left[ \left\{ \vec{r}_i \right\} \right] \right)}.
  \label{eq:gem_thermo_avg}
\end{equation}

\subsection{Pair correlation function}
Let us introduce the vector $\vec{r}=(r^x,r^y,r^z)$ and $\vec{r}_{ij} = \vec{r}_j - \vec{r}_i$. The pair correlation function $\langle \delta(\vec{r} - \vec{r}_{ij}) \rangle$ can be expressed as:
\begin{align}
  \begin{aligned}
    \langle \delta(\vec{r} - \vec{r}_{ij}) \rangle &= \frac{1}{Z} \int \prod \limits_{m=1}^{N} \ud{^3 \vec{r}_m} \delta(\vec{r} - \vec{r}_{ij}) \exp{\left( - \frac{3}{2 b^2} \sum_{m,n} \vec{r}_m \cdot \vec{r}_n \sigma^{-1}_{mn} \right)} \\
    & = \prod \limits_{a=x,y,z} \underbrace{\left[ \frac{1}{z} \int \prod \limits_{m=1}^N \ud{r_m^a} \delta(r^a - r_{ij}^a) \exp{\left( - \frac{3}{2 b^2} \sum_{m,n} r_m^a  r_n^a \sigma^{-1}_{mn} \right)} \right]}_{I(r^a)} ,
  \end{aligned}
  \label{eq:pair_correlation_computation_I}
\end{align}

The integral $I(x)$ can be computed by exponentiating the $\delta$-function:
\begin{align}
  \begin{aligned}
    I(x) & = \frac{1}{z} \int \prod \limits_{m=1}^N \ud{x_m} \int \frac{\ud{k}}{2 \pi} \exp{\left( ik(x - x_{ij})  - \frac{3}{2 b^2} \sum_{m,n} x_m  x_n \sigma^{-1}_{mn} \right)} \\
    & =  \frac{1}{z} \int \frac{\ud{k}}{2 \pi} \exp{(ikx)} \int \ud{^N X} \exp{\left( - \frac{3}{2 b^2} X^T \Sigma^{-1} X  - i k X^T E_{ij} \right)},
  \end{aligned}
\end{align}
where the vector $E_{ij} = E_j - E_i$ and $E_i = (0,\cdots,0,1,0,\cdots,0)$, with the non-zero element being at the index $i$. By performing a first Gaussian integration we obtain:
\begin{align}
  \begin{aligned}
    I(x) & = \int \frac{\ud{k}}{2 \pi} \exp{(ikx)} \exp{\left( - \frac{b^2}{6} k^2 (\sigma_{ii} + \sigma_{jj} - 2 \sigma_{ij}) \right)}.
  \end{aligned}
\end{align}

Finally, by performing a second Gaussian integration and by substituting this result into \cref{eq:pair_correlation_computation_I}, we obtain the expression for the pair correlation function:
\begin{align}
  \langle \delta(\vec{r} - \vec{r}_{ij}) \rangle = \left( \frac{3}{2 \pi \langle r_{ij}^2 \rangle}\right)^{3/2} \exp{\left( - \frac{3}{2} \frac{r^2}{\langle r_{ij}^2 \rangle} \right)} ,
  \label{eq:pair_correlation_computation_final}
\end{align}
where $\langle r_{ij}^2 \rangle = (\sigma_{ii} + \sigma_{jj} - 2 \sigma_{ij}) b^2 $.

\subsection{Form factor dependent contact probability}
\label{subsec:form_factors}
The contact probability between monomers $i$ and $j$ can be expressed as:
\begin{align}
  \begin{aligned}
    c_{ij} &= \langle \mu(r_{ij}) \rangle, \\
    & = \int \ud{^3 \vec{r}} \mu(r) \langle \delta(\vec{r}_{ij} - \vec{r}) \rangle,
  \end{aligned}
  \label{eq:cij_formfac}
\end{align}
where $\mu(r)$ is a form factor. An intuitive choice of form factor is to consider a theta function:
\begin{equation}
  \mu_T(r) = \theta(\xi -r).
  \label{eq:form_factor_theta}
\end{equation}

In the context of Hi-C experiments, this is equivalent to consider that every restriction fragment pair separated by a distance $r < \xi$ can be cross-linked. Or in other words, the probability that restriction fragments separated by a distance $r$ cross-link is
\begin{equation}
  \proba{\text{cross-link between i and j} \mid r_{ij}=r} =
  \begin{cases}
    1 & \text{ if } r < \xi \\
    0 & \text{ otherwise }.
  \end{cases}
\end{equation}

However, formaldehyde, the cross-linking agent used in most Hi-C experiments, can polymerize. It is present in aqueous solution in the form of methylene glycol \ce{HOCH2OH} monomers, but it also exists in the form of oligomers \ce{HO(CH2O)_nH}, where $n$ is a polymerization index. The equilibrium of the polymerization reaction depends on the formaldehyde concentration. For instance, in an aqueous solution with \SI{40}{\percent} mass fraction of formaldehyde at \SI{35}{\degreeCelsius}, the proportion of monomers in solution is only \SI{26.80}{\percent}, the rest being oligomers with $n > 1$ \cite{Jackson1251999,Gunter2000}. This suggests that cross-links between restriction fragments have varying size depending on the formaldehyde oligomer that made the cross-link.

For that reason, the cross-linking probability may be more accurately represented by a function which ensures that most of the cross-links occur for distances $r<\xi$, but which also allows for few cross-links to occur when $r > \xi$. Based on these considerations, it seems natural to consider a Gaussian form factor:
\begin{equation}
  \mu_G(r) = \exp{\left( - \frac{3}{2} \frac{r^2}{\xi^2} \right)},
  \label{eq:form_factor_gaussian}
\end{equation}
or an exponential form factor:
\begin{equation}
  \mu_E(r)= \exp{\left( - \frac{r}{\xi} \right)}.
  \label{eq:form_factor_exponential}
\end{equation}

Let us emphasize that the form factor $\mu(r)$ is not a probability distribution function, so it does not need to be normalized. It should rather be considered as the probability for a Bernoulli random variable. For a pair of restriction fragments separated by a distance $r$, the probability to cross-link is $\mu(r)$ and the probability not to cross-link is $1 - \mu(r)$. Note that $\mu(0)=1$.

\subsection{Contact probabilities of the Gaussian effective model}
From \cref{eq:pair_correlation_computation_final,eq:cij_formfac}, we can compute the contact probability $c_{ij}$ for monomers $i$ and $j$. Substituting $\mu(r)$ by the expression in \cref{eq:form_factor_theta,eq:form_factor_gaussian,eq:form_factor_exponential} we obtain:
\begin{itemize}
  \item For the Gaussian form factor:
    \begin{align}
      \begin{aligned}
        c_{ij} &= F_G(\langle r_{ij}^{2} \rangle) \\
        &= \left( 1 + \frac{\langle r_{ij}^{2} \rangle}{\xi^2} \right)^{-3/2},
      \end{aligned}
      \label{eq:gem_contact_proba_gaussian}
    \end{align}
  \item For the theta form factor:
    \begin{align}
      \begin{aligned}
        c_{ij} &= F_T(\langle r_{ij}^{2} \rangle) \\
        &= \mathrm{erf}{\left( \frac{X}{\sqrt{2}} \right)}  - \sqrt{\frac{2}{\pi}} X \exp{\left( -\frac{X^2}{2} \right)}, \qquad X=\sqrt{\frac{3 \xi^2}{\langle r_{ij}^{2} \rangle}}.
      \end{aligned}
      \label{eq:gem_contact_proba_theta}
    \end{align}
    where we have introduced the standard error function:
    \begin{equation}
      \mathrm{erf}{(x)} = \frac{2}{\sqrt{\pi}} \int \limits_{0}^x \ud{t} e^{-t^2}.
      \label{eq:error_function}
    \end{equation}

  \item For the exponential form factor:
    \begin{align}
      \begin{aligned}
        c_{ij} &= F_E(\langle r_{ij}^{2} \rangle) \\
        &= (1 + Y^2) \left( 1 - \mathrm{erf}{\left( \frac{Y^2}{2} \right)} \right) \exp{\left( \frac{Y^2}{2} \right)} - Y \sqrt{\frac{2}{\pi}}, \qquad Y = X^{-1} = \left( \sqrt{\frac{3 \xi^2}{\langle r_{ij}^{2} \rangle}} \right)^{-1}.
      \end{aligned}
      \label{eq:gem_contact_proba_exponential}
    \end{align}

\end{itemize}

\begin{figure}[!htbp]
  \centering
  \includegraphics[width = 0.5 \linewidth]{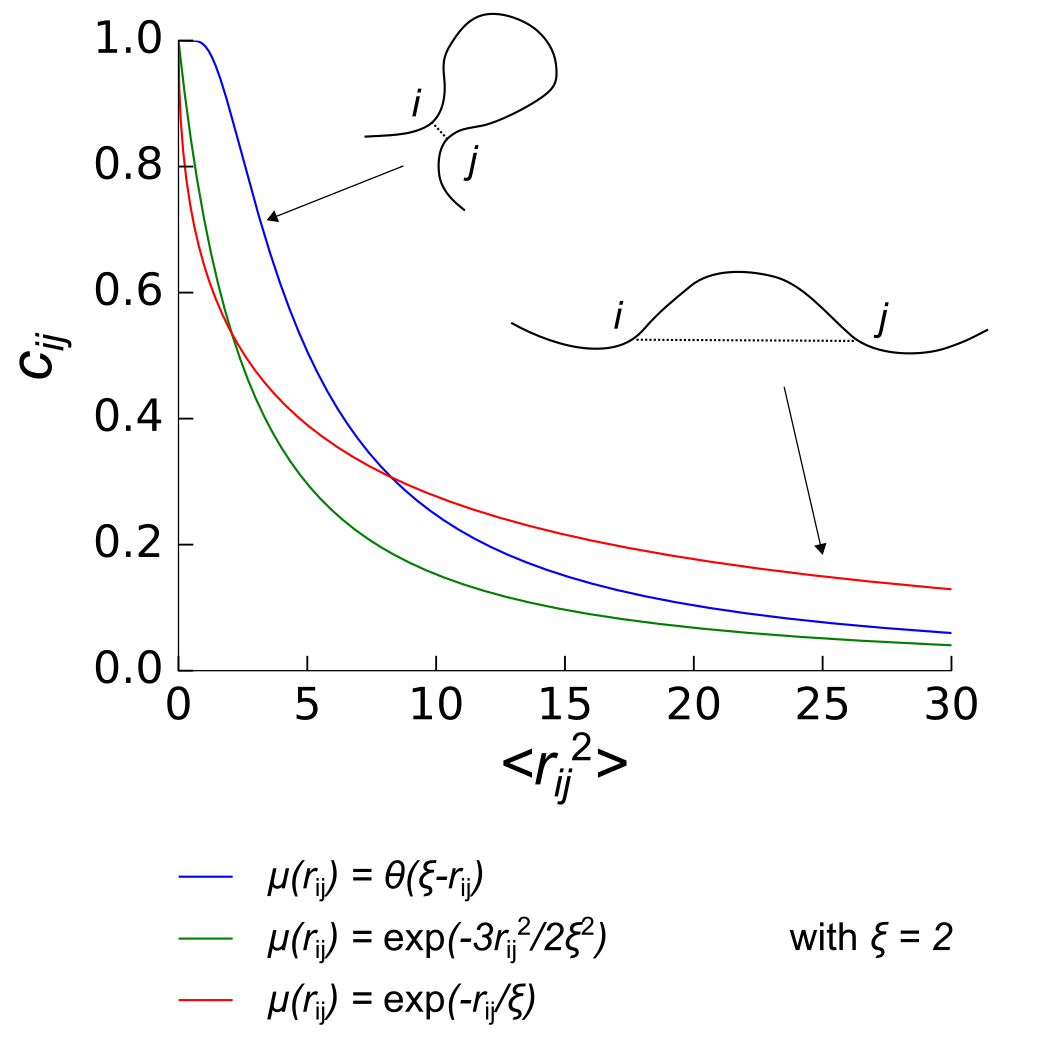}
  \caption{In the GEM, the contact probability $c_{ij}$ is expressed as a function of the mean square distance $\langle r_{ij}^2 \rangle$. This function depends on the form factor $\mu(r_{ij})$ used in the model.}
  \label{fig:gem_form_factors}
\end{figure}

The functional dependence of the contact probability $c_{ij}$ on the average square pair-distance $\langle r_{ij}^2 \rangle$ depends therefore on the choice of the form factor (\cref{fig:gem_form_factors}).

\subsection{Equilibrium properties}
\subsubsection{Radius of gyration}
The radius of gyration of the GEM can be computed from the covariance matrix $\Sigma$. It has the expression:
\begin{equation}
  \langle R_g^2 \rangle = \frac{1}{2 (N+1)^2} \sum_{i,j=0}^N \langle r_{ij}^2 \rangle.
  \label{eq:radius_gyration}
\end{equation}

It can be used to characterize the swelling of the underlying polymer. For instance, we may monitor the ratio $\langle R_g^2 \rangle / \langle R_{g,0}^{2} \rangle$ of the square radius of gyrations of the GEM with respect to the free Gaussian chain (all $k_{ij} = 0$).

\subsubsection{Mean potentials of interaction}
Other quantities of interest include the mean potentials of interaction at equilibrium. For any pair of monomers $i$ and $j$, it is defined as:
\begin{align}
  \begin{aligned}
    \langle e_{ij} \rangle &= \frac{3}{2 b^2} k_{ij} \langle r_{ij}^2 \rangle, \\
    &= - \frac{\partial \ln{Z}}{\partial \ln{k_{ij}}}.
  \end{aligned}
  \label{eq:gem_mean_potentials}
\end{align}

The quantity defined in \cref{eq:gem_mean_potentials}, expressed in $k_\mathrm{B} T$, reflects the state of the polymer. While high energy states are not favoured, they can however occur at thermal equilibrium if they are associated with large conformational entropy.

In addition, the mean potentials of interaction are extensive quantities. For instance, the mean potential of interaction between two groups $A=\lbrace i_1, i_2, \ldots, i_M \rbrace$ and $B = \lbrace j_1, j_2, \ldots, j_{M'}\rbrace$ of monomers is given by:
\begin{equation}
  \langle e_{AB} \rangle = \sum \limits_{(i,j) \in A \times B} e_{ij}.
\end{equation}

\subsection{Illustration}
\begin{figure}[!htbp]
  \centering
  \includegraphics[width = \linewidth]{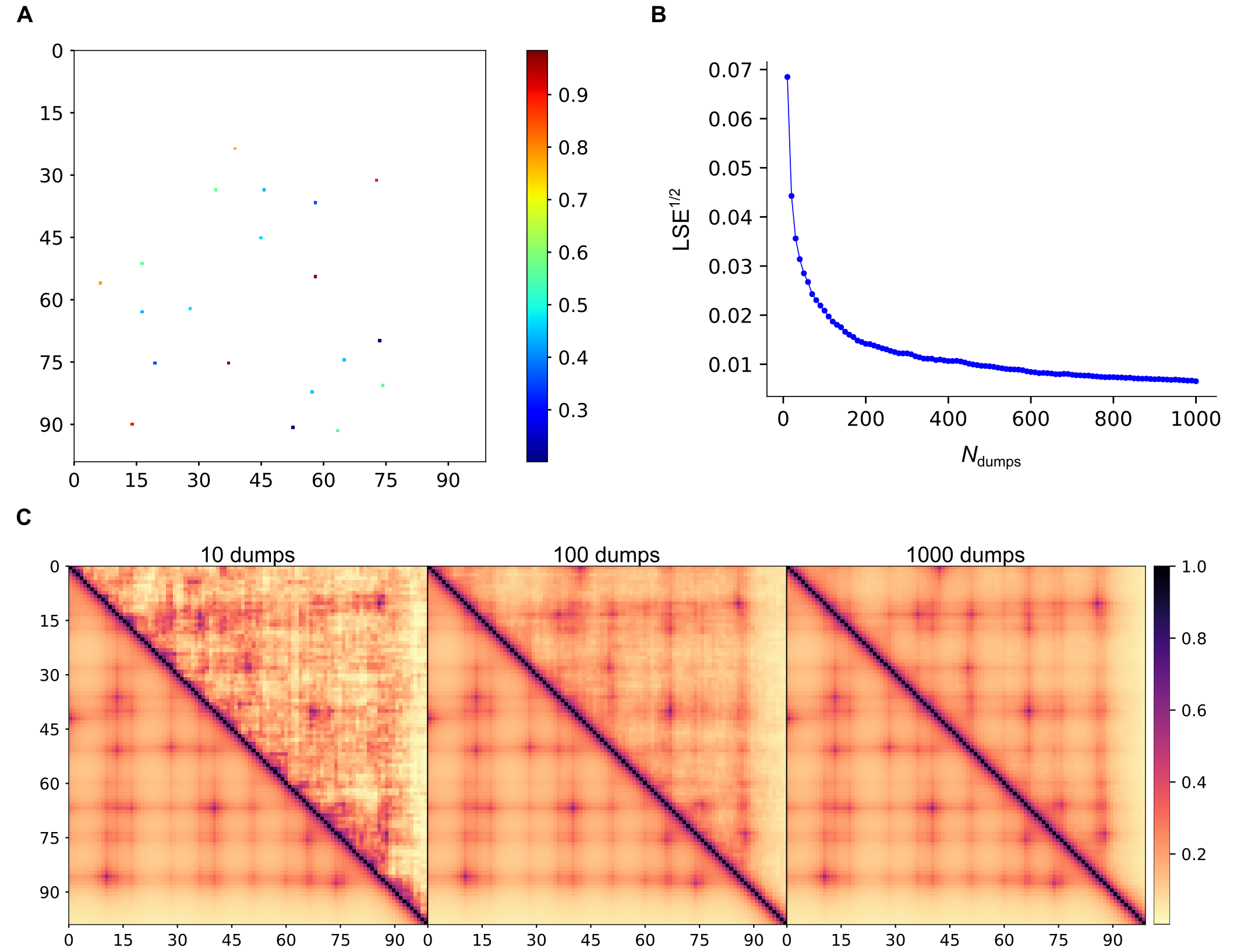}%
  \caption{\textbf{(A)} Arbitrary coupling matrix defining a GEM with $N+1=100$ monomers. \textbf{(B)} Convergence of the contact probability matrix $c_{ij}^{exp}$ computed from a Brownian Dynamics simulations to the GEM contact probability matrix $c_{ij}$, as a function of the number of sampled configurations (we used a threshold $\xi = \num{1.5}$ and a Gaussian form factor). \textbf{(C)} Comparison of the contact probability matrices $c_{ij}^{exp}$ and $c_{ij}$, for 10, 100 and 1000 configurations sampled by BD.}
  \label{fig:gem_cij_comparisons}
\end{figure}

As an example, we considered an arbitrary coupling matrix $[k_{ij}]$, specifying the interactions for a polymer of $N+1=100$ monomers. The coupling matrix was constructed by choosing randomly $M = 10$ pairs $(i,j)$ and by assigning to each coupling a random number $k_{ij} = U$ between $0$ and $1$. Considering a Gaussian form factor with a threshold $\xi = \num{1.5}$, we computed the contact probability of the GEM. We then sampled with Brownian Dynamics simulation configurations in the Boltzmann ensemble for this GEM. To compute the simulated contact probabilities, the average in \cref{eq:cij_formfac} was carried over the sampled configurations. As can be seen in \cref{fig:gem_cij_comparisons}, the simulated contact probabilities converge to the model prediction when the number of sampled configuration increases.

\section{Reconstruction by direct mapping}
\label{sec:gem_directinversion}
\subsection{Method}
In \cref{sec:gem_si}, we have shown that for any GEM, the matrix of contact probabilities is uniquely determined by the matrix of couplings. Reciprocally, for any contact probability matrix $[c_{ij}^{exp}]$ obtained from Hi-C experiments, one can reconstruct the GEM with the same contact probability matrix, $[c_{ij} = c_{ij}^{exp}]$, by computing the corresponding coupling matrix. This can be done as follows:
\begin{enumerate}
  \item Compute the matrix of mean-square distances of the GEM, $[\langle r_{ij}^2 \rangle]$, using the relation:
    \begin{equation}
      \langle r_{ij}^2 \rangle = F^{-1}(c_{ij}),
    \end{equation}
    where $F^{-1}$ is the inverse of one of the maps in \cref{eq:gem_contact_proba_gaussian,eq:gem_contact_proba_theta,eq:gem_contact_proba_exponential}.
  \item Invert the covariance matrix $\Sigma = [\langle \vec{r}_i \cdot \vec{r}_j \rangle]$ and compute the coupling matrix from \cref{eq:gem_inverse_correlation_app,eq:gem_trid_redk_supp}.
\end{enumerate}

In this method, the threshold $\xi$ used in the map $F$ is a free parameter that needs to be adjusted. We chose $\xi$ such that the Euclidean norm of the coupling matrix, $\Vert K \Vert$, is a minimum. This ensures that we select the GEM with the smallest perturbations compared to the free Gaussian chain case.

\begin{figure}[!htb]
  \centering
  \includegraphics[width = \linewidth]{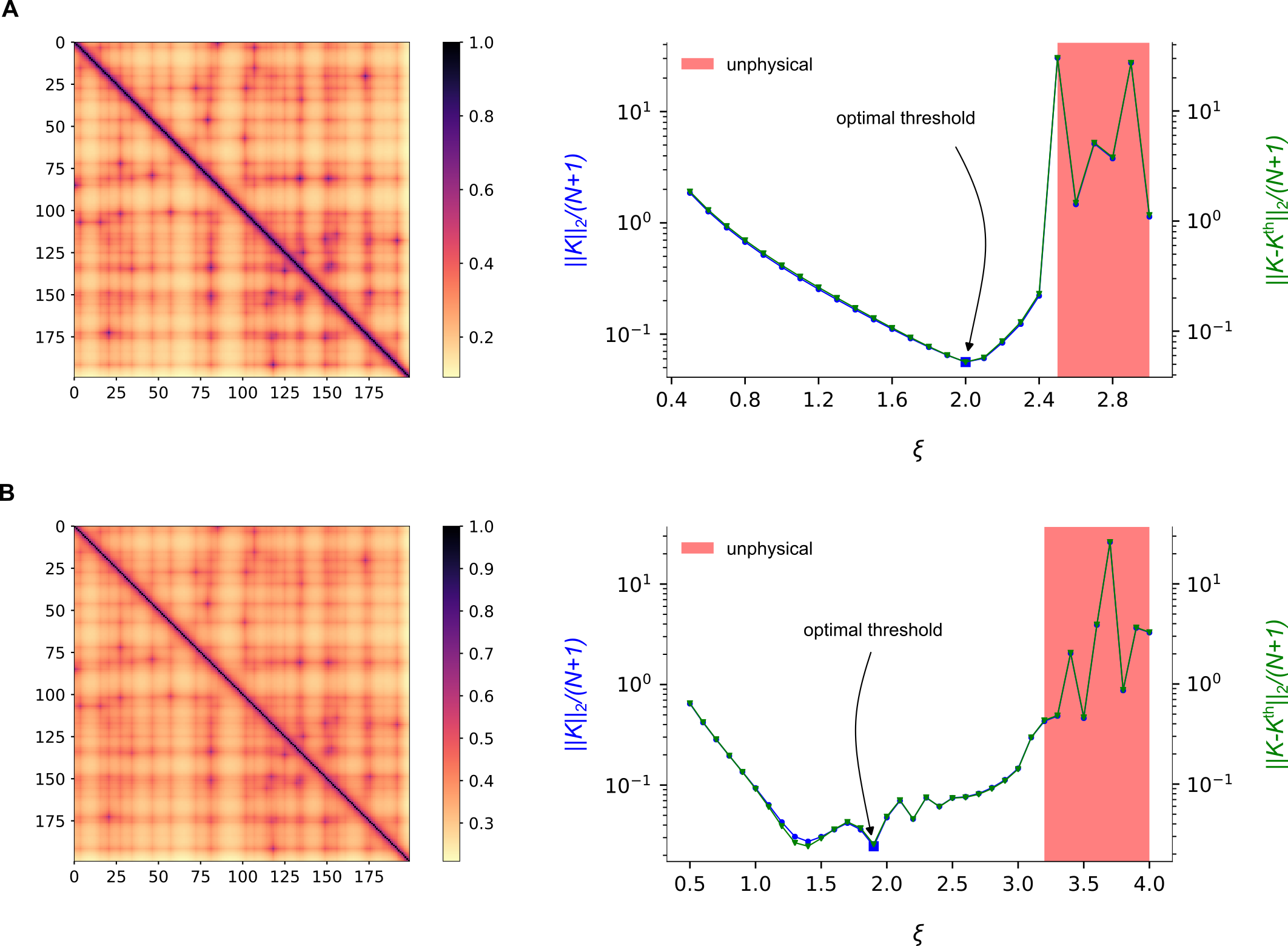}%
  \caption{Reconstruction of a GEM by direct mapping applied to an ``artificial'' contact probability matrix obtained from a Brownian Dynamics (BD) trajectory of an arbitrary GEM with a threshold $\xi^{exp} = \num{2}$ and: \textbf{(A)} a Gaussian or \textbf{(B)} an exponential form factor. The reconstructed coupling matrix is very close to the original one used for the BD simulation. The red area denotes values of the threshold where the reconstructed GEM has a covariance matrix $\Sigma$ with negative eigenvalues.}
  \label{fig:artificial_directmapping}
\end{figure}

As an example, we have applied the reconstruction method by direct mapping to contact probability matrices computed from Brownian Dynamics trajectories of an arbitrary GEM. Namely, we simulated the GEM defined by the coupling matrix $[k_{ij}^{th}]$ in \cref{fig:gem_cij_comparisons}A. The experimental contact probability matrix were computed by carrying the thermodynamical average $c_{ij} = \langle \mu(r_{ij}) \rangle$ over the sampled configurations. We used a threshold $\xi^{exp}=\num{2}$ and either a Gaussian or an exponential form factor. We therefore obtained two ``artificial'' contact probability matrices (see also \cref{fig:artificial_directmapping}):

\begin{longtabu} spread 8pt {|X|X|X|}
\hline
\rowfont{\bfseries} Contact matrix & Form factor & $\xi^{exp}$ \\
\hline
A	& Gaussian	& \num{2.0} \\
\hline
B	& Exponential	& \num{2.0} \\
\hline
\end{longtabu}

In this specific scenario, the true coupling matrix is known, and we can therefore compute the distance between those couplings and the reconstructed ones by monitoring the quantity $\Vert K - K^{th} \Vert$. As can be seen in \cref{fig:artificial_directmapping}, both $\Vert K \Vert$ and $\Vert K - K^{th} \Vert$ are minimum for the same value of the threshold $\xi$ so we use one or the other as proxies to determined the optimal value of the threshold, even when the true coupling matrix is not known or when the input contact probability matrix was not generated from a GEM.

Note that for contact matrix A, the optimal threshold is the same as the threshold used to compute the ``experimental'' contact probabilities, $\xi = \xi^{exp}$. This is because the form factors used for computing the ``experimental'' contact probabilities and for the reconstruction are both Gaussian. For matrix B, the form factor used to compute the ``experimental'' contact probabilities is exponential, and is therefore different from the Gaussian form factor used in the reconstruction. In this case, $\Vert K \Vert$ has several local minima. Yet at the global minimum, the coupling matrix is still reconstructed to a good accuracy.

\subsection{Unphysical GEM and effect of the noise}
\label{subsec:effect_noise}
In \cref{fig:artificial_directmapping}, there is a region where the reconstructed GEM has a covariance matrix $\Sigma$ with negative eigenvalues. When this happens, the corresponding GEM has a non-finite free energy and does not represent a physical system. Unfortunately, when applying this reconstruction by direct mapping to contact probabilities obtained from Hi-C experiments \cite{liebermanaiden2009comprehensive,rao2014a}, this situation was almost systematic. It is therefore desirable to better understand under which conditions such instabilities occur. In particular, we may expect that Hi-C contact matrices contain some noise due to inaccuracies in the measures or biases inherent to the experimental procedure, that lead to such effects.

\begin{figure}[!h]
  \centering
  \includegraphics[width = 0.6 \linewidth]{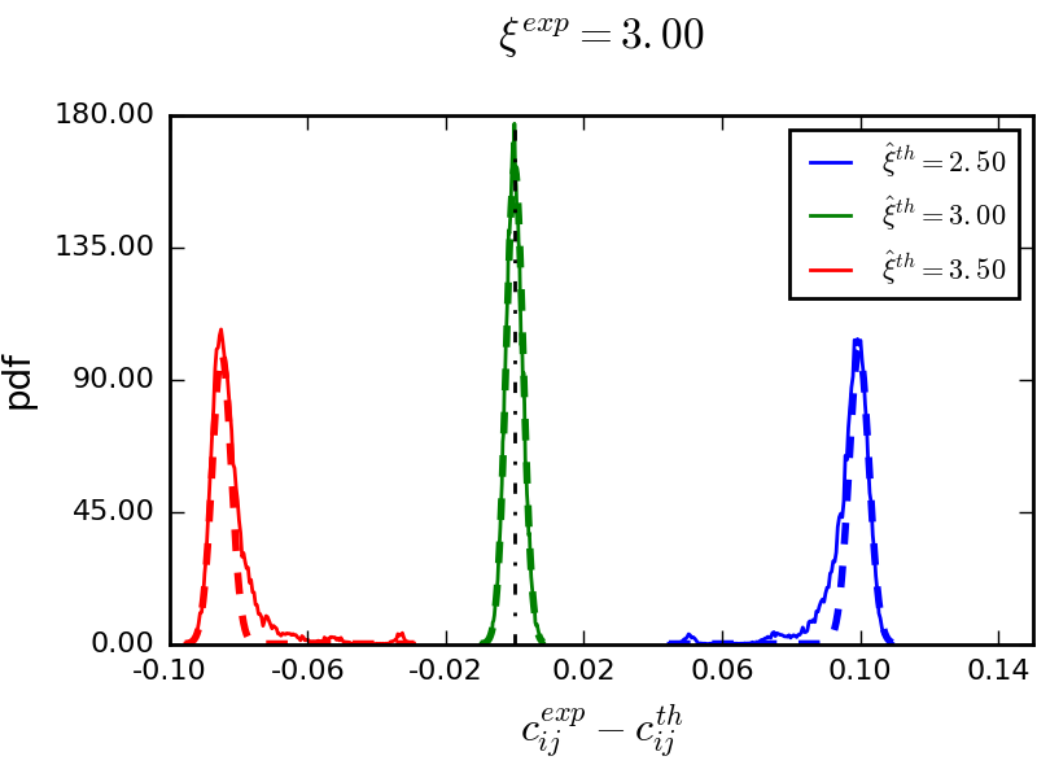}
  \caption{Distribution of the noise $\eta_{ij}=c_{ij}^{exp}-c_{ij}^{th}$, fitted to a Gaussian distribution.}
  \label{fig:gem_noise_gaussianity}
\end{figure}

Let us start from an artificial GEM with arbitrary couplings $K^{th} = [k_{ij}^{th}]$. We compute the associated contact matrix $[c_{ij}^{th}]$, using a threshold $\xi^{th}$ and a form factor $\mu^{th}$. When we perform Brownian Dynamics simulations of this system, we obtain configurations from which we compute the experimental contact matrix $c_{ij}^{exp}$, using a threshold $\xi^{exp}$ and a form factor $\mu^{exp}$. We take $\mu^{th}=\mu^{exp}$ as Gaussian form factors, and we chose $\xi^{exp}=3.00$ to compute the experimental contact probabilities from Brownian Dynamics trajectories. Thermal fluctuations, together with the finite number of such configurations obtained from Brownian Dynamics simulations results in $c_{ij}^{exp} \neq c_{ij}^{th}$. We may therefore write the experimental contact probabilities as:
\begin{equation}
  c_{ij}^{exp} = c_{ij}^{th} + \eta_{ij},
  \label{eq:gem_noise_etaij}
\end{equation}
where $\eta_{ij}$ is a noise with unknown distribution, corrupting the ``true'' contact probabilities. For a chain with $N+1=200$ monomers and $M=20$ non-zeros couplings drawn from a uniform distribution in the interval $[0,1]$, we computed the probability distribution function (pdf) of the difference $c_{ij}^{th} - c_{ij}^{exp}$. We tried different values for the threshold $\xi^{th}$ used in the GEM mapping (\cref{fig:gem_noise_gaussianity}) and obtained that when $\xi^{th}=\xi^{exp}$ the pdf of $\eta_{ij}$ fits well a centered Gaussian distribution.

Consequently, instead of running Brownian Dynamics simulations in order to compute experimental contact matrices $c_{ij}^{exp}$, we may construct pseudo-experimental contact matrices by adding a Gaussian noise with mean and variance given by
\begin{equation}
  \langle \eta_{ij} \rangle = 0, \qquad \langle \eta_{ij}^2 \rangle = \varepsilon^2,
  \label{eq:gem_noise_etaij_mean_var}
\end{equation}
to the theoretical contact matrix $c_{ij}^{th}$. This trick allows us to investigate the stability of the reconstruction method by direct mapping as a function of the noise amplitude $\varepsilon$. Furthermore, it also allows us to explore more values for $M$ than if we had to run systematically a Brownian Dynamics simulation.

Following this observation, we explored the stability of the reconstruction method by direct mapping in the $(\varepsilon,M)$ plane. We considered a large size of polymer with $N+1=1000$. For each value of $M$, we generated a random coupling matrix $k_{ij}^{th}$ by drawing $M$ random variables from a uniform distribution in the interval $[0,1]$ and computed the theoretical contact probabilities $c_{ij}^{th}$ of the corresponding GEM. Then we computed a pseudo-experimental contact probability matrix $c_{ij}^{exp}$ by adding to the theoretical contact probabilities a centered Gaussian noise with standard deviation $\varepsilon$. Following our previous observation, we assumed that the contact probabilities obtained are a good approximation for the experimental contact probabilities that would be obtained by performing a Brownian Dynamics simulation of the GEM. Then we applied the reconstruction procedure to $c_{ij}^{exp}$ using $\xi = \xi^{th}$. We therefore obtained a predicted GEM with couplings $[k_{ij}]$ that we compared to the theoretical couplings by computing the distance:
\begin{align}
  d(\hat{k}_{ij}, k_{ij}^{th}) =  \frac{1}{(N+1)} \left[ \sum \limits_{ij} (\hat{k}_{ij} - k_{ij}^{th})^2 \right]^{1/2},
\end{align}

\begin{figure}[!h]
  \centering
  \includegraphics[width = \linewidth]{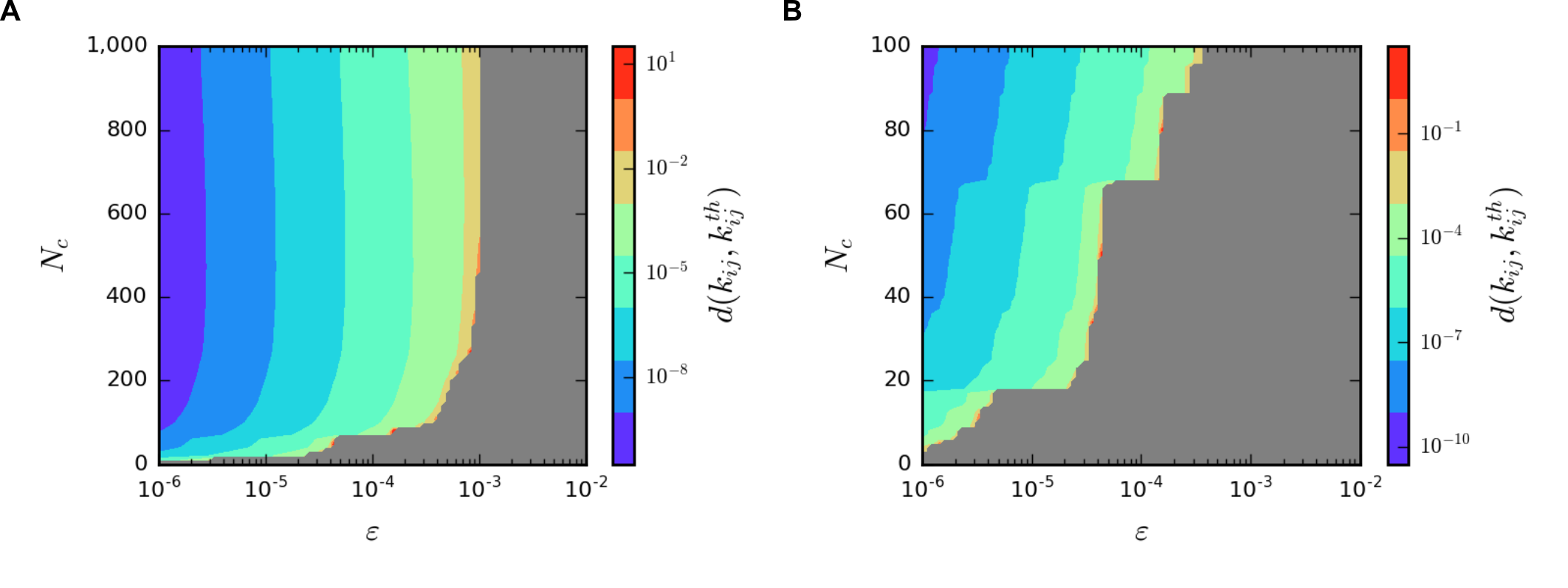}%
  \caption{Performance of the direct reconstruction method when the theoretical contact probabilities $c_{ij}^{th}$ are blurred with a Gaussian noise such that $\langle \eta_{ij} \rangle = 0$ and $\langle \eta_{ij}^2 \rangle = \varepsilon^2$. We used $N+1=1000$. The region in which the predicted couplings $[{k}_{ij}]$ define an unstable GEM was shaded in grey. \textbf{(A)} $M=0,\dots,1000$. \textbf{(B)} Zoom for $M=0,\dots,100$.}
  \label{fig:gem_noise_nc_eta_ij}
\end{figure}

The result of this analysis is shown in \cref{fig:gem_noise_nc_eta_ij}, in which we shaded in grey the region where the reconstructed couplings $[k_{ij}]$ define an unstable GEM with a correlation matrix $\Sigma$ having negative eigenvalues. We observe that for each value of the number of constraints, $M$, there is an upper bound $\overline{\varepsilon}$ on the noise amplitude such that for $\varepsilon > \overline{\varepsilon}$, the direct reconstruction method fails, in the sense that the predicted GEM is unstable. It is remarkable that for $\varepsilon < \overline{\varepsilon}$ the direct reconstruction methods perform very well, with $d(\hat{k}_{ij}, k_{ij}^{th}) \lesssim \num{e-2}$ in the worse cases. Therefore, the reconstruction by direct mapping appears to be robust to noise until some critical value of the noise amplitude is reached. Then the method suddenly starts to fail. We also note that the value of $\overline{\varepsilon}$ seems to depend on the number of constraints of the underlying GEM. In particular, it is clear that the performances of the direct reconstruction method get worse when $M \to 0$. Specifically, for $M=0$, we observe that even blurring the theoretical contacts with a noise of very small amplitude is sufficient to make the reconstruction fail. On the contrary, the value of $ \overline{\varepsilon}$ seems to be maximum for a number of constraints in a range between $M= \num{0.1} N$ and $M = N$.

\clearpage
\section{Reconstruction by LSE minimization}
\label{sec:lse_minimization}

\subsection{Steepest descent approach}
As emphasized in the main text, some coupling matrices can lead to an unstable GEM. More precisely, the covariance matrix $\Sigma$ has negative eigenvalues, so that it does not define a physically admissible model. In order to restrain our study to admissible GEMs, we have used a minimization scheme to find the admissible GEM reproducing as closely as possible an experimental contact probability matrix. The function to minimize is:
\begin{align}
  J(K) = \frac{1}{2} \Vert C(K) - E \Vert^2,
  \label{eq:cost_function_J}
\end{align}
where the matrix $C(K)=[c_{ij}]$ is the matrix of contact probabilities of the Gaussian effective model, and $E=[e_{ij}]$ is the matrix of experimental contact probabilities. The contact probability matrix $C(K)$ is a function of the matrix of couplings $K=[k_{ij}]$. Note that $C$, $E$ and $K$ are indexed with $0 \le i,j \le N$, \textit{i.e.} they are $(N+1)\times(N+1)$ matrices. Here, we used the Frobenius norm, such that for any matrix $A$, $\Vert A \Vert^2 = Tr(A^T A) = \sum \limits_{i,j} a_{ij}^2$.

\begin{figure}[htpb]
  \centering
  \includegraphics[width = 0.9 \linewidth]{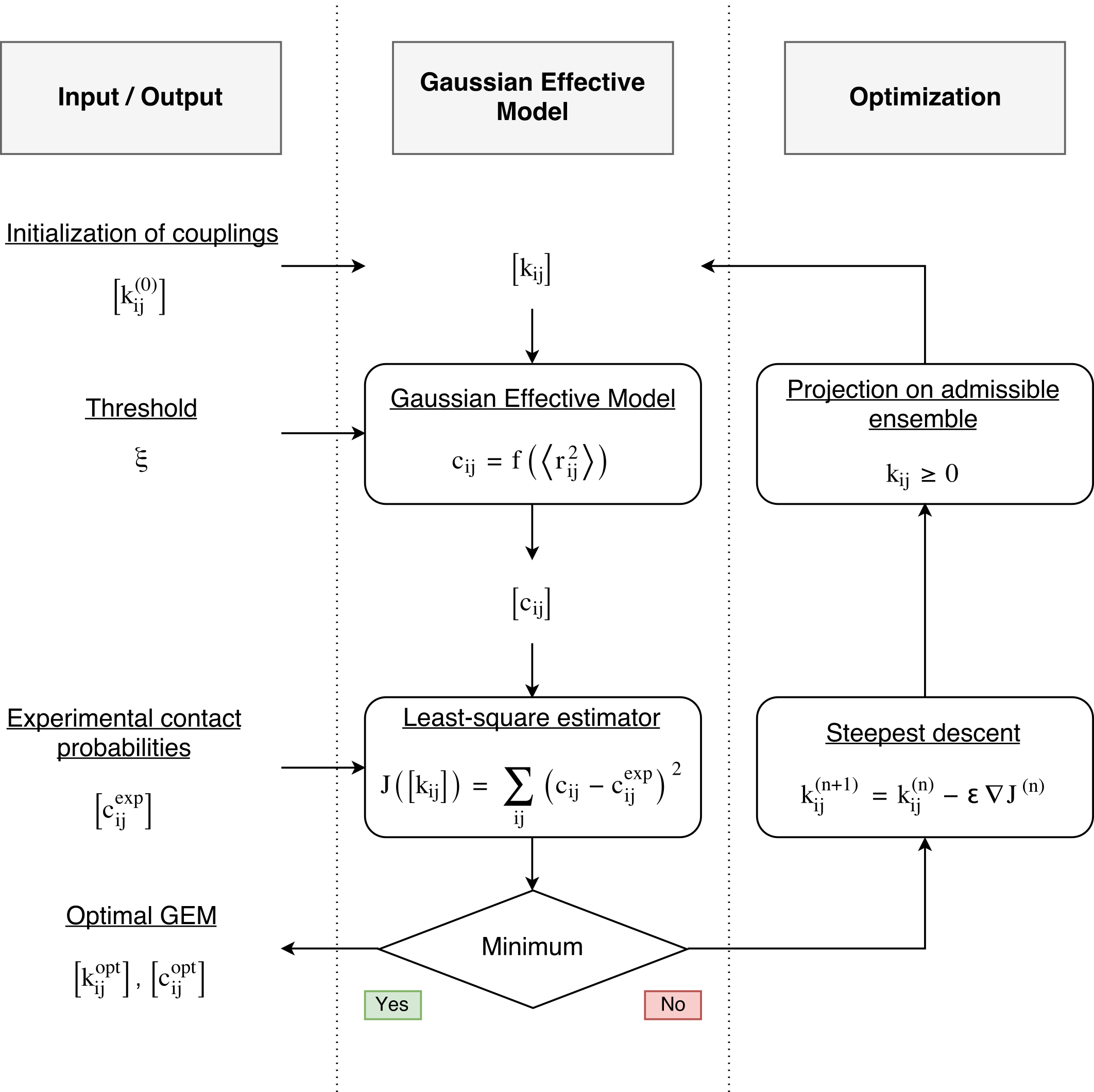}
  \caption{Minimization algorithm to find the Gaussian effective model with the closest contact probability matrix to an experimental contact probability matrix.}
  \label{fig:algorithm}
\end{figure}

In order to minimize $J$ as a function of $K$, under the constraint $K \ge 0$ (\textit{i.e.} all $k_{ij}$ are positive), we implemented a steepest descent method with projection (\cref{fig:algorithm}). At each iteration $n$, the matrix of couplings $K^n$ is updated according to:
\begin{align}
  K' &= K^n - h \frac{\nabla J^n}{\Vert \nabla J^n \Vert}, \label{eq:stp_descent_update} \\
  K^{n+1} &= p_{\mathbb{R}_+}(K'), \label{eq:stp_descent_projection}
\end{align}
where the scalar $h$ is a small time step, and the projection operator $p_{\mathbb{R}_{+}}$ applies the operation $x \leftarrow \max(x,0)$ to all entries of its matrix argument. In practise, the time step was adjusted at each iteration. Namely, if $J^{n+1} > J^n$, then we decreased the time step according to: $h \leftarrow \num{0.1} \times h$. Otherwise, we increased $h$ for the next iteration according to $h \leftarrow 2 \times h$.

We stopped the minimization when the relative variation in the cost function became sufficiently small:
\begin{equation}
  \frac{2|J^{n+1} - J^{n}|}{|J^{n+1}| + |J^{n}|} < \varepsilon_r,
  \label{eq:stp_descent_convergence}
\end{equation}
with typically $\varepsilon_r = \num{1e-9}$.

The minimization scheme that we just described requires to compute the gradient as a function of the $k_{ij}$ variables.

\subsection{Expression of the gradient of the least-square estimator}
We will express $J$ as the composition of several maps, and then use rules of differential calculus to find its differential form $dJ$. Since $J$ takes scalar values, we will then find its gradient as the matrix such that: $dJ_K(H) = Tr(\nabla J(K)^T H)$.

Let us first consider the matrix of reduced couplings $W=[w_{pq}]$, as defined in \cref{eq:gem_trid_redk_supp}, which is indexed with $1 \le p,q \le N$. We may introduce the linear map $\mathcal{A}$ which transforms a coupling matrix in its reduced coupling matrix:
\begin{align}
  \begin{array}{c c c c}
    \mathcal{A}: & \mathbb{R}^{N+1} \times \mathbb{R}^{N+1} & \to &  \mathbb{R}^{N} \times \mathbb{R}^{N} \\
	& K & \mapsto & \mathcal{A} (K) = W.
  \end{array}
\end{align}

Actually, the matrix elements of the reduced couplings can be expressed as:
\begin{align}
  w_{pq} = \sum \limits_{i,j} a_{pqij} k_{ij},
\end{align}
where:
\begin{align}
  a_{pqij} =
  \begin{cases}
    (1 - \delta_{pq}) \left[ - \frac{\delta_{pi} \delta_{qj} + \delta_{pj} \delta_{qi}}{2} \right] + \delta_{pq} \left[ \frac{\delta_{pi} + \delta_{pj}}{2} \right] & \text{if } i \neq j, \\
    0 & \text{otherwise.}
  \end{cases}
\end{align}

Here, $\delta_{pq}=1$ if $p=q$ and $\delta_{pq} = 0$ otherwise. The previous expression ensures that $W$ is a symmetrical matrix. The expression obtained suggests to introduce the tensor $\underline{A} = [a_{pqij}]$ and to use the matrix-vector notation:
\begin{equation}
  \mathcal{A}(K) = \underline{A} K,
\end{equation}
where $K$ is seen at a vector of $\mathbb{R}^{2(N+1)}$ and $\underline{A}$ as a matrix of $\mathbb{R}^{2(N+1)} \times \mathbb{R}^{2(N+1)}$. The differential of $\mathcal{A}$ is expressed as:
\begin{equation}
  \begin{array}{c c c c}
    d\mathcal{A}_K: & \mathbb{R}^{N+1} \times \mathbb{R}^{N+1} & \to &  \mathbb{R}^{N} \times \mathbb{R}^{N} \\
    & H & \mapsto & d\mathcal{A}_K (H) = \underline{A} H.
  \end{array}
\end{equation}

Actually, we may define the map associating to any coupling matrix the associated inverse covariance matrix $\Sigma^{-1} = \tilde{\mathcal{A}}(K) = \mathcal{A}(K) + T$ of a GEM, with $T$ as in \cref{eq:gem_trid_redk_supp}. It is straightforward that $d \tilde{\mathcal{A}} = d \mathcal{A}$.

Next, following \cref{eq:gem_inverse_correlation_app}, we can express the covariance matrix as $\Sigma = \mathcal{I}(W + T)$, where we introduced the inversion operator:
\begin{equation}
  \begin{array}{c c c c}
    \mathcal{I}: & \mathbb{R}^{N} \times \mathbb{R}^{N} & \to &  \mathbb{R}^{N} \times \mathbb{R}^{N} \\
    & X & \mapsto & \mathcal{I} (X) = X^{-1}.
  \end{array}
\end{equation}

The differential of $\mathcal{I}$ at the matrix $X$ is:
\begin{equation}
  \begin{array}{c c c c}
    d\mathcal{I}_X: & \mathbb{R}^{N} \times \mathbb{R}^{N} & \to &  \mathbb{R}^{N} \times \mathbb{R}^{N} \\
    & H & \mapsto & d\mathcal{I}_X (H) = - X^{-1} H X^{-1}.
  \end{array}
\end{equation}

Then, we introduce the matrix of mean square distances $\Gamma = [\gamma_{ij}]$ of a GEM, with $\gamma_{ij} = \langle r_{ij}^2 \rangle$, indexed with $0 \le i,j \le N$. By definition, it is related to the matrix of covariance $\Sigma=[\sigma_{pq}]$:
\begin{align}
  \begin{array}{cccc}
    \gamma_{ij} &=& \sigma_{ii} + \sigma_{jj} - 2 \sigma_{ij} & \text{ for } 0 < i,j \le N, \\
    \gamma_{0j} &=& \sigma_{jj} & \text{ for } 0 < j \le N.
  \end{array}
\end{align}

We now introduce the map:
\begin{equation}
  \begin{array}{c c c c}
    \mathcal{B}: & \mathbb{R}^{N} \times \mathbb{R}^{N} & \to &  \mathbb{R}^{N+1} \times \mathbb{R}^{N+1} \\
    & \Sigma & \mapsto & \mathcal{B}(\Sigma) = \Gamma.
  \end{array}
\end{equation}

Similarly as before, we may express this map in a matrix-vector notation, $\mathcal{B}(\Sigma) = \underline{B} \Sigma$, where the tensor $\underline{B}$ has the elements:
\begin{equation}
  b_{ijpq} = \left( \delta_{ip} + \delta_{jq} \right) \delta_{pq} - 2 \delta_{ip} \delta_{jq}.
\end{equation}

The differential of $\mathcal{B}$ in $\Sigma$ is then expressed as:
\begin{equation}
  \begin{array}{c c c c}
    d\mathcal{B}_\Sigma: & \mathbb{R}^{N} \times \mathbb{R}^{N} & \to &  \mathbb{R}^{N+1} \times \mathbb{R}^{N+1} \\
    & H & \mapsto & d\mathcal{B}_\Sigma (H) = \underline{B} H.
  \end{array}
\end{equation}

The final step of the Gaussian effective model mapping is to express the matrix of contact probabilities $C$ as a function of $\Gamma$. To this end, we introduce the map:
\begin{equation}
  \begin{array}{c c c c}
    \mathcal{F}: & \mathbb{R}^{N+1} \times \mathbb{R}^{N+1} & \to &  \mathbb{R}^{N+1} \times \mathbb{R}^{N+1} \\
    & \Gamma & \mapsto & \mathcal{F} (\Gamma) = C.
  \end{array}
\end{equation}

In the previous expression, the matrix elements of $C$ are given by:
\begin{equation}
  c_{ij} = F(\gamma_{ij}),
\end{equation}
where $F$ is one of \cref{eq:gem_contact_proba_gaussian,eq:gem_contact_proba_theta,eq:gem_contact_proba_exponential}, depending on the form factor used. We can then identify the differential of $\mathcal{F}$ by performing an expansion around $\Gamma$. We obtain:
\begin{equation}
  \begin{array}{c c c c}
    d\mathcal{F}_\Gamma: & \mathbb{R}^{N+1} \times \mathbb{R}^{N+1} & \to &  \mathbb{R}^{N+1} \times \mathbb{R}^{N+1} \\
    & H & \mapsto & d\mathcal{F}_\Gamma (H) = F'(\Gamma) \circ H,
  \end{array}
\end{equation}
where we introduced the Hadamard product such that for any two matrices $(A \circ B) = [ a_{ij} b_{ij}]$, and the short-hand notation $F'(\Gamma) = [F'(\gamma_{ij})]$.

Finally, we introduce the linear form:
\begin{equation}
  \begin{array}{c c c c}
    \mathcal{G}: & \mathbb{R}^{N+1} \times \mathbb{R}^{N+1} & \to &  \mathbb{R} \\
    & C & \mapsto & \mathcal{G} (C) = \frac{1}{2} \Vert C - E \Vert^2.
  \end{array}
\end{equation}

By definition of the Frobenius norm, $\Vert A \Vert^2 = Tr(A^T A)$, we obtain for the differential of $\mathcal{G}$ in $C$:
\begin{equation}
  \begin{array}{c c c c}
    d\mathcal{G}_C: & \mathbb{R}^{N+1} \times \mathbb{R}^{N+1} & \to &  \mathbb{R} \\
  & H & \mapsto & d\mathcal{G}_C (H) = Tr\left[\left( C -E \right)^T H \right].
  \end{array}
\end{equation}

In summary, we have introduced several maps and expressed the cost function to minimize as $J (K) = \mathcal{G} \circ \mathcal{F} \circ \mathcal{B} \circ \mathcal{I} \circ \tilde{\mathcal{A}}(K)$. Using the rules of composition for differential calculus, we obtain the differential of $J$ in $K$:
\begin{align}
  \begin{aligned}
    d J_K (H) &= d \mathcal{G}_{\mathcal{F} \circ \mathcal{B}\circ \mathcal{I}  \circ \tilde{\mathcal{A}}(K)}\circ d \mathcal{F}_{\mathcal{B}\circ \mathcal{I} \circ \tilde{\mathcal{A}}(K)} \circ d \mathcal{B}_{\mathcal{I} \circ \tilde{\mathcal{A}}(K)} \circ d \mathcal{I}_{\tilde{\mathcal{A}}(K)} \circ d \mathcal{A}_K (H), \\
    & = Tr\left[ \nabla J (K)^T H \right],
  \end{aligned}
\end{align}

After calculations, the gradient of $J$ in $K$ reads:
\begin{align}
  \begin{array}{c c c c}
    & \nabla J(K) &=& - \underline{A}^* \left[ (X^{-1})^T Y (X^{-1})^T \right], \\
	\text{with:} & & & \\
	& X & =& \underline{A} K + T \\
	& Y &=& \underline{B}^* \left[ (C - E) \circ F'(\Gamma) \right].
  \end{array}
\end{align}

To obtain the last expression, we introduced the adjoint tensors $\underline{A}^* = [a_{ijpq}^* = a_{pqij}]$ and $\underline{B}^* = [b_{pqij}^*=b_{ijpq}]$. Or writing explicitely all the summations we have:
\begin{align}
  \begin{array}{c c c c}
    & \frac{\mathrm{d}J}{\mathrm{d}k_{ij}} & = & - \sum \limits_{m,n=1}^{N} a_{mn,ij} \sum \limits_{p,q=1}^{N} x^{-1}_{pm} y_{pq} x^{-1}_{nq}, \\
	\text{with:} & & & \\
  & y_{pq} &=& \sum \limits_{k,l=0}^{N} b_{kl,pq} (c_{kl} - e_{kl}) F'(\gamma_{kl}).
  \end{array}
\end{align}

%We also used the following identities:
%\begin{align}
%  \begin{array}{c c c}
%    Tr\left(M^T (\underline{B} N)\right) & = & Tr\left( \underline{B}^* M \right)^T N) \\
%    Tr\left(M^T (N \circ O))\right) &=& Tr\left((M \circ N)^T O\right).
%  \end{array}
%\end{align}
%

\subsection{Computational burden}
The main computational burden in evaluating the cost function $J$ as well as its gradient $\nabla J$ resides in the matrix inversion $\Sigma^{-1} \to \Sigma$, with $O(N^3)$ complexity. In this work, we have used the routines of the Intel\textregistered Math Kernel Library to perform the algebra operations and the matrix inversion. We used the parallel implementation to distribute the computation over 12 processors.

As an alternative to the cost function in \cref{eq:cost_function_J}, we have also considered minimizing:
\begin{align}
  J' = \frac{1}{(N+1)^2} \sum_{i,j} \left( \sum_{k} \sigma_{ik}^{-1} s_{kj} - \delta_{ij} \right)^2,
  \label{eq:quadratic_lse}
\end{align}
where $[s_{ij}]$ is the covariance matrix of the GEM reproducing exactly the experimental contacts $[c_{ij}^{exp}]$, and $[\sigma_{ij}]$ is the covariance matrix of a candidate (stable) GEM with couplings $k_{ij}$. The advantage of this form over the previous one is that it does not require any matrix inversion. More accurately, it is a quadratic function of the $k_{ij}$ variables. Therefore the existence of a minimum satisfying $k_{ij} \ge 0$ is guaranteed and it is unique. Consequently, it is less computationally intensive and the minimum can be found efficiently with descent methods using conjugate directions. We found this form to work very well with contact probability matrix generated from predefined GEM by Brownian Dynamics simulations. However, for Hi-C contact probabilities, we found that it was much less successful. More precisely, the least-square estimator between the contact probabilities of the Hi-C experiment and of the optimal model was not as low.

\section{Brownian dynamics}
\label{sec:brownian_dynamics}

\subsection{Physical model}

In this article, we have performed two types of Brownian Dynamics simulation. The potentials used for each of them are summarized in the following table and discussed in further details below.

\begin{longtabu} to \linewidth {|X[1,l]|X[2,l]|X[2,l]|}
\hline
\rowfont{\bfseries} Potential & BD of GEM & BD of GEM with semi-flexibility and excluded volume \\
\hline
Chain structure & $U_e$ & $U_{fene}$ \\
\hline
GEM couplings & $U_I$ & $U_{I}$ \\
\hline
Bending rigidity & - & $U_{b}$ \\
\hline
Excluded Volume & - & $U_{ev}$ \\
\hhline{|=|=|=|}
Total & $U_e + U_I$ & $U_{fene} + U_b + U_{ev} + U_I$ \\
\hline
\end{longtabu}

\subsubsection{Chain structure}
We modeled the chromosome as a beads-on-string polymer with monomers of size $b$ and coordinates $\vec{r}_i$. The index varies between $i=0$ and $i=N$. The bond vectors are $\vec{u}_i = \vec{r}_i - \vec{r}_{i-1}$.

In the absence of excluded volume, we considered a Gaussian chain for the polymer structure, with potential:
\begin{equation}
  \beta U_e\left[ \lbrace \vec{r}_i \rbrace \right] = \frac{3}{2 b^2} \sum \limits_{i=1}^{N} (\vec{r}_{i} - \vec{r}_{i-1})^2.
\end{equation}

An important property of Gaussian chains is that the mean-square value of the end-to-end vector $\vec{R}_e = \vec{r}_N - \vec{r}_0$ scales linearly with the contour length:
\begin{equation}
  \langle R_e^2 \rangle = b^2 N.
\end{equation}

In reality, approximating a polymer to a Gaussian chain is only valid for weak perturbations, $R_e \ll N b$. Besides, a Gaussian polymer allows the bond distance to fluctuate quite a lot ($\langle u_i^2 \rangle = b^2$). This is problematic in Brownian Dynamics simulations with excluded volume interactions because this would result in possible crossings between different bonds. Therefore, for Brownian Dynamics with excluded volume interactions, we have preferred instead the finitely-extensible non-linear elastic potential (FENE):
\begin{equation}
  U_{fene} \left[ \lbrace \vec{r}_i \rbrace \right] = -\frac{3 k_{e} r_0^2}{2 b^2} \sum \limits_{i=1}^{N} \ln{\left(1 - \dfrac{u_i^2}{r_0^2} \right)},
\end{equation}
where $r_0$ is a distance above which non-linear effects start to appear in the bonds elasticity and $k_e$ is the rigidity constant of the non-linear spring. Note that for $u_i \ll r_0$ we recover the Gaussian chain potential, \textit{i.e.} a linear spring (with $k_e=1 \, k_\mathrm{B} T$). In practical applications we have taken $r_0=1.5 \, b$ and $k_{e}=10 \, k_\mathrm{B} T$ \cite{Kremer50571990}.

\subsubsection{Gaussian effective model interactions}
\begin{figure}%
  \centering%
  \includegraphics[width = \linewidth]{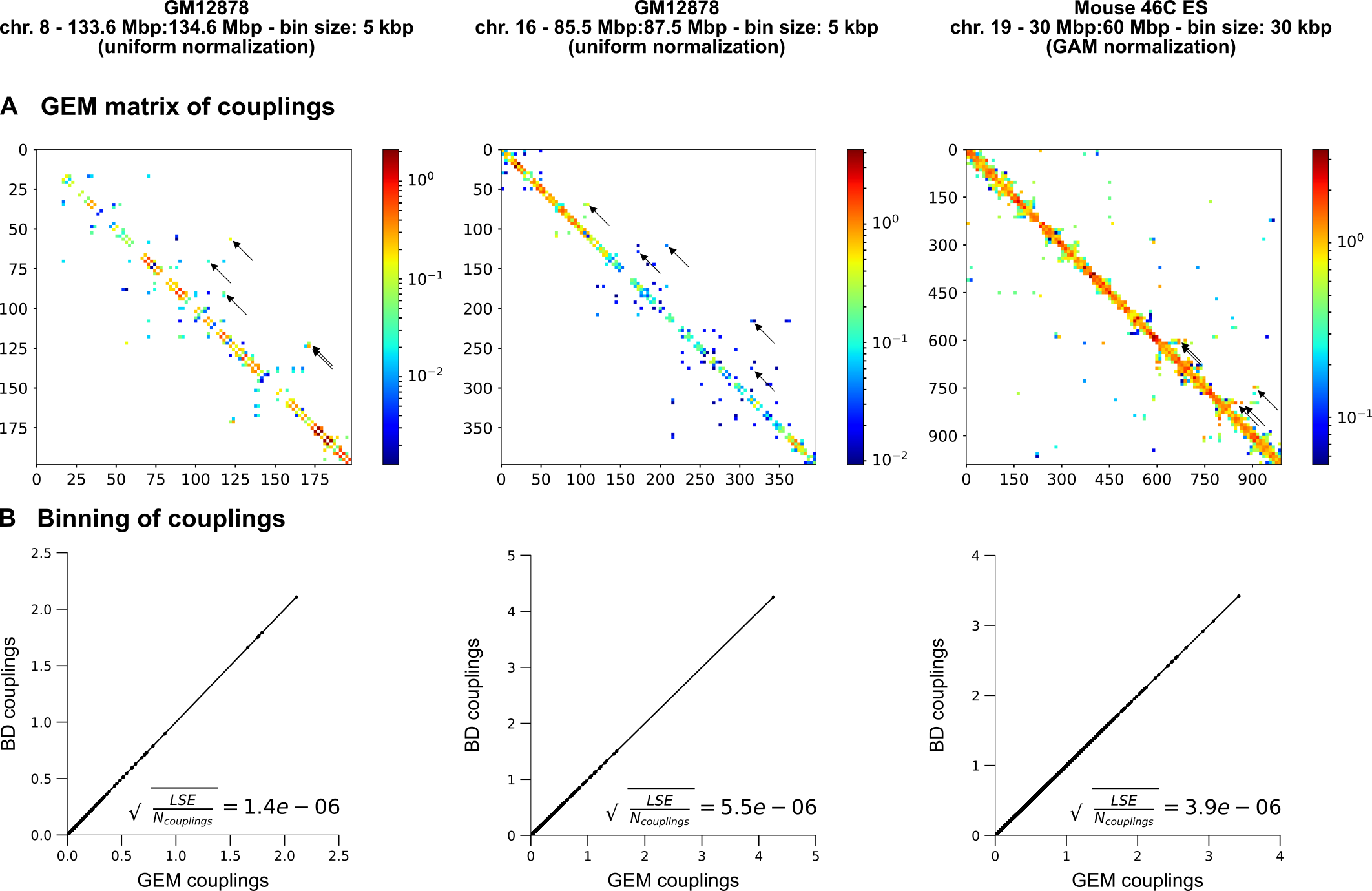}
  \caption{%
    Comparison of GEM couplings with binned couplings used in Brownian Dynamics (BD) simulations.}%
  \label{fig:couplings_gem_sim}%
\end{figure}

Following the model described in the main text, we introduced the GEM interaction potential:
\begin{equation}
  \beta U_I \left[ \left\{ \vec{r}_i \right\}  \right] = \frac{3}{2 b^2} \sum \limits_{0 \le i<j \le N} k_{ij} \left( \vec{r}_i -\vec{r}_j \right)^2,
\end{equation}
where the $k_{ij}$ are the couplings from a GEM. In order to have a reasonable amount of distinct couplings values in the implementation of BD simulations, we binned the GEM couplings. Specifically, we considered \num{1000} bins of same length in the interval $[k_{min},k_{max}]$ where $k_{min}$ (resp. $k_{max}$) is the minimum (resp. maximum) of the reconstructed GEM couplings. Note that we discarded all couplings $k_{ij}$ < \num{0.001}. Despite this binning procedure, the couplings used in the BD simulations remained very close to the reconstructed GEM ones (see \cref{fig:couplings_gem_sim}).

\subsubsection{Bending rigidity}
In reality, the DNA fiber opposes a certain resistance to bending. To model this effect, we used a Kratky-Porod potential:
\begin{equation}
  \beta U_{b}\left[ \left\{ \vec{r}_i \right\} \right] =  l_p \sum \limits_{i=1}^{N-1}\left( 1 -  \cos{\theta_i} \right),
\end{equation}
where $\theta_i$ is the angle between bonds $\vec{u}_i$ and $\vec{u}_{i+1}$.

For a polymer with a Gaussian chain potential plus a bending rigidity potential as defined above, the linear scaling of the mean-square end-to-end distance with the contour length still holds:
\begin{equation}
  \langle R_e^2 \rangle \approx l_K^2 \dfrac{N}{g_K},
\end{equation}
where $l_K = 2 l_p$ is the Kuhn length and $g_K$ is the number of original monomers per Kuhn length. Thus a semi-flexible polymer behaves like a Gaussian chain at large scales, with $N' = N /g_K$ and $b' = l_K$.

\subsubsection{Excluded volume}
A commonly used two-parameter empirical form for describing non-bonded interactions between two neutral (but possibly polarized) particles is the Lennard-Jones, or ``6-12'', potential. For two monomers separated by a distance $r$, it reads:
\begin{equation}
  V_{LJ}(r) = 4 \varepsilon \left( \left( \frac{\sigma}{r} \right)^{12} - \left( \frac{\sigma}{r} \right)^{6} \right),
\end{equation}
where $\varepsilon$ is an energy scale in $k_\mathrm{B} T$ and $\sigma$ is the hard core distance. Here, the interaction still decays as a power law of the distance $r$. A standard method to make this interaction short-range, is to introduce a threshold $r^{th}$ such that for distances $r>r^{th}$ the interaction vanishes. Therefore, in simulations, we have considered the truncated Lennard-Jones potential:
\begin{align}
  U_{ev}(r) =
  \left\lbrace
  \begin{aligned}
    &V_{LJ}(r) - V_{LJ}(r^{th}) & \text{ if } r<r^{th}, \\
    & 0  & \text{ otherwise.}
  \end{aligned}
  \right.
  \label{eq:polymer_lennard_jones_truncated}
\end{align}

We have considered take $\varepsilon = 1 \, k_\mathrm{B} T$, but the hard-core distance may be different from the monomer size (see next below). To model excluded volume interactions, we set $r^{th}=2^{1/6} \sigma$, resulting in $U_{ev}(r) > 0$ for $r<r^{th}$. In particular, this ensures that the repulsive force, $- \partial U_{ev} / \partial r$, vanishes precisely for $r=r^{th}$.

\subsubsection{Numerical values}
In eukaryotes, the interphase chromosome is packed into a fiber with a diameter of \SI{30}{\nm}, which is usually designated as chromatin. It has a linear packing fraction $\nu \approx \SI{100}{bp . \nm^{-1}}$ and persistence length $l_p = \SI{90}{nm}$ \cite{Langowski2412006}. Therefore, the appropriate size for monomers is $\sigma = \SI{30}{nm}$, which correspond to $g = \SI{3000}{bp}$. The persistence length expressed in units of these monomers gives $l_p = 3 \sigma$, and $\sigma$ is also the hard-core distance for excluded volume interactions between monomers.

In the Brownian Dynamics simulations performed in this article, the natural unit of monomer is the Hi-C bin resolution. We have considered specifically $g_1 = \SI{5000}{bp}$ and $g_2 = \SI{30000}{bp}$ with corresponding monomer sizes $b_1$ and $b_2$. The persistence lengths for each case thus read $l_p = \nu l_p / g_1 \, b_1 = 1.8 \, b_1$ and $l_p = \nu l_p / g_2 \, b_2 = 0.3 \, b_2$.

For the first resolution, we may consider that $g_1 \approx g$, meaning that monomers can be represented as impenetrable beads. We thus take for the hard-core distance $\sigma_1 = b_1$. The second resolution however defines monomers much larger than the chromatin fiber diameter. Following the scaling relations introduced above, we may express the monomer sizes as:
\begin{equation}
    b_2^2 \approx \dfrac{g_2}{g_K} l_K^2,
\end{equation}
where $g_K = \SI{18000}{bp}$ is the number of monomers per Kuhn length. We obtain that $b_2 \approx 8 \sigma$. Therefore, we have considered a hard-core distance $\sigma_2 = 0.125 \, b_2$.

We summarize in the following table the values of the different parameters we took for our Brownian Dynamics simulations.

\needspace{5\baselineskip}
\begin{longtabu} to \linewidth {|X[l]|X[l]|X[l]|X[l]|}
\hline
\textbf{Data set} & \makecell[tl]{GM12878\\chromosome 8\\133.6 Mbp:134.6 Mbp\\bin size: 5 kbp\\uniform normalization} & \makecell[tl]{GM12878\\chromosome 16\\85.5 Mbp:87.5 Mbp\\bin size: 5 kbp\\uniform normalization} & \makecell[tl]{Mouse 46C ES\\chromosome 19\\30 Mbp:60 Mbp\\bin size: 30 kbp\\GAM normalization} \\
\hline
\textbf{Gaussian chain} & $b=\num{1}$ & $b=\num{1}$ & $b=\num{1}$ \\
\hline
\textbf{FENE chain} & \makecell[tl]{$b=\num{1}$\\$r_0=\num{1.5}$\\$k_e = \SI{10}{k_\mathrm{B} T}$} & \makecell[tl]{$b=\num{1}$\\$r_0=\num{1.5}$\\$k_e = \SI{10}{k_\mathrm{B} T}$} & \makecell[tl]{$b=\num{8}$\\$r_0=\num{12}$\\$k_e = \SI{10}{k_\mathrm{B} T}$} \\
\hline
\textbf{Bending rigidity} & $l_p=\num{1.8}$ & $l_p=\num{1.8}$ & $l_p=\num{2.4}$ \\
\hline
\textbf{Excluded volume} & \makecell[tl]{$\sigma=\num{1}$\\$r_c=\num{1.1225}$} & \makecell[tl]{$\sigma=\num{1}$\\$r_c=\num{1.1225}$} & \makecell[tl]{$\sigma=\num{1}$\\$r_c=\num{1.1225}$} \\
\hline
\textbf{GEM couplings} & \makecell[tl]{\num{1000} equal sized bins\\$\min{(k_{ij})} \ge \num{e-3}$} & \makecell[tl]{\num{1000} equal sized bins\\$\min{(k_{ij})} \ge \num{e-3}$} & \makecell[tl]{\num{1000} equal sized bins\\$\min{(k_{ij})} \ge \num{e-3}$} \\
\hline
\end{longtabu}

\subsection{Implementation of Brownian Dynamics}
Brownian dynamics simulations are molecular dynamics simulations in which many molecular details are coarse-grained. The classical framework to describe the Brownian motion of a particle is the Langevin equation. For a bead with coordinates $x(t)$ it reads:
\begin{equation}
  m \ddot{x}(t) = - \gamma \dot{x} - \frac{\partial U}{\partial x}(x(t)) + \gamma \eta(t),
  \label{eq:brownian_motion_langevin_equation}
\end{equation}
in which $m$ is the mass of the bead, $\gamma$ is a damping term and $- \partial U / \partial x$ is the force applied to the bead, deriving from a potential $U$. The first two terms in the right-hand side of the above dynamics are deterministic. In addition there is a stochastic term, $\eta(t)$ which represents energy exchanges between the bead and a bath at temperature $T$. More accurately, $\eta$ is an uncorrelated Gaussian random process with two first moments:
\begin{equation}
  \langle \eta(t) \rangle = 0, \qquad \langle \eta(t) \eta(t') \rangle = 2 D \delta(t-t'),
  \label{eq:brownian_motion_noise_moments}
\end{equation}
where $D$ is the diffusion coefficient of the bead. It can be shown that the above dynamics converges to the Boltzmann equilibrium provided that $D$ satisfies the Stokes-Einstein relation:
\begin{equation}
  D=k_\mathrm{B} T / \gamma,
  \label{eq:stokes_einstein}
\end{equation}
where finally from the Stokes' law applied to a bead of diameter $b$ we get $\gamma=3 \pi b \mu$, with $\mu$ being the fluid viscosity.

In order to produce Brownian Dynamics trajectories, the Langevin equation \cref{eq:brownian_motion_langevin_equation} was applied to each bead of our polymer model and integrated numerically with the LAMMPS simulation package \cite{Plimpton19951}. It uses a standard velocity Verlet integration scheme \cite{NumericalRecipes2007}. In practise, this requires the choice of an integration time step, and we chose the value $dt=\num{0.001}$. We also set $\gamma=1$ (in simulation dimensionless units).

The choice of the initial configuration is important, especially when excluded volume is included. Although we can start from an arbitrary configuration respecting excluded volume constraints, the relaxation to the Boltzmann equilibrium can be very slow. To circumvent this problem and generate quickly an initial configuration for a polymer with excluded volume interactions we have used the following procedure.

First, perform a relaxation run without excluded volume nor short-range attractive interactions. This corresponds to the dynamics of an ideal chain and aims at sampling rapidly a large number of configurations to loose the memory of the initial condition.

Second, perform an intermediate run with few iterations (generally \num{e6} iterations) with a soft pair potential:
\begin{equation}
  U_{soft}(r)=A\left( 1 + \cos{\left( \frac{\pi r}{r^{th}} \right)} \right),
  \label{eq:polymer_soft_pair}
\end{equation}
where $r^{th}$ is the same cutoff as in the truncated Lennard-Jones potential from \cref{eq:polymer_lennard_jones_truncated}. The magnitude $A$ is progressively increased from 1 to 60 during the run \cite{Kremer50571990}, so that we obtain in the end a configuration with no overlaps between the beads.

Third, the main run with excluded volume and short-range interactions is performed starting from the configuration without overlaps. Several configurations (generally \num{1000}) are extracted from the resulting trajectory, which sample the Boltzmann ensemble. These configurations can be used to compute equilibrium averages according to the ergodic property of the Boltzmann equilibrium.

It is possible to map the simulation time to the real time. Let us write the diffusion coefficient as $D=b^2 / \tau_B$. During the time $\tau_B$, a bead typically travels through a distance $b$, which is its own size. Consequently $\tau_B$ is the natural unit of time for this diffusive process and is called the Brownian time. In Brownian Dynamics simulations we take $b=1$ and $D=1$ (in dimensionless units), therefore a unit of simulation time correspond to the Brownian time. The diffusion coefficient in the bacterial nucleoid was found to be $D=\SI{10}{\mu \meter^{2} . \second^{-1}}$ \cite{Elowitz1999}. Therefore, for $b=\SI{30}{\nm}$ we find $\tau_B=\SI{90}{\mu \second}$.

% vim: set sw=2 expandtab tabstop=2 foldcolumn=4:
% vim: set spell spelllang=en_us,en_gb:

\newpage
%\printbibliography[heading=bibintoc, title=Supplementary References, section=2, keyword=supplementary]
\printbibliography[heading=bibintoc, title=Supplementary References, segment=2, keyword=supplementary]

\end{document}